\documentclass[aps,twocolumn,prc,superscriptaddress,showpacs,nofootinbib,floatfix,amssymb,amsfonts,amsmath]{revtex4-1}

\usepackage{graphicx}
\usepackage{dcolumn}
\usepackage{bm}

\usepackage{amsmath}    
\usepackage{amsfonts}   
\usepackage{amssymb}
\usepackage{graphicx}   

\begin{document}

\title{From superdeformation to extreme deformation and clusterization
in the $N\sim Z$ nuclei of the $A\sim 40$ mass region.}

\author{D.\ Ray}
\affiliation{Department of Physics and Astronomy, Mississippi
State University, MS 39762}

\author{A.\ V.\ Afanasjev}
\affiliation{Department of Physics and Astronomy, Mississippi
State University, MS 39762}

\date{\today}

\begin{abstract}

 A systematic search for extremely deformed structures in the 
$N\sim Z$ nuclei of the $A\sim 40$ mass region has been performed
for the first time in the framework of covariant density functional
theory. At spin zero such structures are located at high excitation 
energies which prevents their experimental observation. The rotation
acts as a tool to bring these exotic shapes to the yrast line or its 
vicinity so that their observation could become possible with future 
generation of $\gamma-$tracking (or similar) detectors such as 
GRETA and AGATA. The major physical observables of such structures 
(such as transition quadrupole moments as well as kinematic and dynamic 
moments of inertia), the underlying single-particle structure and the 
spins at which they become yrast or near yrast are defined. The search 
for the fingerprints of clusterization and molecular structures is 
performed and the configurations with such features are discussed.
The best candidates for observation of extremely deformed 
structures are identified. For several nuclei in this study (such as 
$^{36}$Ar), the addition of several spin units above currently measured 
maximum spin of $16\hbar$ will inevitably trigger the transition to 
hyper- and megadeformed nuclear shapes.

\end{abstract}


\maketitle

\section{Introduction}
 
  There is a considerable interest to the study of cluster structures
and extremely deformed shapes in light nuclei 
\cite{KK.05,OFE.06,MKKRHT.06,40Ca-AMD.07,RMUO.11,EKNV.12,S-34-mol.14,ZIM.15,IIIMO.15,J.clust.16}. 
Many of these structures are described in terms of clusters, the simplest one being the 
$\alpha$-particle \cite{OFE.06,MKKRHT.06}. Providing a unique insight on the
cluster dynamics inside of nucleus, the initial assumptions about clusters 
represent a limitation of this type of models. Note also that many shell model 
configurations are beyond the reach of the cluster models. It is also important to
remember that the cluster description does not correspond to clearly 
separated $\alpha$-particles, but generates the mean-field states largely 
by antisymmetrization \cite{MKKRHT.06}. In addition, the studies of 
molecular structures, which appear in many extremely deformed configurations, 
have gained considerable interest \cite{MH.04,OFE.06,S-34-mol.14}.

 In recent years, the investigations of exotic cluster configurations have been
undertaken also in the density functional theory (DFT). The advantage of the DFT
framework is the fact that it does not assume the existence of cluster structures; 
the formation of cluster structures proceeds from microscopic single-nucleon degrees 
of freedom via many-body correlations  \cite{EKNV.12,EKNV.14}. As a result, the DFT 
framework allows simultaneous treatment of cluster and mean-field-type states 
\cite{ER.04,RMUO.11,EKNV.12,EKNV.14,YIM.14}. It is important to mention that covariant
(relativistic) energy density functionals (CEDFs) show more pronounced clusterization
of the density distribution as compared with non-relativistic ones because of deeper
single-nucleon potentials \cite{EKNV.12}.

  Let us mention some recent studies of cluster and extremely deformed structures 
in the DFT framework. The  clustering  phenomenon  in  light  stable and  exotic  nuclei  
was  studied  within  the  relativistic  mean  field (RMF)  approach in Ref.\ 
\cite{ASPG.05} and within the Hartree-Fock (HF) approach based on  the Skyrme energy
density functionals (EDF) in 
Ref.\ \cite{RMUO.11}. Linear chain configurations of four $\alpha$-clusters in $^{16}$O 
and the relationship between the stability of such states  and angular momentum were 
investigated using Skyrme cranked HF method in Ref.\  \cite{IMIO.11} and cranked RMF 
(further CRMF) in Ref.\ \cite{YIM.14}. This is an example of the ``rod shaped'' nucleus. 
Another case  of such structures is linear chain of three  $\alpha$ clusters, suggested about 
60 years ago \cite{Mor.56}; it was recently studied in the  CRMF theory in Ref.\ 
\cite{ZIM.15}.  This exotic structure (``Hoyle'' state) plays a crucial role in the 
synthesis of $^{12}$C from three $^{4}$He nuclei in stars \cite{Apj.54}.  The stability 
of rod-shaped structures in highly-excited states of $^{24}$Mg was studied in Ref.\ 
\cite{IIIMO.15} in cranked Skyrme HF calculations.

 The difficulty in investigating cluster and extremely deformed states is that 
they are generally unbound and lie at high excitation energies at low spins
\cite{OFE.06,J.clust.16}. Moreover, they are either formed on the shoulder or in
very shallow minima of potential energy surfaces \cite{EKNV.14,AA.08}; thus, they 
are inherently unstable at low spin. The high density of nucleonic configurations at 
these energies and possible mixing among them is another factor hindering their 
observation with current and future
generations of experimental facilities. Moreover, obtaining unambiguous evidences 
for clustering (such as a transition strengths between different states and the 
structure of the wavefunction) is equally challenging and frequently ambiguous from 
experimental point of view. In addition, the mechanisms of the reactions used in 
experimental studies frequently favor the population of yrast or near-yrast states 
\cite{J.clust.16}.

\begin{figure*}[ht]
\includegraphics[angle=0,width=14.0cm]{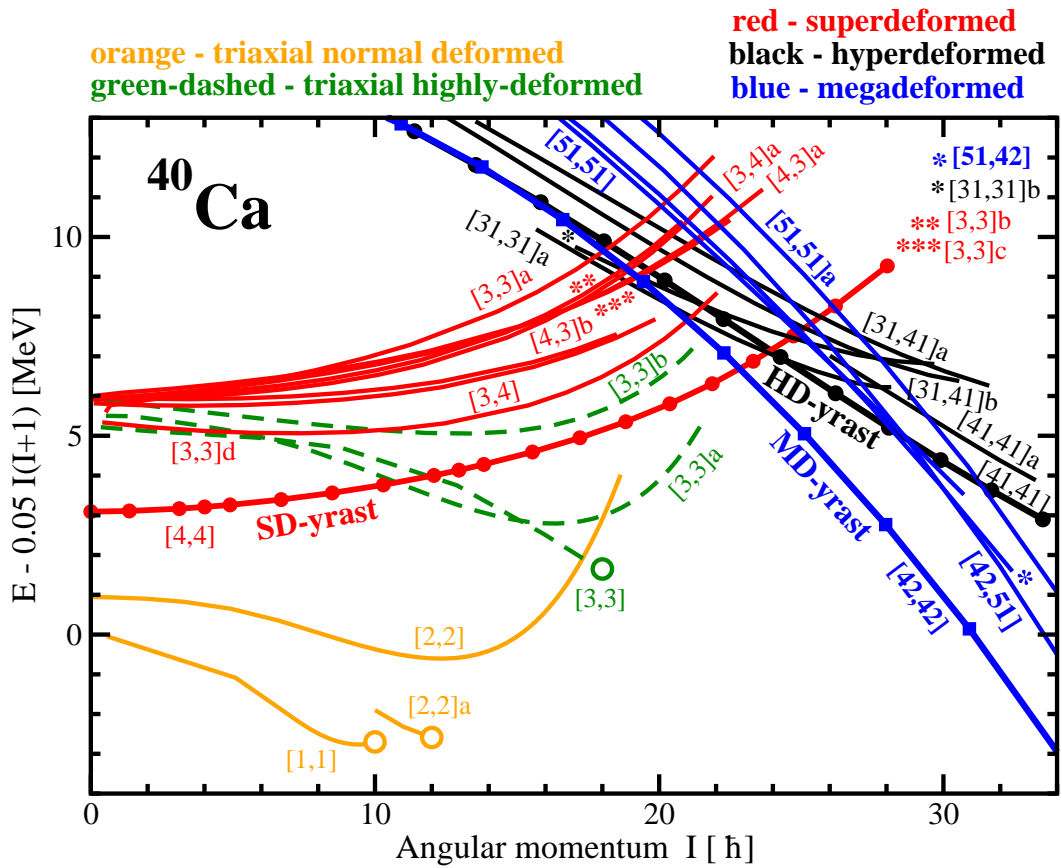}
\caption{(Color online) Energies of the calculated configurations in $^{40}$Ca
relative to a smooth liquid drop reference $AI(I+1)$, with the inertia parameter 
$A=0.05$. This way of the presentation of the results has clear advantages as
compared with the energy versus spin plots, see Sec.\ 4.1 in Ref.\ \cite{PhysRep-SBT}
for details. Different types of configurations are shown by different types of lines. 
The SD, HD and MD configurations, which are yrast in respective deformation 
minima, are shown by thick lines with symbols.}
\label{Ca40-eld}
\end{figure*}

  The rotation of the nucleus could help to overcome these problems in experimental
observation of extremely deformed structures. Two factors are contributing to that.
First, very large deformation configurations (such as super- (SD), hyper- (HD) and 
megadeformed (MD) ones) are favored by rotation at high spins (see, for example, 
the discussion in Refs.\ \cite{DPSD.04,AA.08}). Second, normal- and highly-deformed 
configurations, which are forming yrast or near-yrast structures at low and medium 
spins, have limited angular momentum content. As a consequence, only extreme deformation 
structures (SD, HD, or MD) could be populated above some specific spin values in the 
nuclei of the interest.

  Our systematic search for extremely deformed configurations is focused 
on the $N=Z$ and $N=Z+2$ even-even S ($Z=16$), Ar ($Z=18$), Ca $(Z=40)$, 
Ti $(Z=42)$, Cr $(Z=44)$ (and also on $N=Z+4$ $^{44}$Ca) and odd-odd $N=Z=21$ 
$^{42}$Sc nuclei. The selection of the nuclei is motivated by several factors.
First, the $^{40}$Ca nucleus is a centerpiece of this study because of its 
highly unusual features. This is doubly magic spherical nucleus (in the ground
state) in which normal- and superdeformed configurations based on the 4 particle
- 4 hole (4p-4h) and 8 particle - 8 hole (8p-8h) excitations, respectively, are 
observed at low excitation energies (Ref.\ \cite{Ca40-PRL.01}). Two-dimensional 
alpha cluster model predicts very exotic and highly-deformed configurations in
this nucleus \cite{ZR.93}. Second, in 
this mass region the superdeformation has already been observed in the $^{40}$Ca 
(\cite{Ca40-PRL.01,Ca40-Qt.03}), $^{36}$Ar (\cite{36Ar-SD.PRL.00,Ar36-SD-Qt.01}), 
$^{35}$Cl (\cite{35Cl-SD.13}), $^{40}$Ar (\cite{40Ar-SD.10}) and probably 
$^{28}$Si \cite{Si28.SD.12} nuclei. Moreover, the SD bands have been seen up to 
very high spin of $I=16\hbar$ in some of these nuclei. This is quite important 
fact because according to the results obtained in the present paper further 
modest increase of the spin could lead to the population of extremely deformed 
structures in some of the  nuclei. When populated such structures could be 
observed with the next generation of the $\gamma$-tracking detectors such as 
GRETA and AGATA.
In addition,  the resonance observed at $I\sim 36\hbar$ in the 
$^{24}$Mg+$^{24}$Mg reaction strongly supports the HD shape for a 
compound $^{48}$Cr nucleus formed in this reaction \cite{48Cr.HD.16}.
Moreover, the analysis of light particle energy spectra and
angular correlations in the framework of the statistical model
indicates the onset of large deformations at
high spin in $^{44}$Ti \cite{44Ti.HD-CN.03}.
Third, in experiment the high spin structures in this mass 
region are  better  populated in the $N\sim Z$ nuclei. Note also 
that at present high spin studies are quite active in this mass region 
\cite{Sc42-42.07,Cl-34.14,S-35.14,S-33.14,48Cr.HD.16}.

\begin{figure}[ht]
\includegraphics[width=8.8cm,angle=0]{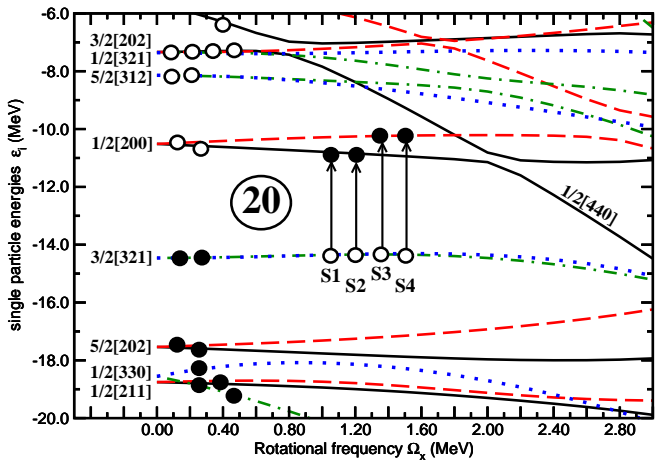}
\caption{(Color online) Neutron single-particle energies (routhians) in the 
self-consistent rotating potential as a function of the rotational frequency 
$\Omega_x$. They are given along the deformation path of the yrast SD configuration  
in $^{40}$Ca. Long-dashed, solid, dot-dashed and dotted lines indicate  
$(\pi=+, r=+i)$, $(\pi=+, r=-i)$, $(\pi=-, r=+i)$ and $(\pi=-, r=-i)$ orbitals, 
respectively.  At $\Omega_x=0.0$ MeV, the single-particle orbitals are labeled by 
the asymptotic quantum numbers $[Nn_z\Lambda]\Omega$ (Nilsson quantum numbers) of 
the dominant component of the wave function. Solid (open) circles indicate the 
orbitals occupied (emptied). The arrows indicate the particle-hole excitations 
leading to excited SD configurations; for these configurations only the changes
(as compared with yrast SD configuration) in the occupation of the orbitals 
are indicated in the figure.}
\label{routh-sd-1}
\end{figure}

\begin{figure}[ht]
\includegraphics[width=8.8cm,angle=0]{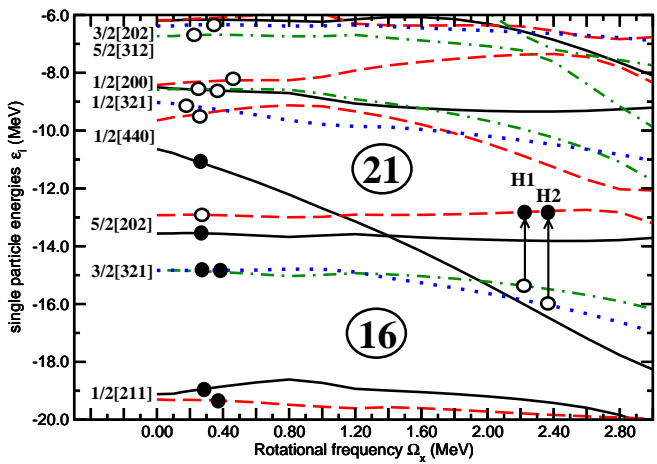}
\caption{(Color online) The same as Fig. \ref{routh-sd-1} but along the 
deformation path of the yrast HD configuration in $^{40}$Ca.  The arrows 
indicate the particle-hole excitations leading to excited HD configurations.}
\label{routh-hd-ca40}
\end{figure}

 There are theoretical studies of deformed structures in these nuclei but they 
either focus on SD structures at high spin or are limited to a shape coexistence
at low spin. For example,  positive-parity states of $^{40}$Ca were studied in 
Ref.\ \cite{40Ca-AMD.07} using antisymmetrized molecular dynamics (AMD) and the 
generator coordinate method (GCM); this is basically alpha-clustering model. The 
coexistence of low-spin normal- and superdeformed states in $^{32}$S, $^{36}$Ar, 
$^{38}$Ar and $^{40}$Ca has been studied in the GCM based on the Skyrme SLy6 
functional in Ref.\ \cite{BFH.03}. The SD and HD rotational bands in the $N=Z$ S, 
Ar, Ca, Ti and Cr nuclei have also been studied in cranked Hartree-Fock (CHF) 
approach based on the Skyrme forces in Refs.\ \cite{YM.00,IMYM.02}. A special 
attention has been paid to the SD structures in $^{32}$S which were studied 
in detail in the CHF frameworks based on Skyrme \cite{YM.00,MDD.00} and Gogny 
\cite{RER.00} forces and cranked relativistic mean field (CRMF) theory 
\cite{Pingst-A30-60}. Exotic and highly-deformed $\alpha$-cluster configurations 
have been predicted long time ago in two-dimensional $\alpha$-cluster model in 
4$N$ nuclei from $^{12}$C to $^{44}$Ti in Ref.\ \cite{ZR.93}. The investigation of 
superdeformation and clustering in these nuclei still remains an active field of 
research within the cluster models (see Refs.\ 
\cite{KK.05,MKKRHT.06,40Ca-AMD.07,40Ar.10,S-34-mol.14}).

  There are several goals behind this study. First, it is imperative to 
understand at which spins extremely deformed configurations are expected 
to become yrast (or come close to the vicinity of the yrast line) and to find 
the best candidates for experimental studies of such structures. This 
requires detailed knowledge of terminating configurations up to their 
terminating states since they form the yrast line at low and medium spins. 
However, the tracing of terminating configurations from low spin up to their 
terminating states is non-trivial problem in density functional theories 
(see Sec.\ 8 in Ref.\ \cite{VALR.05} and Ref.\ \cite{A.08}). To our knowledge, 
such calculations have been done so far only in few nuclei: $^{20}$Ne (in the cranked Skyrme 
HF \cite{FHKW.82} and CRMF \cite{KR.89,TO-rot} frameworks),  $^{48}$Cr 
(in the HFB framework with Gogny forces \cite{CEMPRRZ.95}) and $^{109}$Sb 
(in the CRMF framework \cite{VALR.05}). Note also that in $^{109}$Sb they fail 
to reach the terminating state. With appropriate improvements in the CRMF computer 
code we are able to perform such calculations for the majority  of the 
configurations forming the yrast line at low and medium spins. Second, the
basic properties (such as transition quadrupole moments, dynamic and 
kinematic moments of inertia) of the configurations of interest, which 
could be compared in future with experimental data, are predicted. Third, 
we search for the fingerprints of the clusterization and molecular structures 
via a detailed analysis of the density distributions of the configurations 
under study.

  The paper is organized as follows. Section \ref{sec-theory} 
describes the details of the solutions of the cranked relativistic 
mean field equations. Detailed analysis of the structure of rotational 
spectra of $^{40}$Ca and $^{42}$Sc is presented in Secs.\ \ref{sec-ca40} 
and \ref{sec-sc42}, respectively. 
A special attention is paid to the dependence of
density distributions on the configuration. 
The general features of 
rotational spectra along the yrast line are discussed in Sec.\ 
\ref{sec-general}. Section \ref{sec-other} is devoted to the discussion 
of the appearance of super-, hyper- and megadeformed configurations 
along the yrast line of the $^{32,34}$S, $^{36,38}$Ar, $^{42,44}$Ca, 
$^{44,46}$Ti and $^{48,50}$Cr nuclei and their properties. The 
configurations which reveal the fingerprints of clusterization
and molecular structures in their density distributions are discussed 
in Sec.\ \ref{sec-clus}. The kinematic and dynamic moments of inertia of 
selected SD, HD and MD configurations and their evolution with proton 
and neutron numbers and rotational frequency are considered in Sec.\ 
\ref{sec-j2j1}. Finally,  Section \ref{concl} summarizes the results 
of our work.

\begin{figure}[ht]
\includegraphics[width=8.8cm,angle=0]{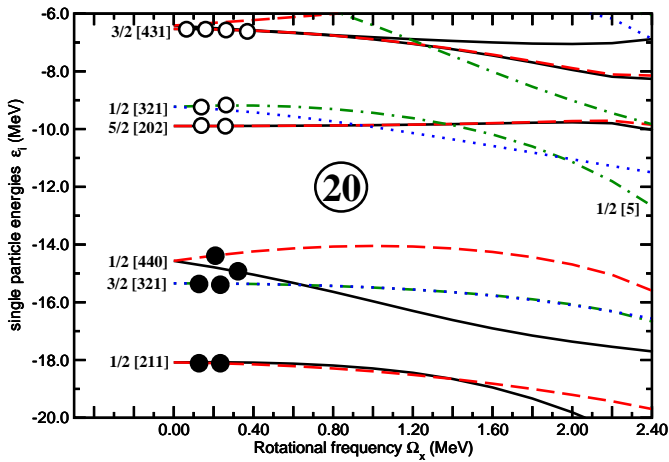}
\caption{(Color online) The same as Fig. \ref{routh-sd-1} but 
along the deformation path of the yrast MD [42,42] configuration 
in $^{40}$Ca.}
\label{routh-md-40ca}
\end{figure}

\section{The details of the theoretical calculations}
\label{sec-theory}

 In the relativistic mean-field (RMF) theory the nucleus is described 
as a system of pointlike nucleons, Dirac spinors, coupled to mesons and to 
the photons \cite{SW.86,Rei.89,VALR.05}. The nucleons interact by the 
exchange of several mesons, namely a scalar meson $\sigma$ and three vector 
particles, $\omega$, $\rho$ and the photon. The CRMF theory 
\cite{KR.89,KR.93,AKR.96} is the extension of the RMF theory 
to the rotating frame in one-dimensional cranking approximation. 
It represents the realization of covariant density functional theory 
(CDFT) for rotating nuclei with no pairing correlations \cite{VALR.05}.
 It has 
successfully been tested in a systematic way on the properties of different 
types of rotational bands in the regime of weak pairing such as normal-deformed
\cite{AF.05}, superdeformed \cite{A60,AKR.96}, as well as smooth terminating
bands \cite{VALR.05} and the bands at the extremes of angular momentum
\cite{ASN.12}.

 The formalism and the applications of the CRMF theory to the description 
of rotating nuclei have recently been reviewed in Ref.\ \cite{A-rev.15} (see 
also Refs.\ \cite{VALR.05,AO.13}). A clear advantage of the CRMF framework
for the description of rotating nuclei is the treatment of time-odd mean
fields which are uniquely defined via the Lorentz covariance \cite{AA.10};
note that these fields substantially affect the properties of rotating 
nuclei \cite{AR.00,TO-rot}. Because the details of the CRMF framework could 
be found in earlier publications (Refs.\ \cite{KR.89,KR.93,AKR.96,AA.08,}), 
we focus here on the features typical for the present study.

  The pairing correlations are neglected in the present calculations. 
There are several reasons behind this choice. First, it is well known
that pairing correlations are quenched by rotation (Coriolis anti-pairing 
effect) \cite{RS.80,VDS.83}. The presence of substantial shell gaps also leads 
to a quenching of pairing correlations \cite{SGBGV.89}. Another mechanism 
of pairing quenching is blocking effect which is active in many 
nucleonic configurations \cite{RS.80}. In a 
given configuration, the pairing is also very weak at the spins close to 
band termination \cite{PhysRep-SBT}. Moreover, the pairing 
drastically decreases after paired band crossings in the proton and 
neutron subsystems \cite{AF.05}; at these spins the results of the
calculations with and without pairing are very similar.

\begin{figure}[ht]
\includegraphics[angle=0,width=8.8cm]{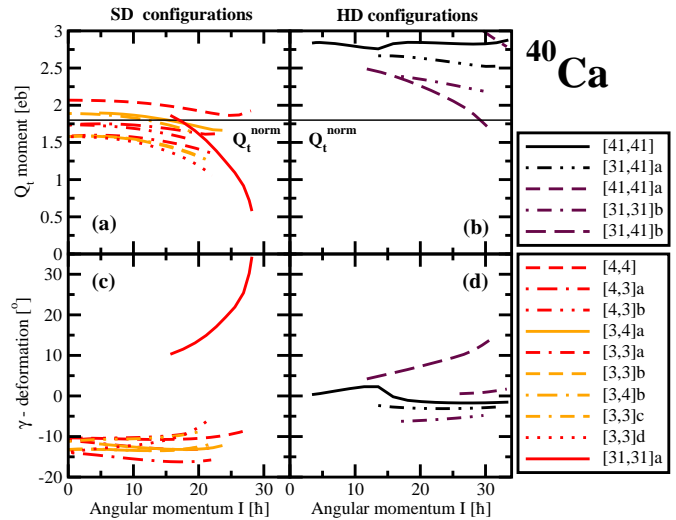}
\caption{(Color online) Calculated transition quadrupole moments
$Q_t$ and $\gamma$-deformations of the yrast and excited SD and HD 
configurations in $^{40}$Ca. The colors of the lines for different
types of configurations  roughly correspond to those used in 
Fig.\ \ref{Ca40-eld}. Red and orange (black and dark brown) are 
used for the SD (HD) configurations.}
\label{Ca40-Qt-gamma}
\end{figure}

 Second, the calculations for blocked configurations within the cranked 
Relativistic Hartree-Bogoliubov (CRHB) framework \cite{CRHB} are frequently 
numerically unstable \cite{AO.13}. This is a common problem for self-consistent 
Hartree-Bogoliubov or Hartree-Fock-Bogoliubov calculations which appears both 
in relativistic and non-relativistic frameworks \cite{DABRS.15}. On the contrary, 
these problems are much less frequent in unpaired CRMF calculations (see Ref.\ \cite{AA.08}). 
Even then it is not always possible to trace the configuration in the desired spin 
range. This typically takes place when (i) the local minimum is not deep enough for the 
solution (unconstrained in quadrupole moments) to stay in it during convergence 
process and (ii) occupied and unoccupied single-particle orbitals with the same 
quantum numbers come close in energy and start to interact.

  Based on previous experience in $^{40}$Ca (Ref.\ \cite{Ca40-PRL.01}), $^{48}$Cr 
(Ref.\ \cite{PhysRep-SBT}) and somewhat heavier $N\sim Z$ $A=58-80$ nuclei (Refs.\ 
\cite{A60,AF.05}), we estimate that the pairing becomes quite small and thus not very 
important above $I\sim 10\hbar$ in the nuclei of interest. This is exactly the spin 
range on which the current study is focused. Note also that the comparison of 
the CRHB and CRMF results for a few configurations in $^{40}$Ca presented at the 
end of Sect.\ \ref{sec-ca40} supports this conclusion.

  In the current study, we restrict ourselves to reflection
symmetric shapes since previous calculations in the cranked Hartree-Fock 
approach with Skyrme forces \cite{IMYM.02} showed that odd-multipole 
(octupole, . . .) deformations play a very limited role in extremely 
deformed configurations of the mass region under study.

 The CRMF equations are solved in the basis of an anisotropic three-dimensional 
harmonic oscillator in Cartesian coordinates characterized by the deformation 
parameters $\beta_0$ and $\gamma$ and oscillator frequency $\hbar \omega_0 = 41 A^{−1/3}$ 
MeV, for details see Refs.\ \cite{AKR.96,KR.89}. The truncation of basis is performed 
in such a way that all states belonging to the major shells up to
$N_F=14$ fermionic shells for the Dirac spinors and up to $N_B=20$ bosonic 
shells for the meson fields are taken into account. This truncation scheme provides
sufficient numerical accuracy (see Ref.\ \cite{AA.08} for details).

\begin{figure}[ht]
\includegraphics[angle=0,width=8.8cm]{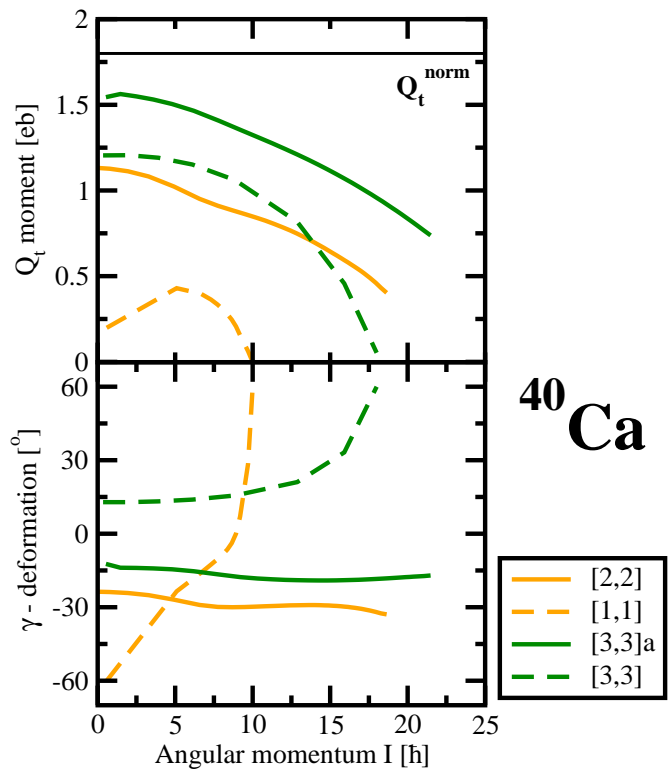}
\caption{(Color online) The same as Fig.\ \ref{Ca40-Qt-gamma} but for the
normal and highly-deformed triaxial configurations in $^{40}$Ca.
}
\label{Ca40-qt-gamma-ND}
\end{figure}

 The CRMF calculations have been performed with the NL3* functional \cite{NL3*}
which is state-of-the-art functional for nonlinear meson-nucleon coupling model
\cite{AARR.14}. It is globally tested for ground state observables in even-even 
nuclei \cite{AARR.14} and systematically tested for physical observables related 
to excited states in heavy nuclei \cite{AS.11,AO.13,AAR.10}. The CRMF and CRHB 
calculations with the NL3* CEDF provide a very successful description of different 
types of rotational bands \cite{NL3*,ASN.12,AO.13} both at low and high spins.

 The quadrupole deformation $\beta_{2}$ is defined in self-consistent 
calculations from calculated quadrupole moments using the simple 
relation \cite{A250,HG.07,SGP.05}
\begin{equation}
\beta_{2}= \frac{1}{XR^{2}} \sqrt{ \frac{5 \pi}{9}} Q^{X}_{0}
\label{def-def}
\end{equation}
where $R = 1.2A^{1/3}$ fm is the radius and $Q^{X}_{0}$ is a 
quadrupole moment of the $X$-th (sub)system expressed in fm$^{2}$. 
Here $X$ refers either to proton ($X=Z$) or neutron ($X=N$) 
subsystem or represents total nuclear system $(X=A)$. However this 
expression neglects the higher powers of $\beta_{2}$ and higher 
multipolarity deformations $\beta_{4}$, $\beta_{6}$,... \cite{NZ-def}, 
which have an important role at very large deformations.

\begin{figure}[ht]
\includegraphics[angle=0,width=8.8cm]{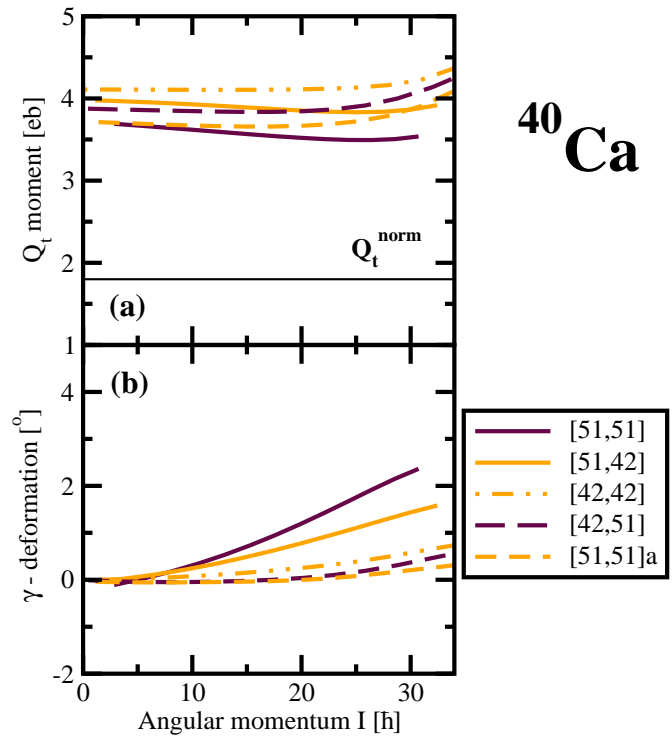}
\caption{(Color online) The same as Fig.\ \ref{Ca40-Qt-gamma} but for the
MD configurations in $^{40}$Ca.
}
\label{Ca40-qt-gamma-MD}
\end{figure}

 Because the definition of the deformation is model dependent \cite{NZ-def} and 
deformation parameters are not experimentally measurable quantities, we prefer 
to use transition quadrupole moment $Q_{t}$ for the description of deformation 
properties of the SD, HD and MD bands. This is an experimentally measurable 
quantity and thus in future our predictions can be directly compared with the 
experimental results. The deformation properties of the yrast SD band in 
$^{40}$Ca \cite{Ca40-PRL.01} are used as a reference. The measured transition 
quadrupole moment of this band is $Q^{exp}_{t} = 1.8^{+0.35}_{-0.29}$ $e$b 
\cite{Ca40-PRL.01}. Note that the CRMF calculations with the NL3* 
functional come close to experiment only slightly overestimating an experimental 
value (see Fig.\ \ref{Ca40-Qt-gamma} below). Thus we use $Q_{t}^{exp} = 1.8$ 
$e$b in $^{40}$Ca as a reference point. Note that the SD band in $^{40}$Ca is
the most deformed SD band among observed SD bands in this mass region.

 Using this value we introduce the normalized transition quadrupole moment 
$Q_{t}^{norm}(Z,A)$ in the $(Z,A)$ system 
\begin{equation}
Q^{norm}_{t}(Z,A)=\frac{ZA^{2/3}}{129.96}\,\,\,e{\rm b}
\end{equation}
  This is similar to what has been done in Ref.\ \cite{AA.08} in the analysis 
of the HD configurations in medium mass region.  This equation is based on the 
ratio $Q_{t}^{norm}(Z,A)/Q_{t}(^{40}{\rm Ca})$ calculated using Eq.\ (\ref{def-def}) 
under the assumption that the $\beta_{2}$ values in the $(Z,A)$ system and in 
$^{40}$Ca are the same.

 The band will be described as HD if its calculated $Q_{t}$ value exceeds 
$Q_{t}^{norm}(Z,A)$ by approximately 50\%. This definition of HD is similar to 
the one employed in Ref.\ \cite{AA.08}. Following suggestion of Ref.\ 
\cite{DPSD.04}, we describe even more deformed bands as megadeformed. The band 
is classified as MD when its calculated $Q_{t}$ value is approximately twice of 
$Q_{t}^{norm}(Z,A)$ or higher.

\begin{table}[h]
\begin{center}
\caption{   The semi-axis ratios of the density distributions of 
             the indicated configurations. They are defined only
             for plotted density distrubutions (see, for example,
             Fig.\ \ref{density-Ca40}). The semi-axis ratios are extracted
             at $\rho_p=0.04$ fm$^{-3}$ which roughly corresponds
             to a half of proton density in the central part of
             nucleus. The type of configuration (SD=superdeformed,
             HD=hyperdeformed and MD=megadeformed) is shown in 
             column 3. 
\label{table-ratio}
}
\begin{tabular}{|c|c|c|c|} \hline
 Nucleus      &   Configuration, spin  & Type &  Semi-axis ratio \\ \hline
   1          &          2             &   3  &      4           \\ \hline 
$^{40}$Ca      &    [4,4], $I=12$       &  SD  &  2.05            \\
              &    [41,41], $I=24$     &  HD  &  2.27            \\ 
              &    [42,42], $I=25$     &  MD  &  2.90            \\ \hline
$^{42}$Sc      &    [41,41], $I=22$     &  HD  &  2.23            \\
              &    [52,52], $I=25$     &  MD  &  2.65            \\      
              &    [421,421], $I=31$   &  MD  &  3.40            \\
              &    [421,421], $I=40$   &  MD  &  3.64            \\  \hline
$^{42}$Ca      &    [4,4]a, $I=21$      &  SD  &  2.17            \\
              &    [62,42], $I=0$      &  MD  &  2.72            \\ 
              &    [62,42], $I=16$     &  MD  &  2.79            \\ \hline
$^{44}$Ca      &    [62,42], $I=27$     &  MD  &  2.39            \\   \hline
$^{44}$Ti      &    [41,41], $I=25$     &  SD  &  2.03            \\
              &    [62,62], $I=0$      &  MD  &  2.70            \\ 
              &    [62,62], $I=32$     &  MD  &  2.88            \\ \hline
$^{46}$Ti      &    [62,51], $I=26$     &  SD  &  1.75            \\   
              &    [62,42], $I=28$     &  HD  &  2.40            \\   \hline
$^{48}$Cr      &    [62,62], $I=0$      &  HD  &  2.24            \\
              &    [62,62], $I=28$     &  HD  &  2.39            \\ \hline
$^{50}$Cr      &    [62,62], $I=31$     &  HD  &  2.27           \\ \hline
$^{36}$Ar      &    [2,2], $I=4$        &  SD  &  1.9             \\
              &    [4,4], $I=16$       &  HD  &  2.21            \\
              &    [31,31], $I=21$     &  MD  &  2.56            \\
              &    [41,41], $I=30$     &  MD  &  2.64            \\  \hline
$^{38}$Ar      &    [3,2]a, $I=12$      &  SD  &  1.91            \\
              &    [42,31], $I=24$     &  HD  &  2.27            \\
              &    [42,31], $I=32$     &  MD  &  2.74            \\  \hline
$^{32}$S       &    [2,2], $I=12$       &  SD  &  2.09            \\
              &    [21,21], $I=31$     &  HD  &  2.15            \\ \hline
$^{34}$S       &    [2,1], $I=14$       &  SD  &  1.32            \\
              &    [31,21], $I=21$     &  HD  &  2.32            \\ \hline
\end{tabular}
\end{center}
\end{table}

 Single-particle orbitals are labeled by $\Omega[Nn_{z}\Lambda](r=\pm i)$. 
$\Omega[Nn_{z}\Lambda]$  are the asymptotic quantum numbers (Nilsson 
quantum numbers) of the dominant component of the wave function at 
$\Omega_{x} = 0.0$ MeV and $r$ is the signature of the orbital.

  Because the pairing correlations are neglected, the intrinsic 
structure of the configurations of interest can be described by 
means of the dominant single-particle components of the 
intruder/hyperintruder/megaintruder states occupied. Thus, 
the calculated configurations will be labeled by shorthand 
[$n_{1}$($n_{2}$)($n_{3}$),$p_{1}$($p_{2}$)($p_{3}$)] labels, 
where $n_{1}$, $n_{2}$ and $n_{3}$ are the number of neutrons 
in the $N=3$, 4 and 5 intruder/hyperintruder/megaintruder orbitals 
and $p_{1}$, $p_{2}$ and $p_{3}$ are the number of protons in the 
$N=3$, 4 and 5 intruder/hyperintruder/megaintruder orbitals.  
The $N=5$ megaintruder orbitals are not occupied in the HD 
configurations. As a consequence, the labels $n_{3}$ and $p_{3}$ 
will be omitted in the labeling of such configurations. Moreover, 
the  $N=4$ and $N=5$ orbitals are not occupied in the SD 
configurations. So, in those configurations the $n_{2}$, $n_{3}$ 
and $p_{2}$, $p_{3}$ labels will be omitted. An additional letter 
{\it{(a,b,c, ...)}} at the end of the shorthand label is used to 
distinguish the configurations which have the same occupation 
of the intruder/hyperintruder/megaintruder orbitals (the same
[$n_{1}$($n_{2}$)($n_{3}$),$p_{1}$($p_{2}$)($p_{3}$)] label)
but differ in the occupation of non-intruder orbitals.

\begin{figure*}[ht]
\includegraphics[angle=0,width=8.8cm]{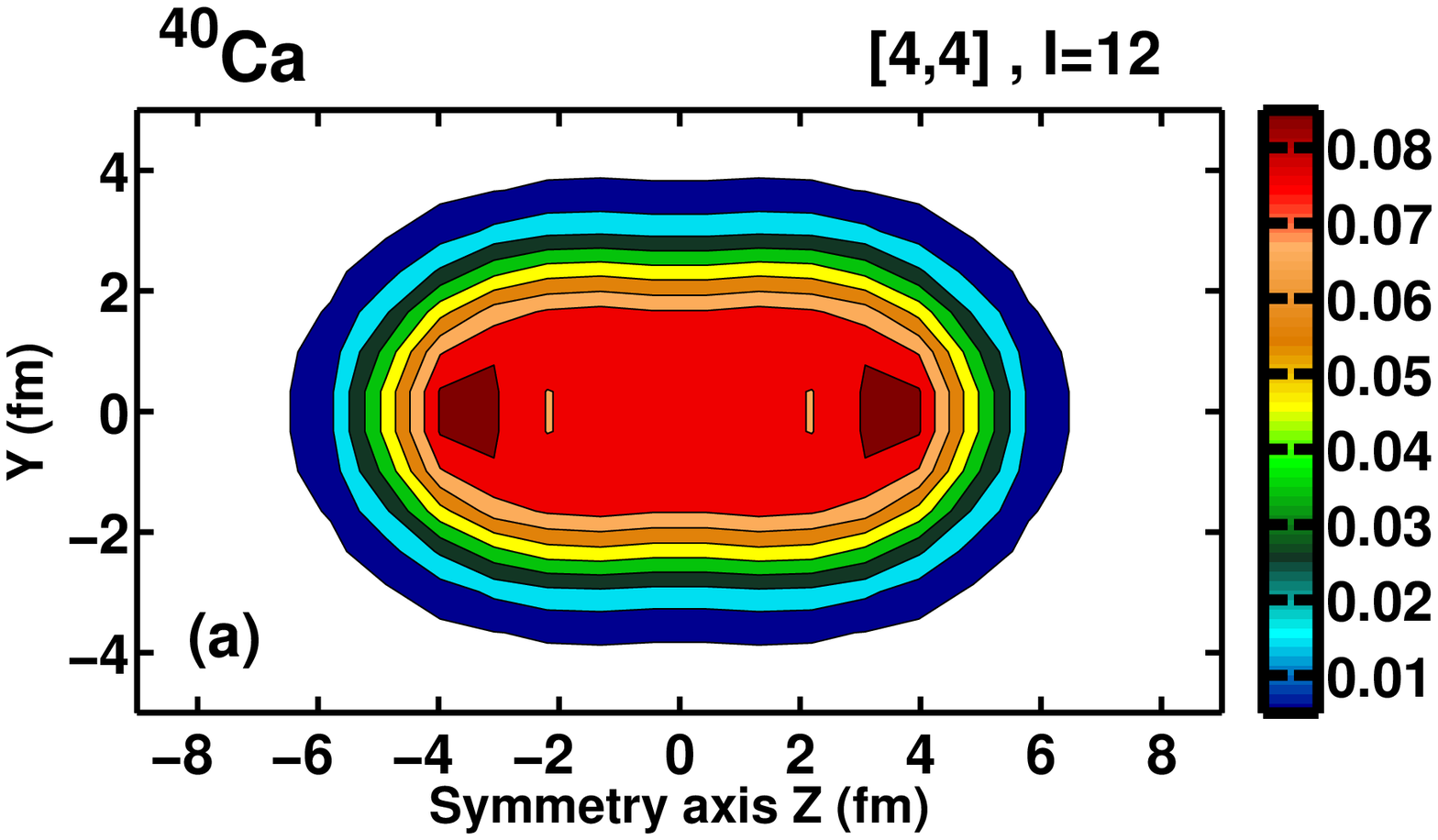}
\includegraphics[angle=0,width=8.8cm]{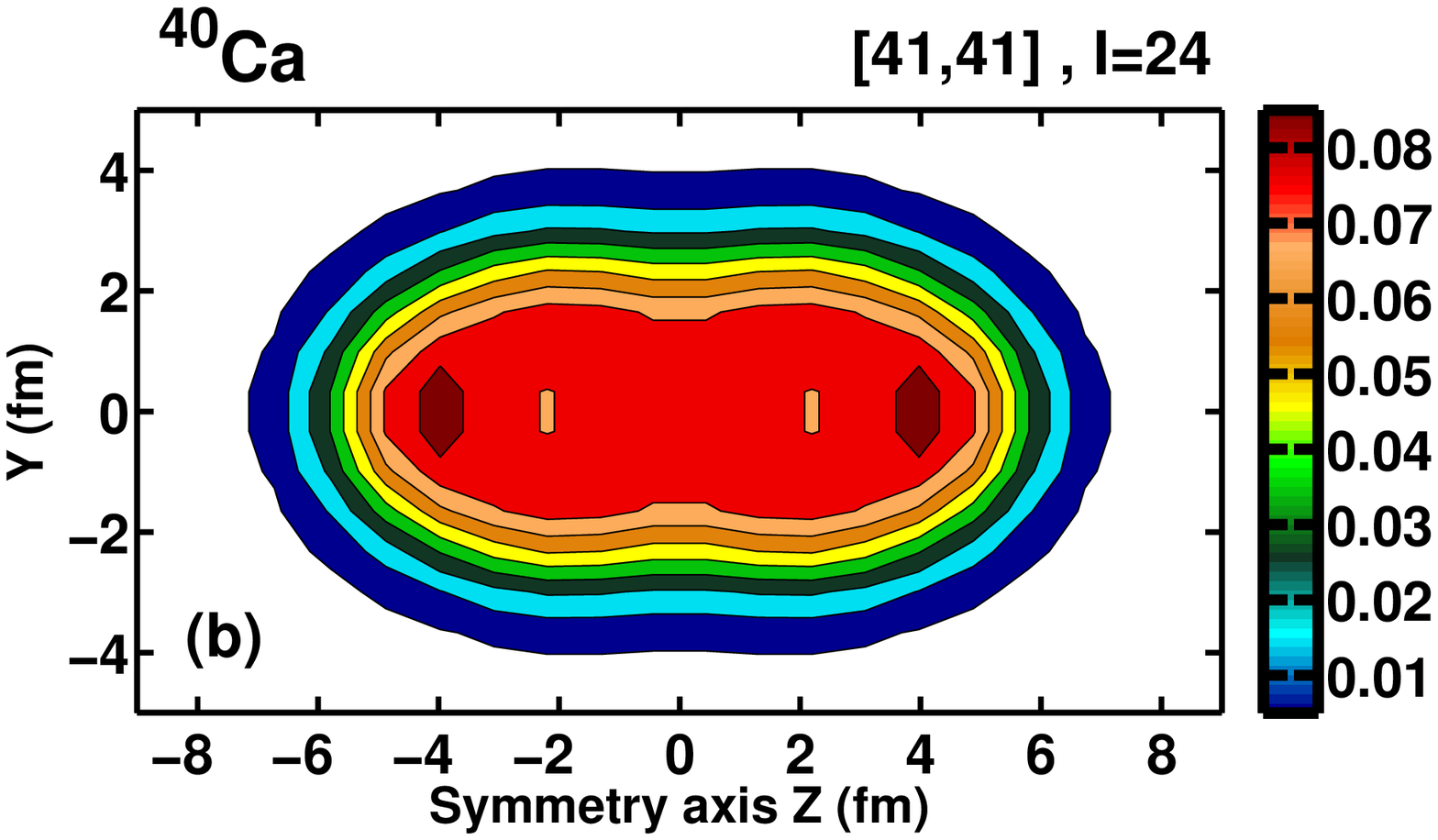}
\includegraphics[angle=0,width=8.8cm]{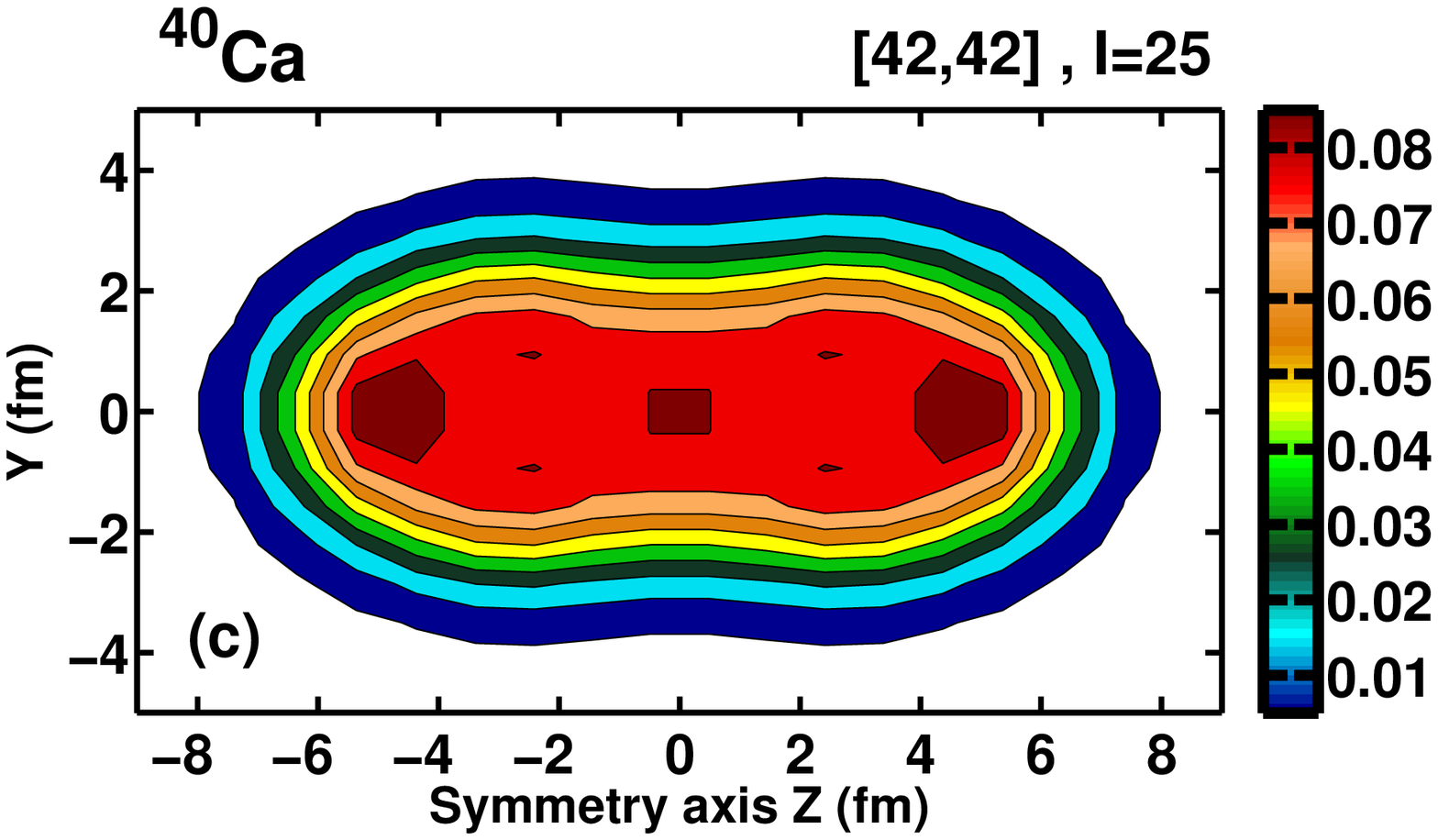}
\caption{(Color online) The self-consistent proton density $\rho_p(y,z)$ as a 
function of $y$ and $z$ coordinates for the indicated configurations in $^{40}$Ca 
at specified spin values. The equidensity lines are shown in steps of 0.01 fm$^{-3}$
starting from $\rho_p(y,z)=0.01$ fm$^{-3}$. 
}
\label{density-Ca40}
\end{figure*}

\begin{figure}[ht]
\includegraphics[angle=0,width=8.8cm]{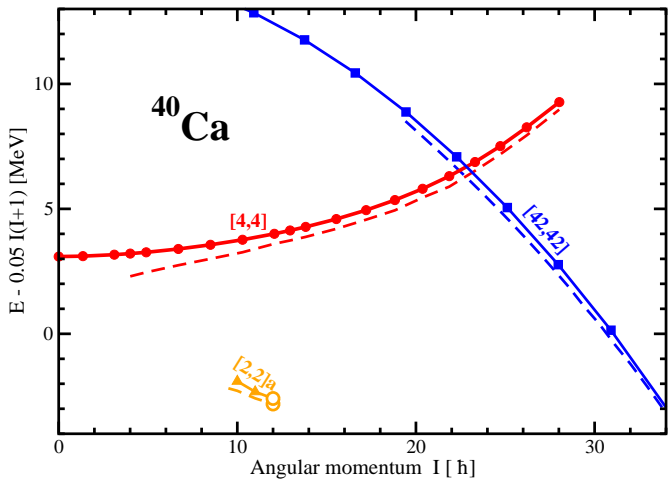}
\caption{(Color online) The comparison of the results of the calculations
with and without pairing for the configurations of $^{40}$Ca which do 
not require blocking procedure in the CRHB calculations. The results of
the calculations without pairing are shown by solid lines with symbols.
The results for paired analogs of unpaired configurations are shown by 
dashed lines of the same color.}
\label{Ca40-eld-CRHB}
\end{figure}

\section{The $^{40}$Ca nucleus}
\label{sec-ca40}

 $^{40}$Ca is a doubly magic spherical nucleus with 20 neutrons and 20 protons. 
Three lowest shells with $N=0$, 1 and 2 are occupied in its spherical 
ground 
state with $I=0^+$. Higher spin states are formed by particle-hole excitations 
from the $N=2$ shell into $f_{7/2} (N=3)$ subshell across the respective 
$Z=20$ and $N=20$ spherical shell gaps. This leads to a formation of 
complicated high-spin level scheme which includes spherical states and
deformed, terminating and superdeformed rotational structures 
\cite{Ca40-PRL.01,Ca40-Qt.03,Ca-40.07}. In experiment, they extend up to 
spin $I=16^+$.

 The results of the CRMF calculations for deformed configurations forming 
the yrast line are shown in Fig.\ \ref{Ca40-eld}. Different colors are used
to indicate different classes of the bands. Note that low-spin spherical
solutions are not shown here since we are interested in high-spin behavior
of this nucleus.

  The lowest deformed configuration [1,1] is based on simultaneous 
excitations of proton and neutron from the $d_{3/2}$ spherical subshell
into the $f_{7/2}$ subshell across the $Z=20$ and $N=20$ spherical gaps. 
It has small quadrupole deformation of $\beta_2\sim 0.16$ and 
$\gamma \sim -24^{\circ}$ at $I=4\hbar$ and terminates at $I=10^+$ in a 
terminating state with 
the structure $\pi (f_{7/2})^1_{3.5} (d_{3/2})^{-1}_{1.5} \otimes \nu  
(f_{7/2})^1_{3.5} (d_{3/2})^{-1}_{1.5}$ and near-spherical shape. Additional 
excitations of the proton and neutron across the  $Z=20$ and $N=20$ 
spherical gaps lead to a more deformed [2,2] configuration which
has $\beta_2\sim 0.32$ and $\gamma \sim -30^{\circ}$ at $I=10\hbar$.
It is expected to terminate at $I_{max}=20\hbar$ with the terminating
state built at high energy cost and located high above the yrast line. 
However, we were not able to trace this configuration up to termination 
in the calculations. Next excitations of
proton and neutron across the  $Z=20$ and $N=20$ spherical gaps lead 
to even more deformed [3,3] configurations which are located close
to each other up to spin $I=16\hbar$ (see Fig.\ \ref{Ca40-eld}).
The configuration which terminates
at spin $I=18\hbar$ is located in positive $\gamma$ minimum of potential
energy surfaces and has $\beta_2\sim 0.47$ 
and $\gamma \sim 21^{\circ}$ at $I=12\hbar$.  The structure of terminating
state is $\pi (f_{7/2})^3_{7.5} (d_{3/2})^{-3}_{1.5} \otimes \nu  
(f_{7/2})^3_{7.5} (d_{3/2})^{-3}_{1.5}$. Another [3,3] configuration is 
located in the negative $\gamma$ minimum of potential energy surfaces
and is expected to terminate at $I=24^+$. Similar to the [2,2] configuration 
we were not able to trace it up to terminating state which is expected to
be located high above the yrast line.

  Subsequent particle-hole excitations lead to an increase of the deformation
of the configurations resulting in the formation of superdeformed rotational
bands. Note that the bands with such deformation do not terminate in the 
single-particle terminating states (see Sec.\ 2.5 of Ref.\ 
\cite{PhysRep-SBT}). The yrast SD configuration [4,4] is characterized by
large SD shell gap at particle number 20 both in the proton and neutron 
subsystems (Fig.\ \ref{routh-sd-1}). All single-particle states below
these gaps are occupied in the [4,4] configuration. Note that apart of 
the Coulomb shift in energy the proton routhian diagram is similar to 
the neutron one shown in Fig.\ \ref{routh-sd-1}. The [4,4] configuration 
is only yrast at $I=22\hbar$; it is located above the yrast line at lower 
spin in agreement with the experiment \cite{Ca40-PRL.01}. The accuracy of 
the description of 
the experimental data (dynamic and kinematic moments of inertia, transition
quadrupole moments) is similar to the one obtained with the NL1 CEDF;
the results obtained with this functional are compared with experiment
in Ref.\ \cite{Ca40-PRL.01}.
  
  Starting from the yrast SD configuration [4,4] there are two ways to 
build excited configurations. The first one is by exciting particles from 
the $3/2[321](r=\pm i)$ orbitals into the $1/2[200](r=\pm i)$ orbitals; they 
are shown as the S1$-$S4 excitations in Fig.\ \ref{routh-sd-1}. 
 The combination of proton and neutron excitations of this type leads to
the [3,3] configurations. If the proton (neutron) excitations of this type
are combined with the neutron (proton) configuration of the yrast SD band,
the [3,4] and [4,3] configurations are created. These configurations are
excited with respect to the yrast [4,4] SD configuration; some of them are
shown by red lines in Fig.\ \ref{Ca40-eld}. Note that due 
to the similarity of the proton and neutron routhian diagrams some of these
excited configurations are degenerated (or nearly degenerated) in 
energy. In addition, we show only some of highly excited SD 
configurations for the sake of clarity. An important feature is quite 
large energy gap between the yrast [4,4] and lowest excited [3,3]d
SD configurations. Such a situation favors the observation of the
yrast SD band since the feeding intensity is concentrated on it (see 
discussion in Refs.\ \cite{AA.08,AA-HD.09}).

\begin{figure}[ht]
\includegraphics[angle=0,width=8.8cm]{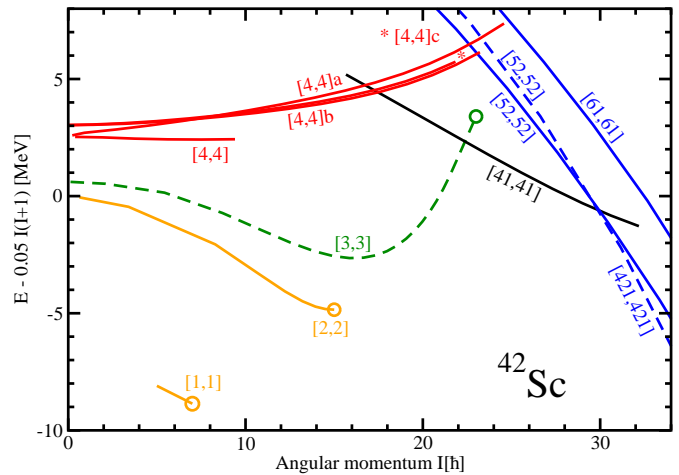}
\caption{(Color online) The same as Fig.\ \ref{Ca40-eld} but for 
$^{42}$Sc.}
\label{Sc42-eld}
\end{figure}

  Alternatively, one can excite the particle from either the $5/2[202](r=-i)$ 
or $5/2[202](r=+i)$ orbital to the lowest in energy hyperintruder 
$1/2[440](r=-i)$ orbital emerging from hyperintruder $N=4$ shell; the 
occupation corresponding to such a configuration is shown on left side of 
Fig.\ \ref{routh-hd-ca40}. The combination of the proton and neutron 
excitations of this kind leads to four-fold degenerate [41,41] HD
configurations. This degeneracy is due to very small signature splitting
of the configurations built on opposite signatures of the $5/2[202]$
orbitals and the combination of proton and neutron configurations of this
kind.

  At first glance this statement is in contradiction with Fig.\ 
\ref{routh-hd-ca40} where there is a substantial energy splitting 
between the $r=-i$ and $r=+i$ branches of the $5/2[202]$ orbital
which are almost parallel in rotational frequency. This feature 
is the consequence of non-pairwise occupation of the opposite 
signatures of some orbitals which leads to the presence of 
nucleonic currents at rotational frequency $\Omega_x=0.0$ MeV
(see Sec.\ IVA in Ref.\ \cite{AA.10}). The occupied orbital is 
always more bound than its  unoccupied time-reversal counterpart.
So the change of the signature of occupied $5/2[202]$ state 
(from $r=-i$ in Fig.\ \ref{routh-hd-ca40} to $r=+i$) will
only inverse the relative positions of both signatures of this
orbital so that the total energy of the configurations built
on the holes in the $5/2[202](r=-i)$ and $5/2[202](r=+i)$ orbitals
will be almost the same.

  The [41,41] configurations are the lowest in energy among the 
HD configurations at spins above $I=24\hbar$ (Fig.\ 
\ref{Ca40-eld}). The excited HD configurations [31,31]a and [31,31]b 
(Fig.\ \ref{Ca40-eld}) are formed as the combination of the H1 
and H2 excitations (shown in Fig.\ \ref{routh-hd-ca40}) in the proton 
and neutron subsystems. The [31,41]a and [31,41]b configurations are
based on the H1 and H2 excitations in the neutron subsystem
and the proton configuration of the yrast [41,41] HD configuration.
The [41,31]a and [41,31]b configurations (not shown in Fig.\ \ref{Ca40-eld}), 
based on the H1 and H2 excitations in the proton subsystem and the neutron 
configuration of the yrast [41,41] HD configuration, are located at the 
energies which are similar to the ones of the [31,41]a and [31,41]b 
configurations.

  The HD configurations never become yrast in $^{40}$Ca. However, such
configurations compete with megadeformed ones for yrast status
in neigbouring nuclei (see, for example, Secs.\ \ref{sec-sc42}
and \ref{Ca42-section}). That was a reason for a quite detailed discussion 
of their structure.

\begin{figure}[ht]
\includegraphics[angle=0,width=8.8cm]{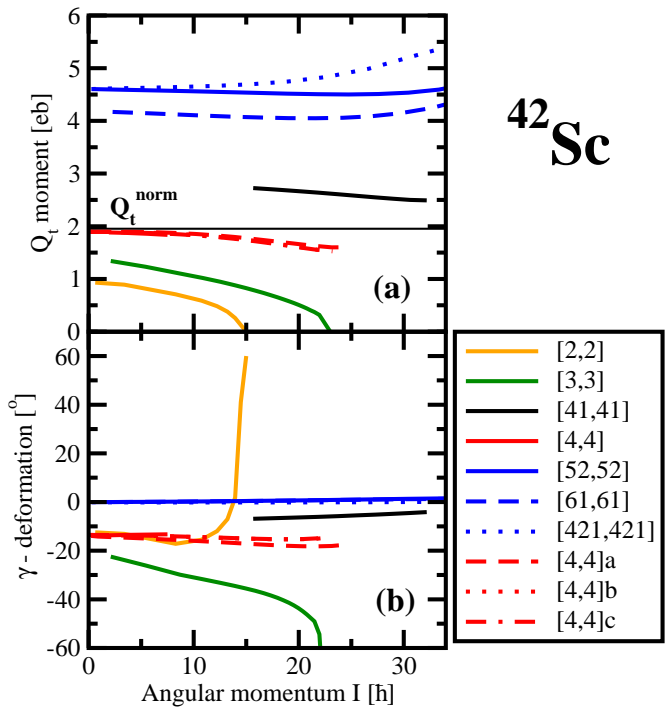}
\caption{(Color online) The same as Fig.\ \ref{Ca40-Qt-gamma} but for 
$^{42}$Sc.} 
\label{Sc42-beta-gamma}
\end{figure}

  The additional occupation of the $N=4$ proton and neutron orbitals leads
to the [42,42] MD configuration which is yrast at spin above
$I=23\hbar$ (Fig.\ \ref{Ca40-eld}). It is characterized by large (around 
3 MeV) MD $Z=20$ and $N=20$ shell gaps (Fig.\ \ref{routh-md-40ca}). Thus, this 
configuration can be considered as doubly magic megadeformed configuration.
Indeed, excited MD configurations (such as [42,51], [51,51], [51,42] etc)
are located at excitation energy of more than 2 MeV with respect to the yrast 
MD configuration (Fig.\ \ref{Ca40-eld}). The fact that the yrast MD 
configuration [42,42] is separated from the excited configurations by 
a such large energy gap should make its observation in experiment easier. 
This is because of the concentration of feeding intensity on the yrast
MD configuration in such a situation (see discussion in Refs.\ 
\cite{AA.08,AA-HD.09}).

 Calculated transition quadrupole moments $Q_t$ and $\gamma$-deformations
of the normal- and highly-deformed triaxial, SD, HD and MD configurations are shown 
in Figs.\ \ref{Ca40-Qt-gamma}, \ref{Ca40-qt-gamma-ND} and \ref{Ca40-qt-gamma-MD}.
The configurations which are yrast in local deformation minimum, namely,
SD [4,4], HD [41,41] and  MD [42,42] have the largest transition quadrupole 
moment among the calculated SD, HD, and MD configurations, respectively. This is 
because particle-hole excitations leading to excited configurations reduce the 
number  of occupied  deformation driving orbitals.

  Note that most of the calculated SD configurations have 
$\gamma \sim -12^{\circ}$. The only exception is the unusual [31,31]a configuration 
which has large positive $\gamma$-deformation rapidly increasing with spin. It has 
some  similarities with the HD configurations. First, it involves the $N=4$ proton 
and neutron. Second, its slope in the $E-E_{RLD}$ plot is similar to the one of the
HD configurations (see Fig.\ \ref{Ca40-eld}). However, it has substantially 
smaller $Q_t$ values as compared with the HD configurations.

\begin{figure*}[ht]
\includegraphics[angle=0,width=8.8cm]{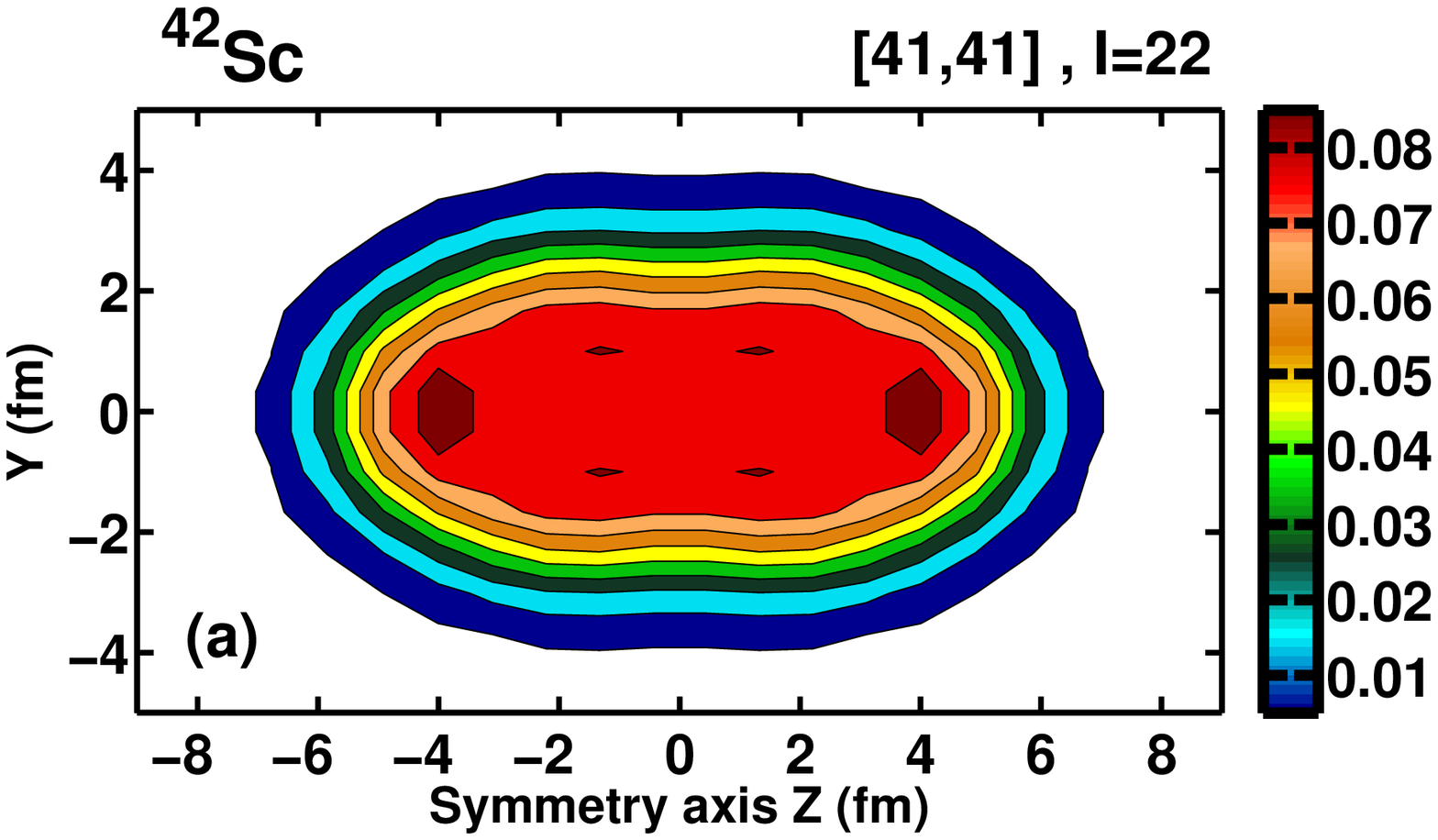}
\includegraphics[angle=0,width=8.8cm]{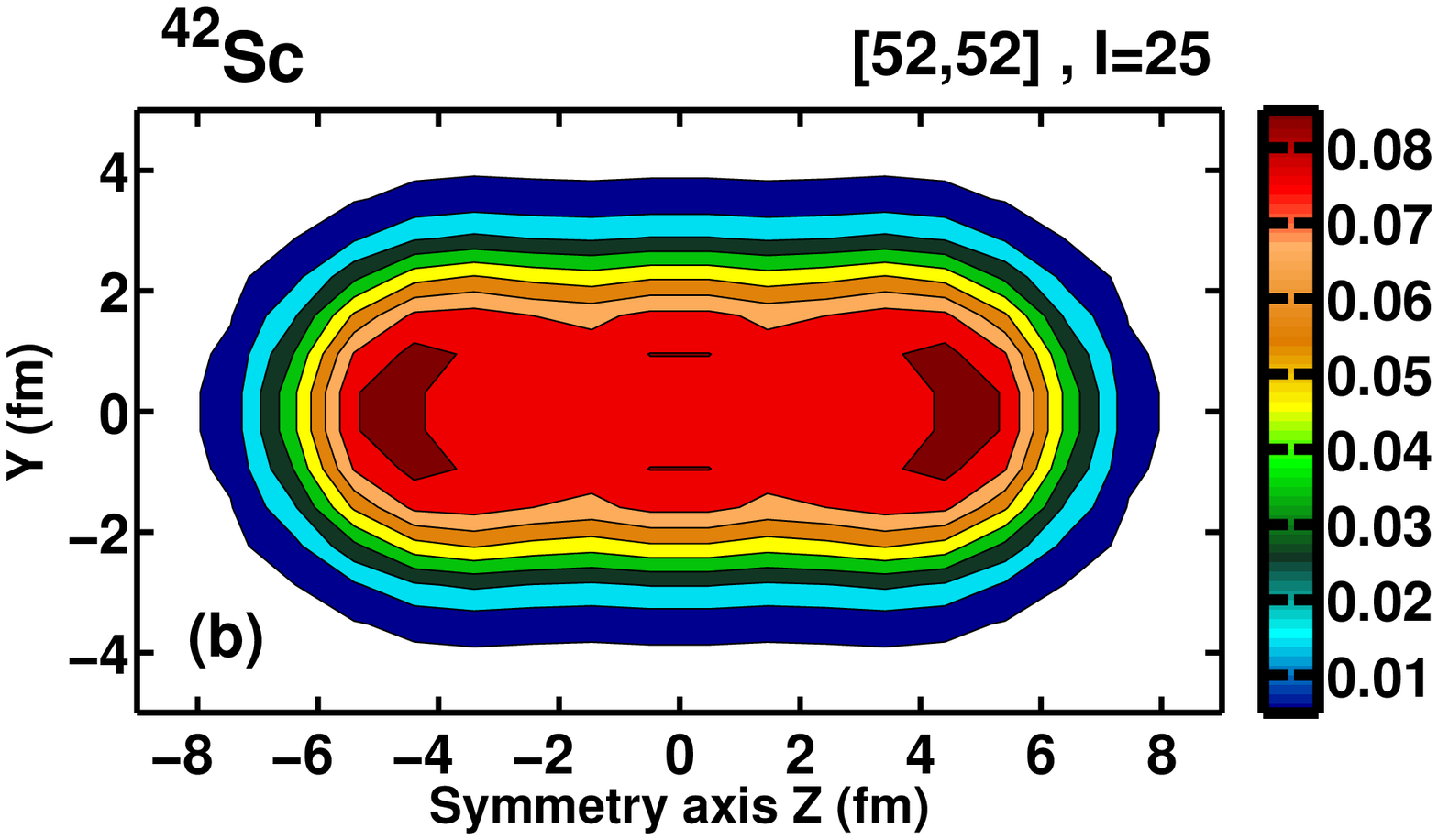}
\includegraphics[angle=0,width=8.8cm]{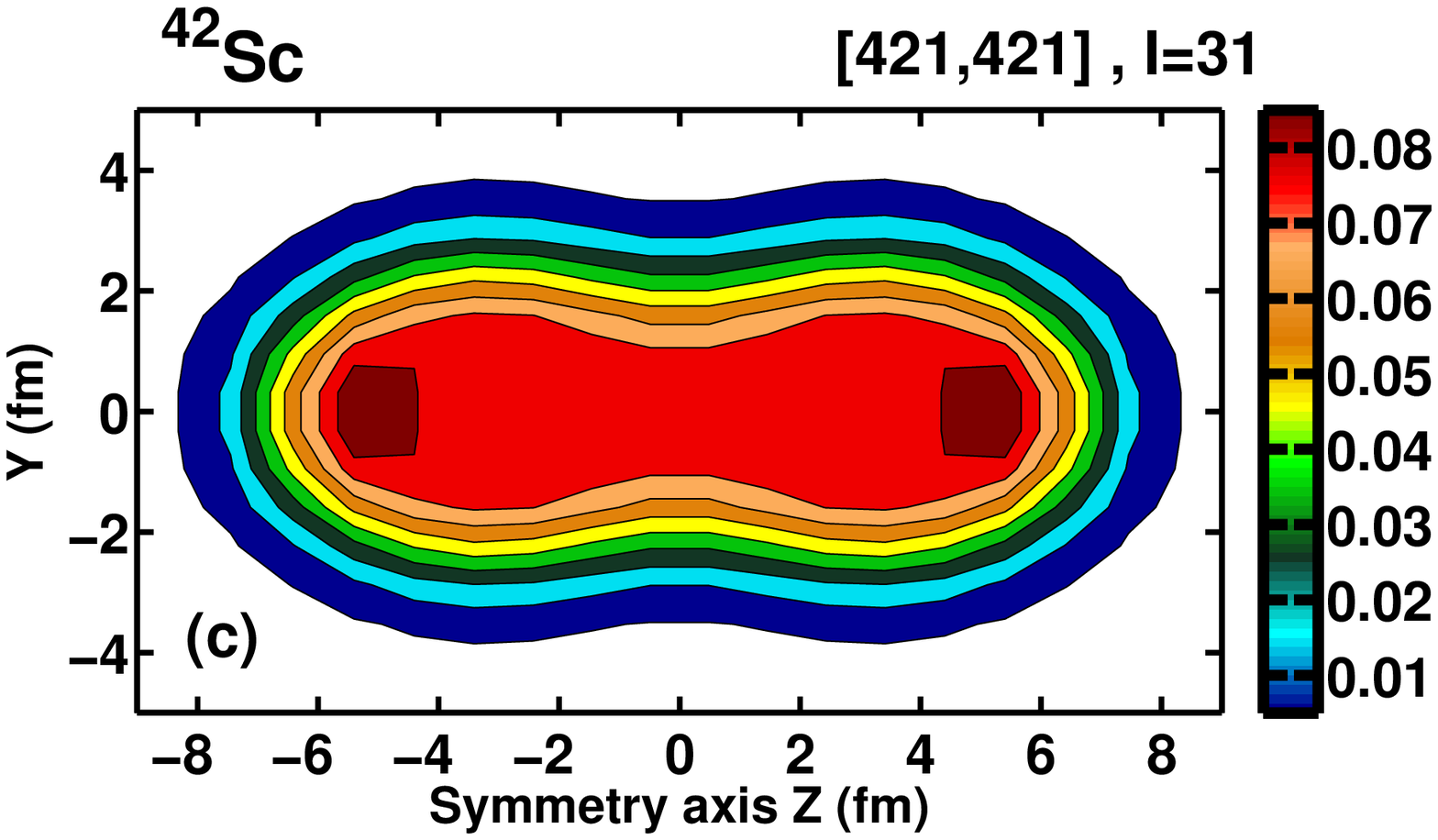}
\includegraphics[angle=0,width=8.8cm]{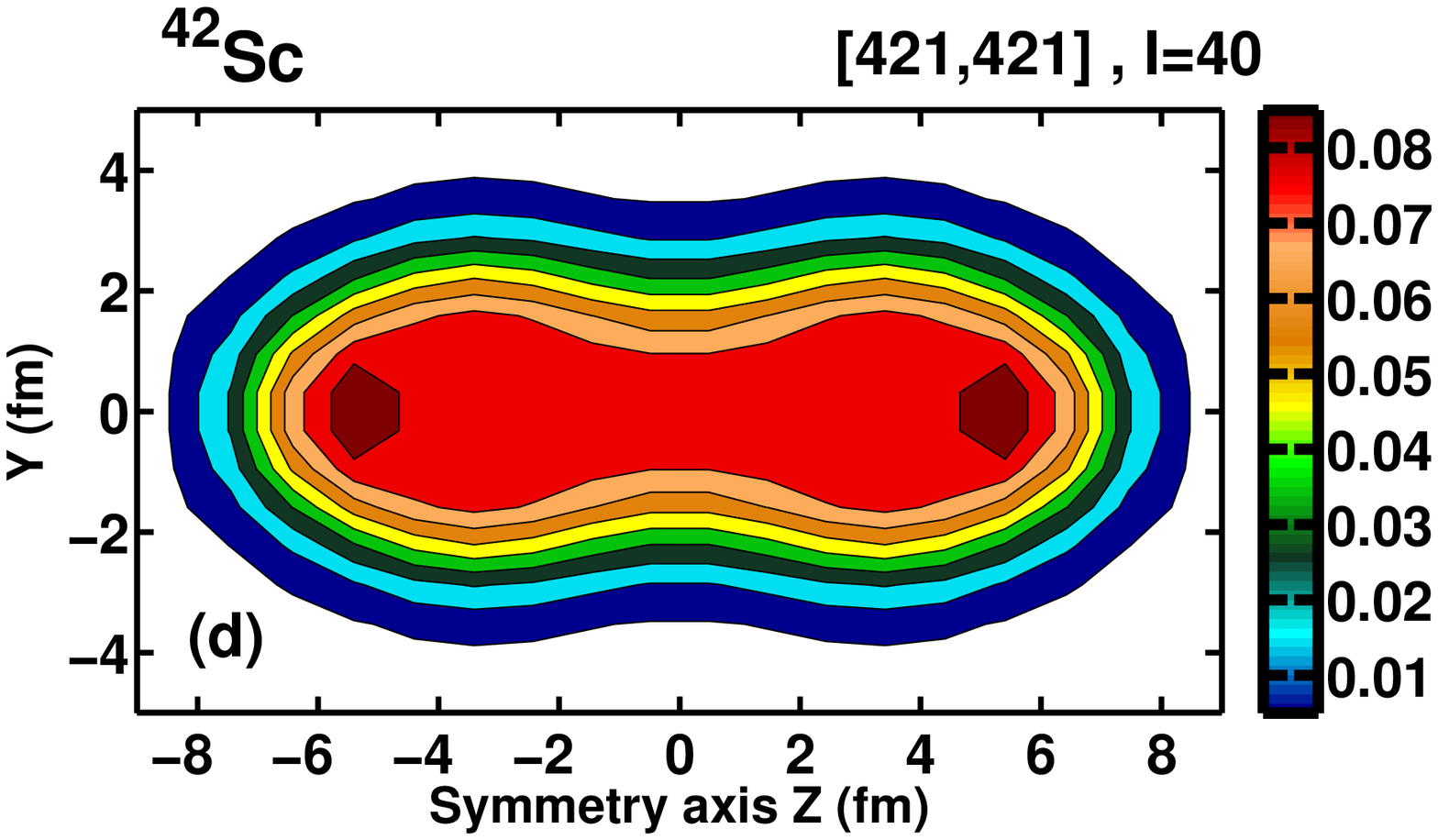}
\caption{(Color online) The same as Fig.\ \ref{density-Ca40} 
          but for $^{42}$Sc.}
\label{density-42Sc}
\end{figure*}

  While the calculated $Q_t$ and $\gamma$ values cluster for the
SD configurations (Figs.\ \ref{Ca40-Qt-gamma}a and c), they are scattered for
the HD configurations  (Figs.\ \ref{Ca40-Qt-gamma}b and d). This suggests that
the potential energy surfaces are much softer in the HD minimum as compared 
with the SD one. Indeed, in the HD minimum a single particle-hole excitation 
induces much larger changes in the $Q_t$ and $\gamma$ values than in the SD 
one. On the contrary, the MD configurations show the clusterization of the 
calculated $Q_t$ and $\gamma$ values which is similar to the one observed 
in the SD minimum (Fig.\ \ref{Ca40-qt-gamma-MD}).

  The most deformed HD configuration ([41,41]) has $Q_t$ values which are by
roughly 40\% larger than the ones for most deformed SD configuration ([4,4)])
(Fig.\ \ref{Ca40-Qt-gamma}a and c). Yrast MD configuration [42,42] has the
$Q_t$ values which are larger by roughly 45\% and 105\% than the ones for
most deformed HD and SD configurations (compare Fig.\ \ref{Ca40-qt-gamma-MD}a
and Fig.\ \ref{Ca40-Qt-gamma}a and c).

  The self-consistent proton densities of the yrast SD, HD and MD 
configurations are shown in Fig.\ \ref{density-Ca40} at indicated spin 
values. The stretching of nuclear shape is definitely more pronounced 
in the HD [41,41] and especially in the MD [42,42] configurations. 
Indeed, the semi-major to semi-minor axis ratio is 2.05, 2.27 and 2.9 
for the densities of the SD [4,4], HD [41,41] and MD [42,42] 
configurations. Note that the changes in the semi-axis ratio on going
from one type of configurations to another are substantially smaller
than relevant changes in the $Q_t$ values discussed above. The densities 
of the [41,41] HD configuration show some indications of the development 
of neck and these indications become much more pronounced in the MD 
[42,42] configuration.

  Fig.\ \ref{Ca40-eld-CRHB} compares the results of the calculations 
with and without pairing for few selected configurations in $^{40}$Ca. The calculations 
with pairing are performed in the cranked relativistic Hartree Bogoliubov 
(CRHB) framework of Ref.\ \cite{CRHB}. In these calculations the Lipkin-Nogami 
method is employed for an approximate particle number projection and the Gogny 
D1S force is used in pairing channel. The presence of pairing correlations 
leads to an additional binding. However, above $I=10\hbar$ this additional 
binding is rather modest (around 0.5 MeV or less) and it is similar for 
different calculated configurations. As a result, the general structure of 
the calculated configurations in the $E-E_{RLD}$ plot above this spin is 
only weakly affected by the presence of pairing. Similar effect has already
been seen in the case of $^{72}$Kr (Ref.\ \cite{72Kr-qt}).

  Note that among  large number of the configurations obtained in
unpaired calculations and presented in Fig.\ \ref{Ca40-eld} only these 
three configurations (terminating [2,2]a, superdeformed [4,4] and
hyperdeformed [42,42]), in which signature partner orbitals are pairwise 
occupied (see Figs.\ \ref{routh-sd-1} and \ref{routh-md-40ca}), can 
be calculated in the CRHB framework without blocking 
procedure. Particle-hole excitations leading to excited configurations
remove pairwise occupation of the signature partner orbitals. As a
result, the blocking procedure has to be employed for the calculation
of such configurations in the CRHB framework. For example, the blocking 
of two particles is needed if the configuration label contains at least 
one odd number in either proton or neutron subsystem. If the configuration 
label contains odd number in both proton and neutron subsystems, then 
the blocking of four particles (two in proton subsystem and two in neutron 
subsystem) is needed. Such calculations are inherently unstable \cite{AO.13,Rb74}. 
On the other hand, the blocking leads to an additional reduction of the impact of
pairing correlations on physical observables (see Ref.\ \cite{AO.13}).
As a result, even smaller effect of pairing (as compared with the one
shown in Fig.\ \ref{Ca40-eld-CRHB}) is expected on binding energies 
of the configurations of Fig.\ \ref{Ca40-eld} not shown in Fig.\ 
\ref{Ca40-eld-CRHB}.

\section{$^{42}$Sc nucleus}
\label{sec-sc42}

 $^{42}$Sc nucleus is formed by an addition of one proton and one
neutron to $^{40}$Ca. This is only odd-odd nucleus considered in the
present paper. The configurations forming the yrast line of $^{42}$Sc 
are shown in Fig.\ \ref{Sc42-eld}. The [1,1] configuration is built in 
valence space; it terminates at $I=7^+$.  The [2,2] configuration
is an analog of the [1,1] configuration in $^{40}$Ca but with an extra 
proton and extra neutron placed into the orbitals emerging from the 
$f_{7/2}$ spherical subshell. As a consequence, it has substantially
larger deformation and maximum spin within the configuration than the
[1,1] configuration in $^{40}$Ca. At spin $I=4\hbar$, the deformation
of the [2,2] configuration is $\beta_2\sim 0.27$ and $\gamma \sim 
-15^{\circ}$. It terminates at $I=15^+$ in a terminating state with 
the structure $\pi (f_{7/2})^2_{6.0} (d_{3/2})^{-1}_{1.5} \otimes \nu  
(f_{7/2})^2_{6.0} (d_{3/2})^{-1}_{1.5}$ and near-spherical shape with 
$\beta_2 \sim 0.05$.

\begin{table}[h]
\begin{center}
\caption{The maximum spin (in $\hbar$) which could be built within the 
         configuration of given type. The asterisk is used to indicate
         the configurations which involve the hole(s) in the $d_{5/2}$
         orbital(s). The SD configurations are not included into this 
         table. See text for the discussion of the details. 
\label{table-Imax}
}
\begin{tabular}{|c|c|c|c|} \hline
 Nucleus      &   valence space  & 2p-2h     &  4p-4h      \\ \hline
   1          &          2         &   3                 &      4               \\ \hline 
$^{40}$Ca      & [0,0], $I_{max}=0$  & [1,1], $I_{max}=10$  &  [2,2], $I_{max}=20$*  \\
$^{42}$Ca      & [2,0], $I_{max}=6$  & [3,1], $I_{max}=14$  & [4,2], $I_{max}=24$*   \\
$^{44}$Ca      & [4,0], $I_{max}=8$  & [5,1], $I_{max}=14$  & [6,2], $I_{max}=20$*   \\ 
              &                    &                     &                      \\
$^{42}$Sc      & [1,1], $I_{max}=7$  & [2,2], $I_{max}=15$  & [3,3], $I_{max}=23$*   \\
              &                    &                     &                      \\
$^{44}$Ti      & [2,2], $I_{max}=12$ & [3,3], $I_{max}=18$  & [4,4], $I_{max}=24$*   \\
$^{46}$Ti      & [4,2], $I_{max}=14$ & [5,3], $I_{max}=19$  & [6,4], $I_{max}=22$*   \\
              &                     &                    &                 \\
$^{48}$Cr      & [4,4], $I_{max}=16$ & [5,5], $I_{max}=20$*  & [6,6], $I_{max}=20$*  \\
$^{50}$Cr      & [6,4], $I_{max}=14$ & [7,5], $I_{max}=16$*  &             \\
              &                    &                      &                     \\
$^{36}$Ar      & [0,0], $I_{max}=8$*  & [1,1], $I_{max}=16$*  &                    \\
$^{38}$Ar      & [0,0], $I_{max}=4$*  & [1,1], $I_{max}=14$*  &                    \\
              &                     &                     &                     \\
$^{32}$S       & [0,0], $I_{max}=12$* & [1,1], $I_{max}=20$* &             \\
$^{34}$S       & [0,0], $I_{max}=10$* & [1,1], $I_{max}=19$* &             \\ \hline
\end{tabular}
\end{center}
\end{table}

  Additional excitations of the proton and neutron across the  $Z=20$ 
and $N=20$ spherical gaps lead to a more deformed [3,3] configuration 
which has $\beta_2\sim 0.37$ and $\gamma \sim -31^{\circ}$ at 
$I=10\hbar$. It is expected to terminate at $I_{max}=23\hbar$ with the 
terminating state built at high energy cost and located above the yrast 
line. However, we were able to trace this configuration in the calculations
only up to $I\approx 22\hbar$ (one $\hbar$ short of termination).

  The lowest four SD configurations [4,4] in $^{42}$Sc are formed from the
yrast SD configuration [4,4] in $^{40}$Ca by addition of the proton and neutron 
to the $1/2[200](r=\pm i)$ orbitals located above the $Z=20$ and $N=20$ 
SD shell gaps (see Fig.\ \ref{routh-sd-1}).  Their deformation properties 
are summarized in Fig.\ \ref{Sc42-beta-gamma}. Similar to the SD bands in 
$^{40}$Ca, they are located in the $\gamma \sim -12^{\circ}$ minimum of 
potential energy surfaces. Note that the lowest [4,4] SD configuration 
undergoes unpaired band crossing (due to the crossing of the $1/2[400](r=-i)$ 
and  $1/2[200](r=-i)$ orbitals seen in Fig.\ \ref{routh-sd-1}) which 
leads to the [41,41] HD configuration.

  At spin above $I=22\hbar$, the HD configuration [41,41] becomes 
the lowest in energy. In this configuration, all single-particle states 
below the $Z=21$ and $N=21$ HD shell gap (Fig.\ \ref{routh-hd-ca40}) 
are occupied. So contrary to the yrast HD bands in $^{40}$Ca, which 
are degenerate in energy, the yrast HD line in $^{42}$Sc is represented 
by a single strongly decoupled branch of the [41,41] configuration.

\begin{figure}[ht]
\includegraphics[angle=0,width=8.8cm]{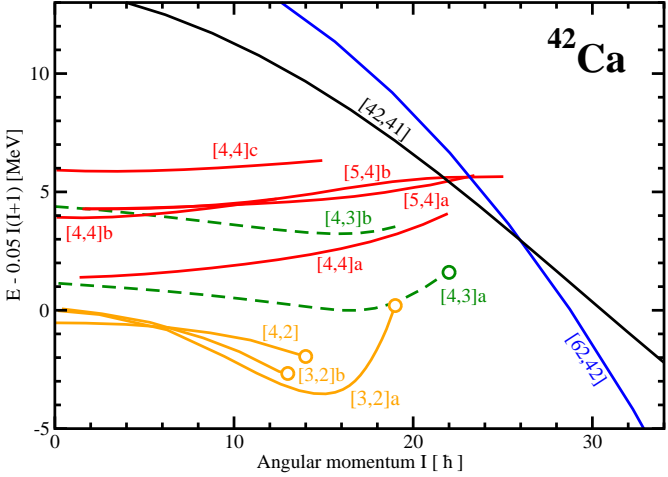}
\caption{(Color online) The same as Fig.\ \ref{Ca40-eld} but for 
$^{42}$Ca.}
\label{Ca42-eld}
\end{figure}

  At even higher spin (above $I=30\hbar$), the yrast line is formed by
the megadeformed configuration [421,421] (Fig.\ \ref{Sc42-eld}). This
configuration contains the proton and neutron in the lowest megaintruder 
$N=5$ orbital above the unpaired band crossing at $\Omega_x \sim 1.80$ 
MeV (above $I=31\hbar$). 
At  lower spin the structure of this MD configuration is [52,52]; this 
is a result of unpaired band crossing emerging from the interaction of the
lowest megaintruder $(N=5)(r=+i)$ orbital with the $1/2[321](r=+i)$ 
orbital taking place at $\Omega \sim 1.8$ MeV (Fig.\ \ref{routh-md-40ca}). 
Note that this band crossing is blocked in the closely lying [52,52] MD 
configuration, shown by solid blue line in Fig.\ \ref{Sc42-eld}, in which 
the 21th proton and 21th neutron are placed into the $1/2[321](r=-i)$ orbital 
located above the $Z=20$ and $N=20$ MD shell gaps.

 Proton density distributions for the HD configuration [41,41] and MD 
configurations [52,52] and [421,421]  are shown in Fig.\ \ref{density-42Sc}.
The major semi-axis ratio of the proton density distribution increases only 
moderately (from 2.23 to 2.65 [see Table \ref{table-ratio}]) on going from 
the [41,41] configuration to the [52,52] one. However, this transition 
triggers drastic change in transition quadrupole moment $Q_t$; it is 
increased from $Q_t\sim 2.65$ $e$b for the [41,41] configuration to 
$Q_t\sim 4.5$ $e$b for the [52,52] configuration (see Fig.\ 
\ref{Sc42-beta-gamma}). The occupation of the 
megaintruder proton and neutron $N=5$ orbitals leading to the MD [421,421] 
configuration creates both additional elongation of the proton density and 
neck in this density distribution (see bottom panels of 
Fig.\ \ref{density-42Sc}). The [421,421] configuration is the most
elongated structure studied in the present paper. 
%
%
Three-dimensional representation of its
density distribution is shown in Fig.\ 1a of Supplemental Material
(Ref.\ \cite{Sup-A40}).
This density distribution has large semi-axis ratio of 3.40 at $I=31$ 
which is increasing with spin (Table \ref{table-ratio}). In part, this large 
value is a consequence of the development of the neck which leads to small 
semi-axis in the direction perpendicular to elongation.
Note that despite large difference in the semi-axis ratio (3.40 for the
[421,421] configuration and 2.65 for the [52,52] one), the $Q_t$ value
of the [421,421] configuration ($Q_t\sim 5.2$ $e$b at $I=31\hbar$) is only
by 15\% larger that the one for the [52,52] configuration (see Fig.\ 
\ref{Sc42-beta-gamma}). These examples clearly indicate that
there is no simple relation between the semi-axis ratio of the proton 
density distribution and transition quadrupole moments.

\begin{figure}[ht]
\includegraphics[angle=0,width=8.8cm]{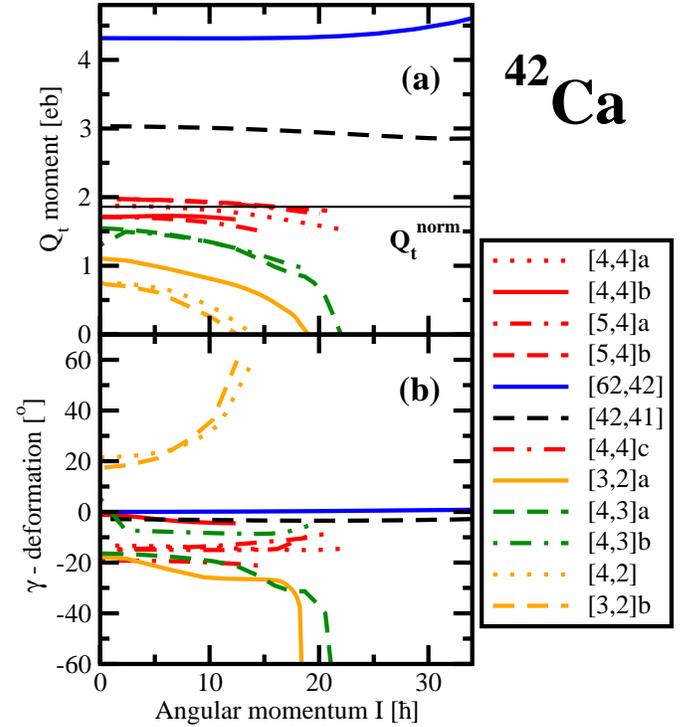}
\caption{(Color online) The same as Fig.\ \ref{Ca40-Qt-gamma} but for $^{42}$Ca.}
\label{Ca42-beta-gamma}
\end{figure}

\section{The general features of high-spin spectra}
\label{sec-general}

 The discussion of low-spin spectra in Secs.\ \ref{sec-ca40}
and \ref{sec-sc42} clearly shows the importance  of particle-hole 
excitations across the $N=20$ and $Z=20$ spherical 
shell gaps in building angular momentum and deformation. The most striking
example is $^{40}$Ca in which only the ground state could be built in
the valence space. Here the valence space is defined as the configuration
space which does not involve particle-hole excitations across either
the $N=20$ and $Z=20$ spherical shell gaps or the $N=28$ and $Z=28$ spherical 
shell gaps.

 The maximum spin which could be built in the valence space
of nuclei is summarized in Table \ref{table-Imax}. It is defined with 
respect of spherical $^{40}$Ca core by the 
occupation of proton and neutron $f_{7/2}$ orbitals in the Ca, 
Sc, Ti and Cr nuclei and by the proton and neutron holes in 
the $N=2$ $d_{3/2}$, $s_{1/2}$ and $d_{5/2}$ orbitals in the Ar 
and S nuclei. One can see that the maximum spin increases on 
approaching the middle of the $f_{7/2}$ subshell where it 
reaches the maximum value of $I_{max}=16\hbar$ in the 
$\pi (f_{7/2})^4_{8.0} \otimes \nu (f_{7/2})^4_{8.0}$ configuration
of $^{48}$Cr. Note that the addition of two neutrons to this configuration 
decreases the maximum spin which could be built in the valence 
space of $^{50}$Cr (see Table \ref{table-Imax}).

 Table \ref{table-Imax} also illustrates how the maximum spin, which
could be built within the configuration, changes when particle-hole
excitations across the spherical $Z=20$ and $N=20$ spherical shell
gaps are involved. Here 2p-2h configurations are defined as the
configurations which involve the excitations of one proton and one
neutron across the respective shell gaps.  The excitations of two
protons and two neutrons across these gaps lead to the 4p-4h 
configurations. The impact of these excitations on the maximum
spin depend on how many occupied $f_{7/2}$ orbitals in the Ca, Sc, 
Ti and Cr nuclei (or holes in the $N=2$ orbitals of the Ar and S 
nuclei) the nucleus has in its valence space. One can see that these 
excitations increase drastically the maximum spin within the 
configurations of $^{40}$Ca but have limited impact on maximum spin 
in $^{48}$Cr (Table \ref{table-Imax}).

  The analysis of the $^{40}$Ca and $^{42}$Sc nuclei clearly 
indicates that subsequent particle-hole excitations lead to the 
yrast or near-yrast SD, HD and MD configurations at the spins 
which are either similar to the maximum spins which could be built 
within the 2p-2h and 4p-4h configurations or slightly above them. 
Note that the nuclei in these 2p-2h and 4p-4h configurations 
could at most be described as highly-deformed.

  The importance of these 2p-2h and 4p-4h configurations lies
in the fact that at low and medium spins they dominate the yrast 
line and  thus are expected to be populated in experiment with 
high intensity. The observation of the SD, HD and MD configurations
requires that these bands are either yrast or close to yrast
at the spins where the feeding of the bands takes place. This is 
especially critical for the HD and MD bands since in most of the 
nuclei they have completely different slope in the $E-E_{RLD}$ 
plots as compared with the bands of smaller deformation. As 
a result, their excitation energies with respect to the yrast 
line grow up very rapidly with decreasing spin below the 
spins where the HD and MD bands are yrast or near yrast. This 
factor will limit the spin range in which they can be observed 
in future experiments to the spin range in which these bands 
are either yrast or close to yrast and few states below this 
spin range.

  Considering the limitations of experimental facilities to 
observe high spin states in light nuclei, it is imperative 
that expected candidates for the SD, HD and MD bands become yrast 
(or close to yrast) at the spins which are not far away from 
currently measured. Indeed, with current generation of  
detectors the SD bands in $^{36}$Ar \cite{36Ar-SD.PRL.00} and 
$^{40}$Ca \cite{Ca40-PRL.01} and ground state band in $^{48}$Cr 
\cite{Cr48-DSAM.98} are observed up to $I=16\hbar$ which represents 
the highest spin measured in this mass region. The advent 
of $\gamma$-tracking detectors such as GRETA and AGATA will 
increase the spin up to which the measurements could be performed 
but this increment in spin is not expected to be drastic.

  Note that the results discussed in this section only illustrate
the general features of rotating nuclei and provide some crude 
estimates of the competition of terminating and extremely deformed 
configurations. Indeed, detailed calculations are needed to define 
the properties of such bands and
the spins at which extremely deformed configurations become 
yrast. For the sake of simplicity, we also do not discuss here possible 
excitations across the spherical $N=28$ and $Z=28$ shell gaps. 
Terminating configurations based on such excitations compete with 
the SD, HD and MD configurations only in $^{46}$Ti and Cr nuclei
(see Secs.\ \ref{sec-ti46}, \ref{sec-cr48} and \ref{sec-cr50} below).

  The examples of the $^{40}$Ca and $^{42}$Sc nuclei discussed above 
once more confirm that rotating nuclei are the best laboratories 
to study the shape coexistence. Indeed, starting from either 
spherical or weakly deformed ground states by means of subsequent 
particle-hole excitations one can built any shape (prolate, oblate, 
triaxial, super-, hyper- and megadeformed as well as cluster and/or 
molecular structures [see Sec.\ \ref{sec-clus} below for a discussion of 
latter structures/shapes]) in the same nucleus.

\section{Other nuclei in the neigbourhood of $^{40}$Ca}
\label{sec-other}
   
 The results for other nuclei studied in this paper will be presented
in this section. In the calculations of terminating structures at
low and medium spins we concentrate on the configurations which define
the general structure of the yrast line and the spins at which the 
transition to extremely deformed configurations takes place. Apart from
few interesting cases, we do not discuss them in detail. The main focus
of this section is on the super-, hyper- and megadeformed rotational 
configurations and, in particular, on the ones which potentially show 
the features of clusterization and molecular structures. In order to 
provide the guidance for future experiments, we present 
the $(E-E_{RLD})$ plots and the figures with transition quadrupole moments 
and $\gamma$-deformations for each nucleus. In addition, the proton 
density plots are provided 
for the yrast or near-yrast configurations which could be measured in future 
experiments. The goals behind that is to see the evolution of the density 
distribution with configuration and nucleus and to find interesting candidates 
for clusterization and molecular structures.  Note that some graphical 
results of the calculations are provided in the Supplemental Material with 
this article as Ref.\ \cite{Sup-A40}.

\subsection{$^{42}$Ca nucleus}
\label{Ca42-section}

 The energies of calculated configurations are shown in Fig.\ \ref{Ca42-eld}. 
The calculated transition quadrupole moments and $\gamma$-deformations of these 
configurations are displayed in Fig.\ \ref{Ca42-beta-gamma}. Note that this nucleus 
has two extra neutrons as compared with $^{40}$Ca which affects the structure of
the configurations.


 The SD configurations are represented by the [4,4]a, [4,4]b, [4,4]c, [5,4]a 
and [5,4]b configurations (Fig.\ \ref{Ca42-eld}). The yrast [4,4]a SD configuration 
in $^{42}$Ca is formed by an addition of two extra neutrons in the 1/2[200] 
orbitals (located above the $N=20$ SD  shell gap [Fig.\ \ref{routh-sd-1}]) 
to the yrast [4,4] SD configuration of $^{40}$Ca. Note that similar to $^{40}$Ca 
the SD configurations in $^{42}$Ca are triaxial with $\gamma \sim -12^{\circ}$.

  At spin $I\sim 23\hbar$, the expected continuation of the SD [4,4] configuration
is crossed by the HD [42,41] configuration which is formed from the yrast HD [41,41] 
configuration in $^{40}$Ca (Fig.\ \ref{routh-hd-ca40}) by adding one neutron into the 
$1/2[440](r=+i)$ orbital and another into the $5/2[202](r=+i)$ orbital. The HD 
[42,41] configuration has near prolate shape with $Q_t$ value which exceed by 
50\% the $Q_t$ values which are typical for the SD bands (Fig.\ 
\ref{Ca42-beta-gamma}).

\begin{figure}[ht]
\includegraphics[angle=0,width=8.8cm]{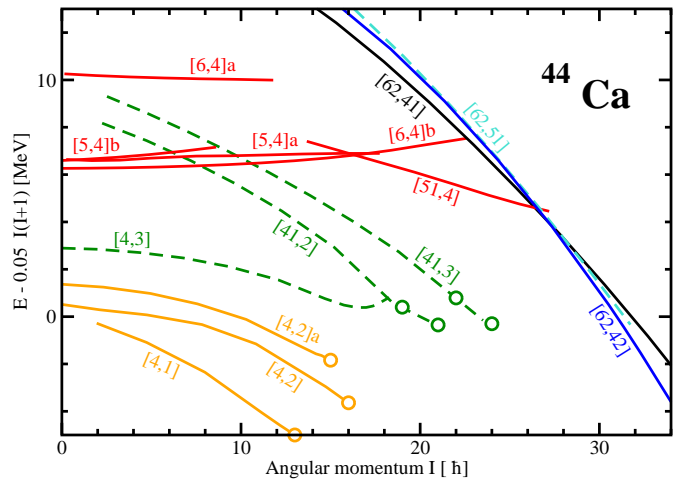}
\caption{(Color online) The same as Fig.\ \ref{Ca40-eld} but for 
$^{44}$Ca. Note that for some configurations several aligned 
(terminating) states can be formed; the reasons for their
formation are discussed in Sec.\ 6.6 of Ref.\ \cite{PhysRep-SBT}.}
\label{Ca44-eld}
\end{figure}

   Axially symmetric MD configuration [62,42] becomes yrast above $I=26\hbar$ 
(Fig.\ \ref{Ca42-eld}). At these spins, its transition quadrupole moment is by 
$\sim 50\%$ larger than that of the HD [42,41] configuration (Fig.\ 
\ref{Ca42-beta-gamma}).

Proton density distributions of the yrast SD [4,4]a and MD [62,42] 
configurations are shown in Fig.\ 2a of supplemental material (Ref.\ 
\cite{Sup-A40}) and Figs.\ \ref{density-rot}c and d below, respectively.  
The major semi-axis ratio is 2.17 and 2.79 for these configurations (Table 
\ref{table-ratio}). However, the $Q_t$ values of the latter configuration 
are by a factor of more than 2 larger than those of the former one 
(Fig.\ \ref{Ca42-beta-gamma}).

\subsection{The $^{44}$Ca nucleus}
\label{Ca44-sect}

  The SD configurations are represented by the [6,4] and [51,4] configurations
(Fig.\ \ref{Ca44-eld}). However, up to spin $I\sim 22\hbar$ the total yrast line 
is formed by the normal- and highly-deformed triaxial terminating configurations 
and these SD configurations are located at high excitation energy with respect 
to the total yrast line. Only around $I\sim 24\hbar$, the [51,4] SD configuration 
becomes yrast. However, already at spin $I=26\hbar$ and above the yrast line is 
formed by closely lying [62,41] HD and [62,42] MD configurations. 

The calculated values of the transition quadrupole moment $Q_t$ of these 
configurations are shown in Fig.\ 3 of supplemental material (Ref.\ 
\cite{Sup-A40}).
Proton density distribution of the MD [62,42] 
configuration is displayed in Fig.\ 2b in the supplement
to present paper (Ref.\ \cite{Sup-A40}).
The [62,51] configuration,
shown by dashed cyan line in Fig.\ \ref{Ca44-eld}, is lying closely in energy to 
these two configurations. Its calculated $Q_t$ values are in the between the ones 
for the HD and MD configurations 
(see Fig.\ 3 in the supplement).

  The yrast line of $^{44}$Ca shows that with increasing neutron number up to
$N=24$ (which leads to the placement of the neutron Fermi level in the middle of
deformed single-particle states emerging from spherical $1f_{7/2}$ subshell) it
become energetically favorable to excite the neutron across the spherical $N=28$
shell gap. Such excitation leads to the occupation of the lowest $1g_{9/2}$ 
neutron orbital (which carries substantial single-particle angular momentum) and, in the 
case of $^{44}$Ca, to the formation of the [41,2] and [41,3] configurations which 
terminate at spins $I=21, 22$ and 24$\hbar$ (Fig.\ \ref{Ca44-eld}). As a  
consequence, the yrast line could be built by either normal- or highly-deformed 
terminating configurations up to higher spins in $^{44}$Ca as compared with lighter 
Ca isotopes (compare Figs.\ \ref{Ca44-eld}, \ref{Ca42-eld} and \ref{Ca40-eld}).  As 
a result, the observation of the SD, HD and MD configurations would be more difficult 
in heavier Ca isotopes as compared with $^{40}$Ca. Note that this mechanism of neutron 
excitations across the $N=28$ spherical shell gap affects also the yrast line in
$^{46}$Ti (configuration [41,4], Fig.\ \ref{Ti46-eld}). Similar proton 
excitation across the $Z=28$ spherical shell gap become active in the Cr 
isotopes. Indeed the configurations of the type [*1,*1] built on simultaneous
neutron and proton excitations across the $N=28$ and $Z=28$ spherical shell
gaps are active in the creation of the yrast line at medium spin in $^{48,50}$Cr (see 
Figs.\ \ref{Cr48-eld} and \ref{Cr50-eld} below).

\begin{figure}[ht]
\includegraphics[angle=0,width=8.8cm]{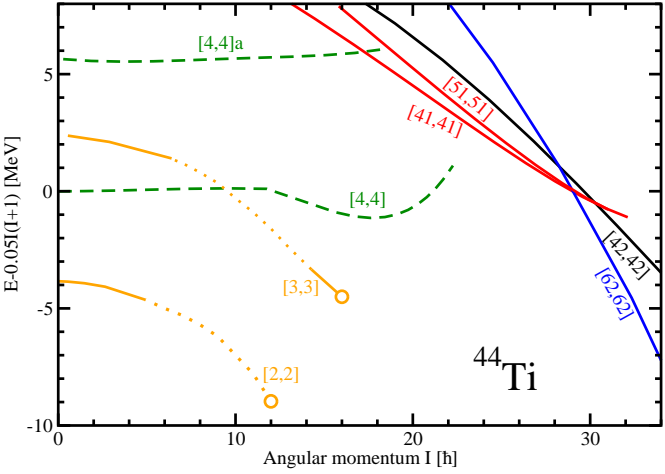}
\caption{ (Color online) The same as Fig.\ \protect\ref{Ca40-eld} but for 
          $^{44}$Ti. Note that is was not possible to trace the middle
          parts of the [2,2] and [3,3] configurations in the 
          calculations. Thus, they are shown by dotted lines.}
\label{Ti44-eld}
\end{figure}

\begin{figure}[ht]
\includegraphics[angle=0,width=8.8cm]{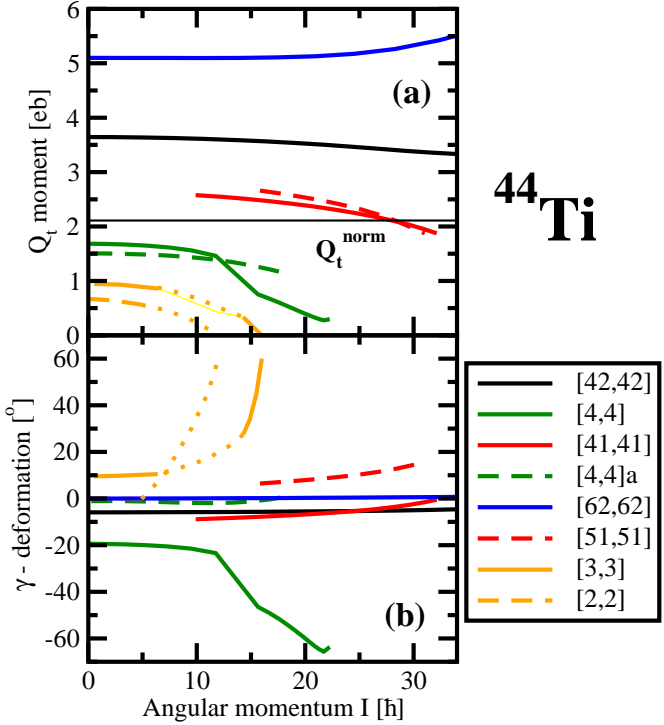}
\caption{ (Color online) The same as Fig.\ \protect\ref{Ca40-Qt-gamma} but for 
         $^{44}$Ti.}
\label{Ti44-qt-gamma}
\end{figure}

\begin{figure}[ht]
\includegraphics[angle=0,width=8.8cm]{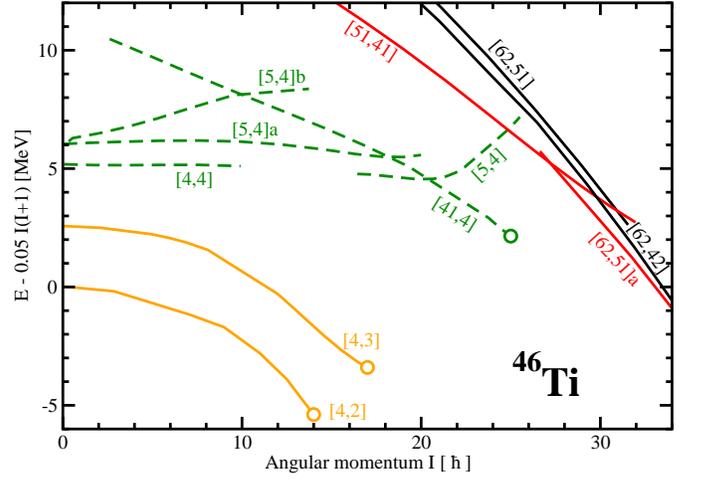}
\caption{ (Color online) The same as Fig.\ \ref{Ca40-eld} but for 
          $^{46}$Ti.}
\label{Ti46-eld}
\end{figure}

\subsection{The $^{44}$Ti nucleus}

 The SD configurations are represented by the [41,41] and [51,51]
configurations (see Figs.\ \ref{Ti44-eld} and \ref{Ti44-qt-gamma}). 
These configurations are either yrast or very close to yrast line above 
$I=24\hbar$ after the crossing with expected continuation of the [4,4]
configuration. The
[4,4] and [4,4]a configurations may also be considered as SD at low spin
since they are located at the borderline between the SD and highly-deformed
bands. The increase of the proton and 
neutron numbers to 22 leads to a decrease of the impact of the hyperintruder 
$N=4$ orbitals. Indeed, the [41,41] and [51,51] configurations have transition 
quadrupole moments $Q_t$ which are only slightly above the one typical for the 
SD configurations (Fig.\ \ref{Ti44-qt-gamma}).

 Only the occupation of additional proton and neutron $N=4$ hyperintruder 
orbitals leads to the configuration [42,42] (see Fig.\ 
\ref{Ti44-eld}) which is truly hyperdeformed (Fig.\ \ref{Ti44-qt-gamma}).
However, the HD configurations are never yrast in this nucleus.
At spin $I\sim 29\hbar$, the MD [62,62] configuration becomes yrast. 
Proton density distribution of this configuration (see Figs.\ \ref{density-rot}e 
and f below) could be compared with the one for the SD [41,41] configuration 
(see Fig. 2c in the supplemental material (Ref.\ \cite{Sup-A40})). Its three 
dimensional representation is shown in Fig.\ 1b of the supplement to this 
manuscript.
The $Q_t$ value of the MD configuration is larger than the one for the SD 
configuration by a factor of approximately 2.5 (Fig.\ \ref{Ti44-qt-gamma}). On 
the other hand, the difference in the ratio of major semi-axis of the density 
distribution of these two configurations is smaller (the major semi-axis ratio
is 2.88 for the [62,62] configuration and 2.03 for the [42,42] configuration 
[Table \ref{table-ratio}]).

\subsection{The $^{46}$Ti nucleus}
\label{sec-ti46}

\begin{figure}[ht]
\includegraphics[angle=0,width=8.8cm]{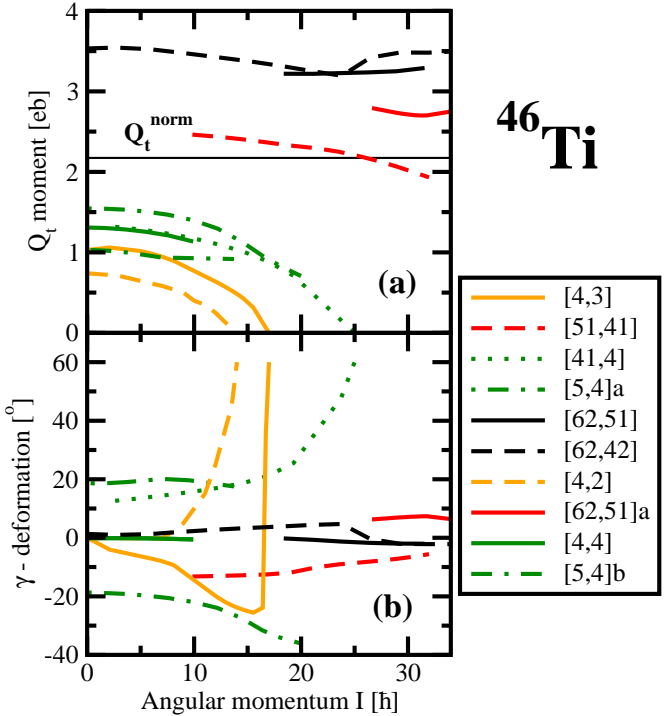}
\caption{(Color online) The same as Fig.\ \ref{Ca40-Qt-gamma} but for 
         $^{46}$Ti.}
\label{Ti46-beta-gamma}
\end{figure}

  As discussed in Sec.\ \ref{Ca44-sect}, the increase of neutron number
to $N=24$ leads to a situation in which the particle-hole excitations 
across the $N=28$ spherical shell gap create the configurations which 
contribute to the yrast line at medium spin (the [4,4] configuration terminating 
at $I=25\hbar$, Fig.\ \ref{Ti46-eld}). At higher spin the SD [62,51]a 
configuration becomes yrast. The HD [62,42] configuration is only slightly 
excited in energy with respect to this configuration. Proton density distributions 
of these two configurations are shown 
in Figs.\ 2d and e of supplemental material (Ref.\ \cite{Sup-A40}). 
Note that 
the MD configurations are not energetically favored in this  nucleus and they do 
not show up in the vicinity of the yrast line in the spin range of interest. The 
calculated $Q_t$ and $\gamma$-deformation values of the configurations displayed 
in Fig.\ \ref{Ti46-eld} are summarized in Fig.\ \ref{Ti46-beta-gamma}.

\subsection{$^{48}$Cr nucleus}
\label{sec-cr48}

\begin{figure}[ht]
\includegraphics[angle=0,width=8.8cm]{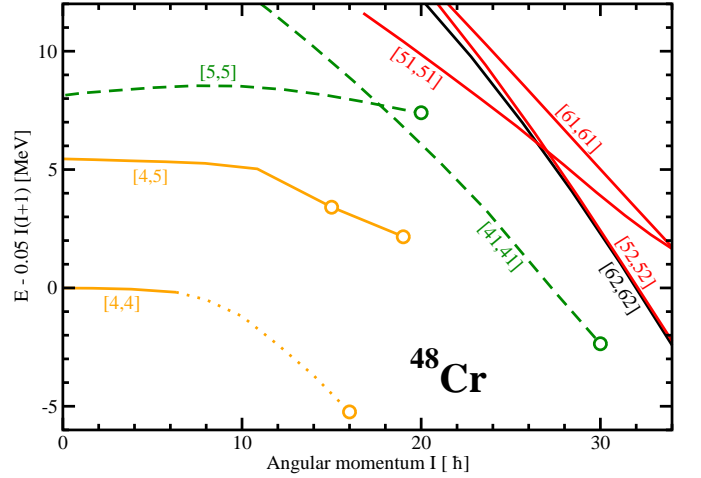}
\caption{(Color online) The same as Fig.\ \ref{Ca40-eld} but for 
          $^{48}$Cr.}
\label{Cr48-eld}
\end{figure}

  The valence space terminating [4,4] configuration is forming the yrast line up 
to $I=16\hbar$ (Fig.\ \ref{Cr48-eld}). This band has been observed in experiment 
up to its termination in Ref.\ \cite{Cr48-DSAM.98}.
Particle-hole excitations across the $Z=20$ and $N=20$ spherical shell gaps lead 
to only marginal increase of angular momentum content of the configurations 
but cost a lot of energy (see, for example, the configurations [4,5] and [5,5] in
Fig.\ \ref{Cr48-eld} and Table \ref{table-Imax}). Higher spin configurations
are built at a reasonable energy cost by particle-hole excitations across 
the $Z=28$ and $N=28$ spherical shell gaps. Such excitations lead both to 
terminating and SD/HD configurations. The first type of configurations is
represented by the [41,41] one which terminates at $I=30\hbar$. The SD [52,52] 
and HD [62,62] configurations lying at similar energies become the lowest 
configurations at spin $I\geq 30\hbar$ (Fig.\ \ref{Cr48-eld}).
Note that the resonance observed at $I\sim 36\hbar$ in the $^{24}$Mg+$^{24}$Mg 
reaction strongly supports a HD shape for a compound $^{48}$Cr nucleus formed
in this reaction \cite{48Cr.HD.16}. The evolution of proton density distribution 
with spin in the [62,62] configuration is shown in Figs.\ 2f and g of the 
supplemental material (Ref.\ \cite{Sup-A40}). The $Q_t$ and $\gamma$-deformation 
values are summarized in Fig.\ 5 of the supplemental material.

\subsection{$^{50}$Cr nucleus}
\label{sec-cr50}
 
\begin{figure}[htp]
\includegraphics[angle=0,width=8.8cm]{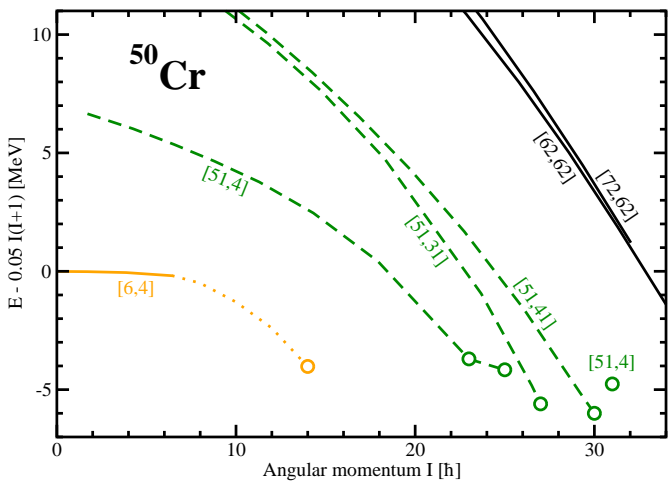}
\caption{(Color online) The same as Fig.\ \ref{Ca40-eld} but for 
          $^{50}$Cr.}
\label{Cr50-eld}
\end{figure}

  The calculated configurations are shown in Fig.\ \ref{Cr50-eld}. Below spin
$I=14\hbar$, the yrast line is built from valence space [6,4] configuration. 
Higher spin terminating configurations ([51,4], [51,31] and [51,41]) are build 
by means of particle-hole excitations across the $Z=28$ and $N=28$ spherical 
shell gaps. They dominate the yrast line up to $I=31\hbar$. At even higher spin 
closely lying HD [62,62] and [72,62] configurations are either yrast or close 
to yrast. Transition quadrupole moments $Q_t$ and $\gamma$-deformations of the 
calculated configurations are summarized in Fig.\ 6 of the supplemental 
material (Ref.\ \cite{Sup-A40}). An example of proton density distribution is shown 
in Fig.\ 2h of the supplemental material for the HD [62,62] configuration at 
$I=31\hbar$. Note that neither superdeformed nor megadeformed configurations show 
up in the vicinity of the yrast line in this nucleus in the spin range of interest.

\subsection{$^{36}$Ar nucleus}

\begin{figure}[htp]
\includegraphics[angle=0,width=8.8cm]{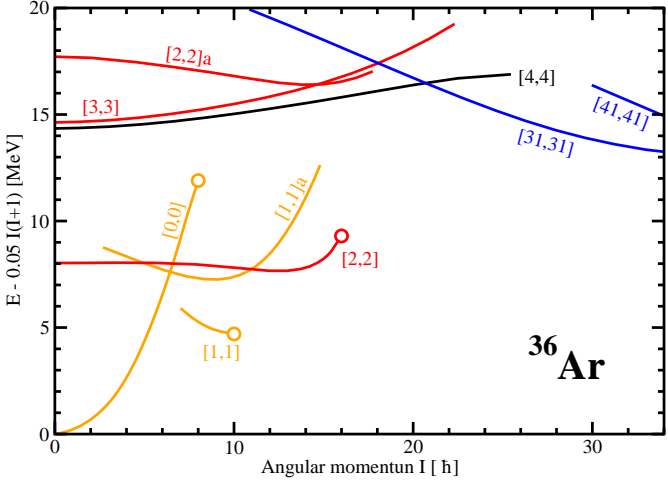}
\caption{(Color online) The same as Fig.\ \ref{Ca40-eld} but for 
          $^{36}$Ar.}
\label{Ar36-eld}
\end{figure}

  The maximum spin which could be built in the valence space of this nucleus is
quite limited, namely, $8\hbar$ in the [0,0] configuration (Fig.\ \ref{Ar36-eld}
and Table \ref{table-Imax}). Particle-hole excitations leading to the [1,1] 
configurations increase this spin up to $16\hbar$ (configuration [1,1]a in Fig.\ 
\ref{Ar36-eld}).  Subsequent particle-hole excitations generate the [2,2] 
configurations the maximum spin within which is $20\hbar$ (see Table 
\ref{table-Imax}).

\begin{figure}[htp]
\includegraphics[angle=0,width=8.8cm]{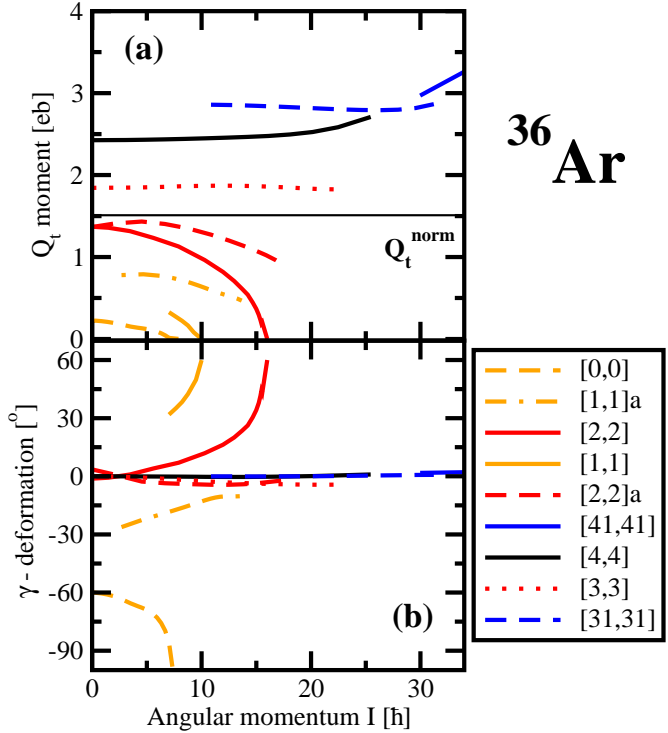}
\caption{(Color online) The same as Fig.\ \ref{Ca40-Qt-gamma} but for 
          $^{36}$Ar.}
\label{Ar36-qt-gamma}
\end{figure}

  One of such configurations terminating at 
$I=16\hbar$ is assigned to the SD band observed in Refs.\ 
\cite{36Ar-SD.PRL.00,Ar36-SD-Qt.01}. Its properties have been studied earlier
within the spherical shell model \cite{Ar36-SM.05} and cranked Nilsson-Strutinsky 
approach \cite{36Ar-SD.PRL.00,Ar36-SD-Qt.01}. Based on the calculated deformation 
properties, this configuration could be considered as SD only at low spin. This is
because with increasing spin its transition quadrupole moment $Q_t$ is decreasing
rapidly and $\gamma$-deformation is increasing up to $\gamma=60^{\circ}$ (Fig.\ 
\ref{Ar36-qt-gamma}). Terminating state of this configuration/band is reached at 
$I=16^+$ both in theory (Fig.\ \ref{Ar36-eld}) and in experiment (Refs.\ 
\cite{36Ar-SD.PRL.00,Ar36-SD-Qt.01}). From our point of view, the classification 
of this band as highly-deformed triaxial is more appropriate but we label it as
SD in Fig.\ \ref{Ca40-eld} following the classification established in the 
literature.

  At spin above $I=16\hbar$, the yrast line is built from the HD [4,4] and MD 
[31,31] configurations. It is easy to understand the structure of these 
configurations from the routhian plot for the [42,42] MD configuration in 
$^{40}$Ca (Fig.\ \ref{routh-md-40ca}). The [4,4] configuration in $^{36}$Ar 
is built by removing two protons and two neutrons in the $N=4$ orbitals from 
the MD [42,42] configuration in $^{40}$Ca. The [31,31] configurations in 
$^{36}$Ar are built by removing one proton and one neutron in the $N=4$ orbital 
from the [42,42] configuration in $^{40}$Ca and another proton and another neutron 
from the $3/2[321]$ orbital of the same configuration. Since opposite signature 
branches of the $3/2[321]$ orbital are either degenerate in energy (as in Fig.\ 
\ref{routh-md-40ca}) or have very small energy splitting, four [31,31] configurations 
are calculated at close energies. For simplicity, we show only the lowest one
in Fig.\ \ref{Ar36-eld}.

  Note that in many calculations these extremely deformed structures do not 
form a stable minimum in the potential energy surfaces at spin zero (see,
for example, Fig.\ 2 
in Ref.\ \cite{EKNV.14}). Thus, the rotation helps to stabilize this minimum. This 
is similar to the situation with the stabilization of hyperdeformation at high 
spin in medium mass nuclei (Ref.\ \cite{AA.08}).

\begin{figure*}[htp]
\includegraphics[angle=0,width=8.8cm]{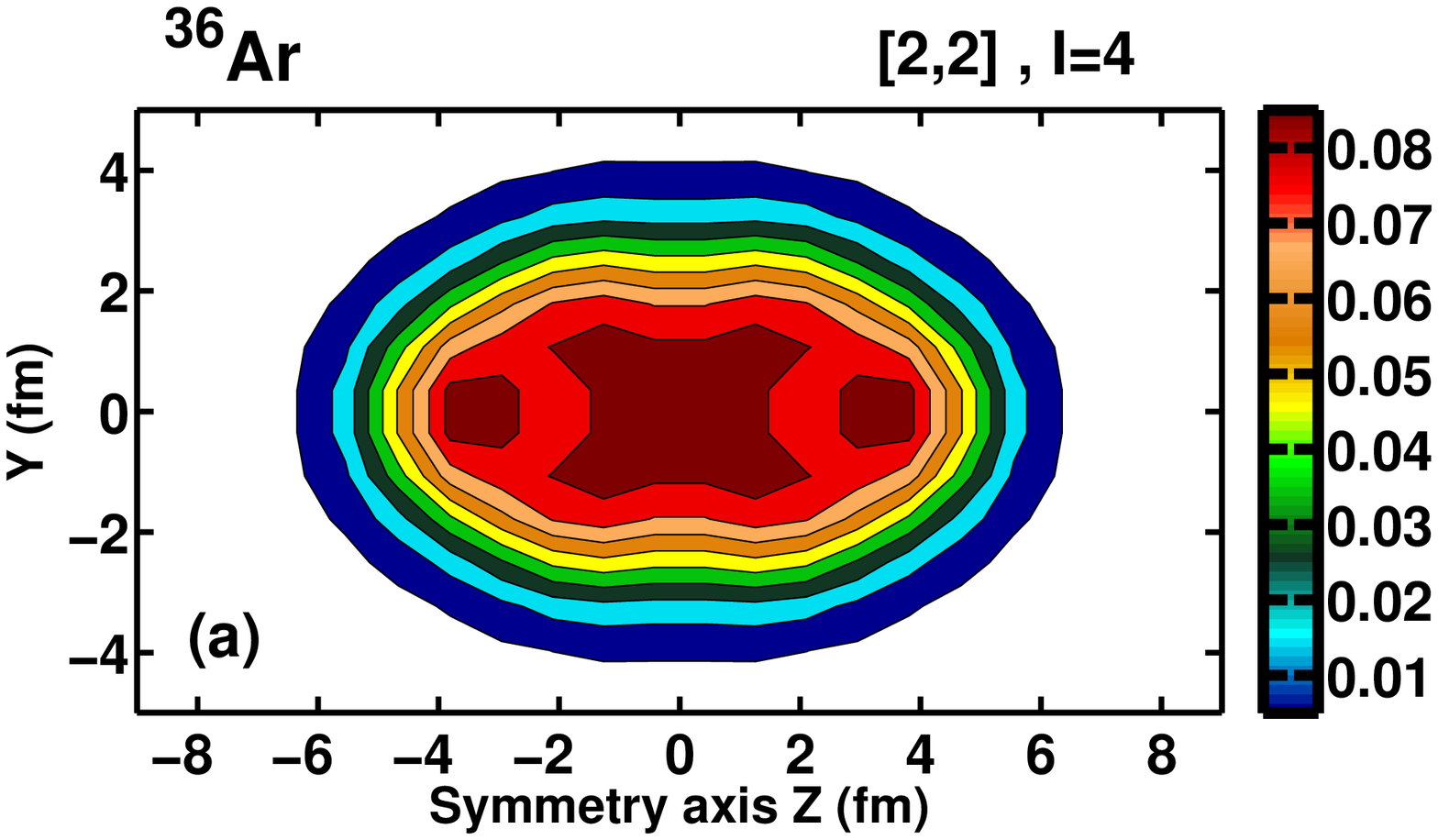}
\includegraphics[angle=0,width=8.8cm]{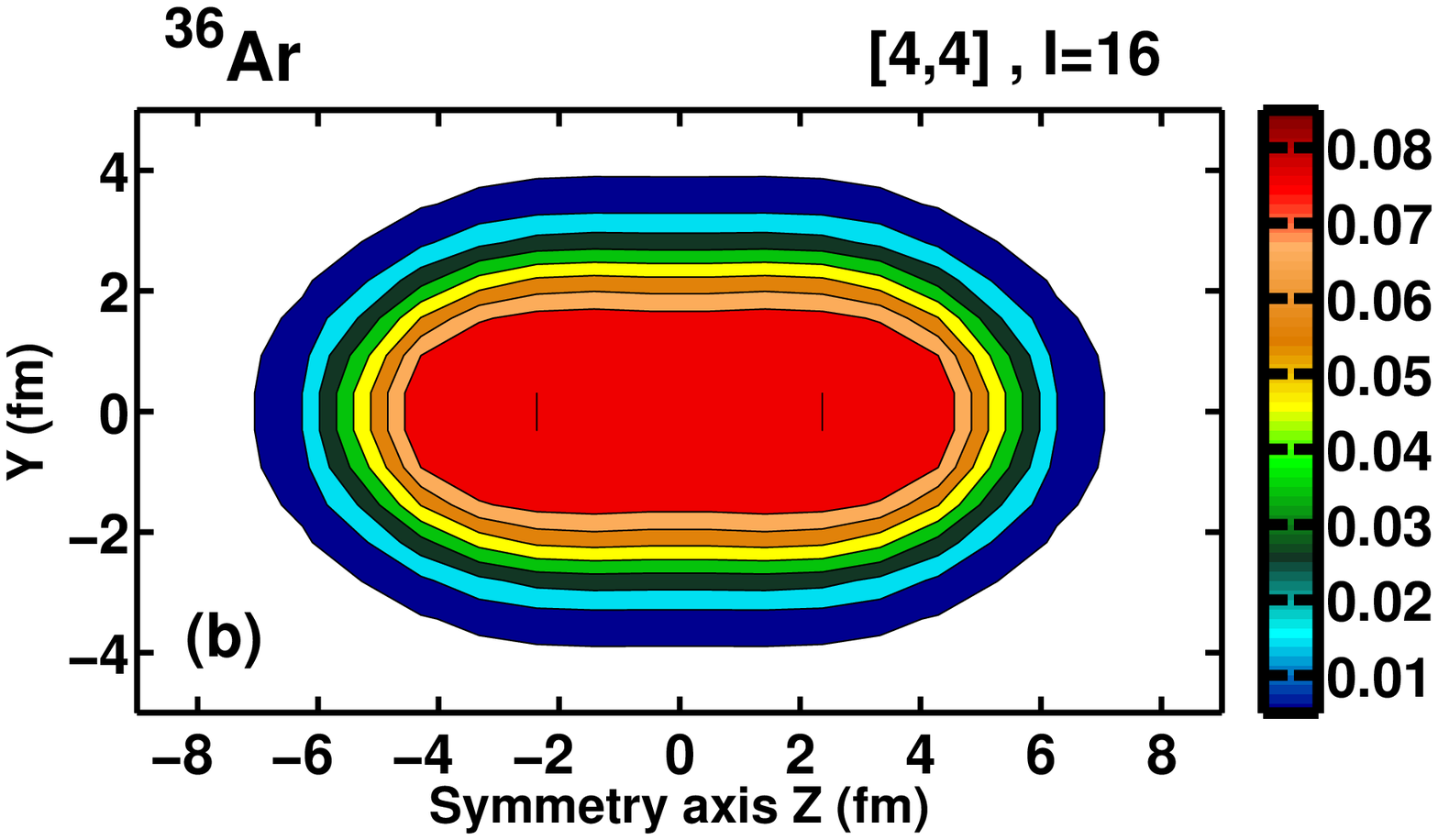}
\includegraphics[angle=0,width=8.8cm]{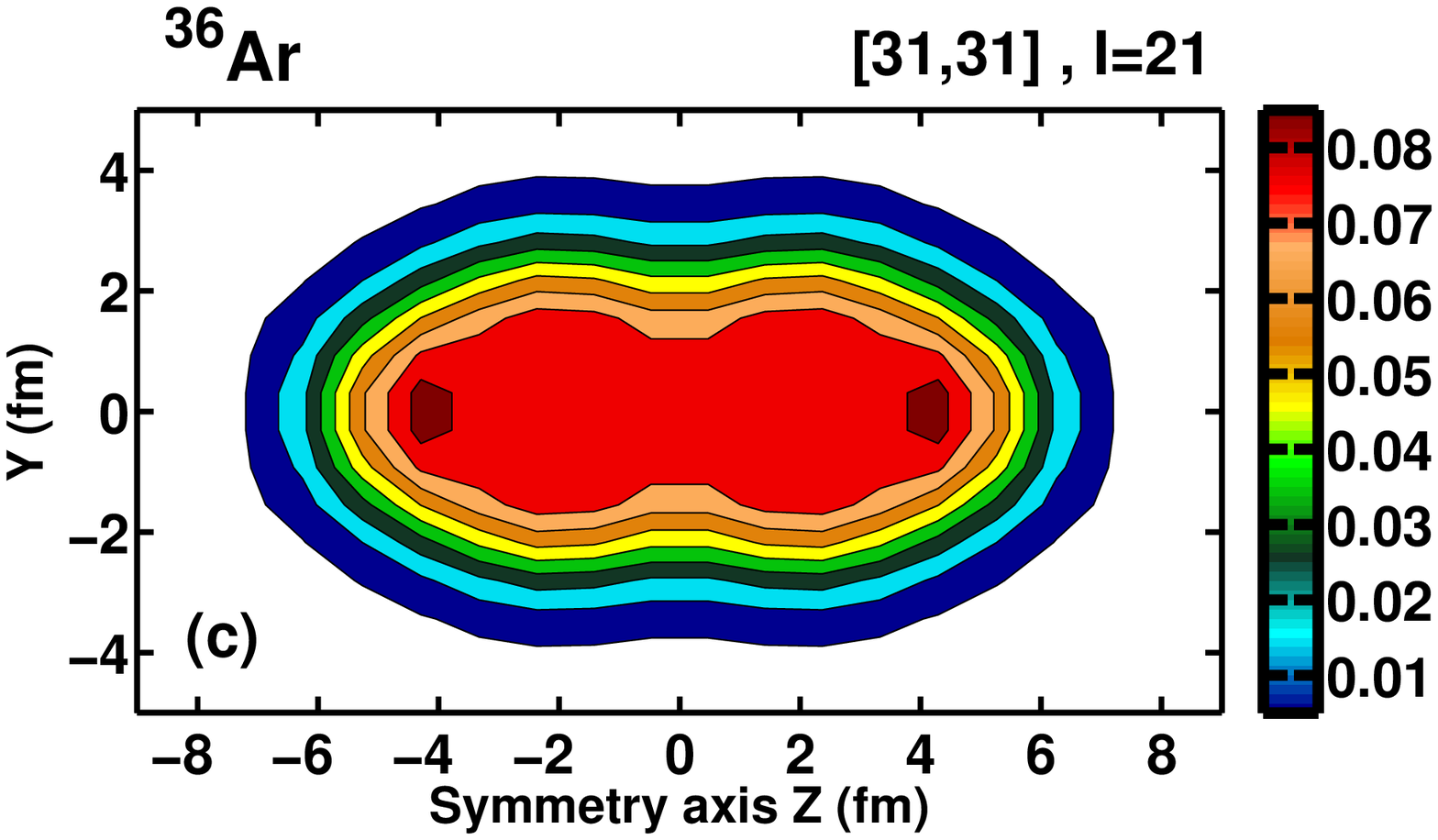}
\includegraphics[angle=0,width=8.8cm]{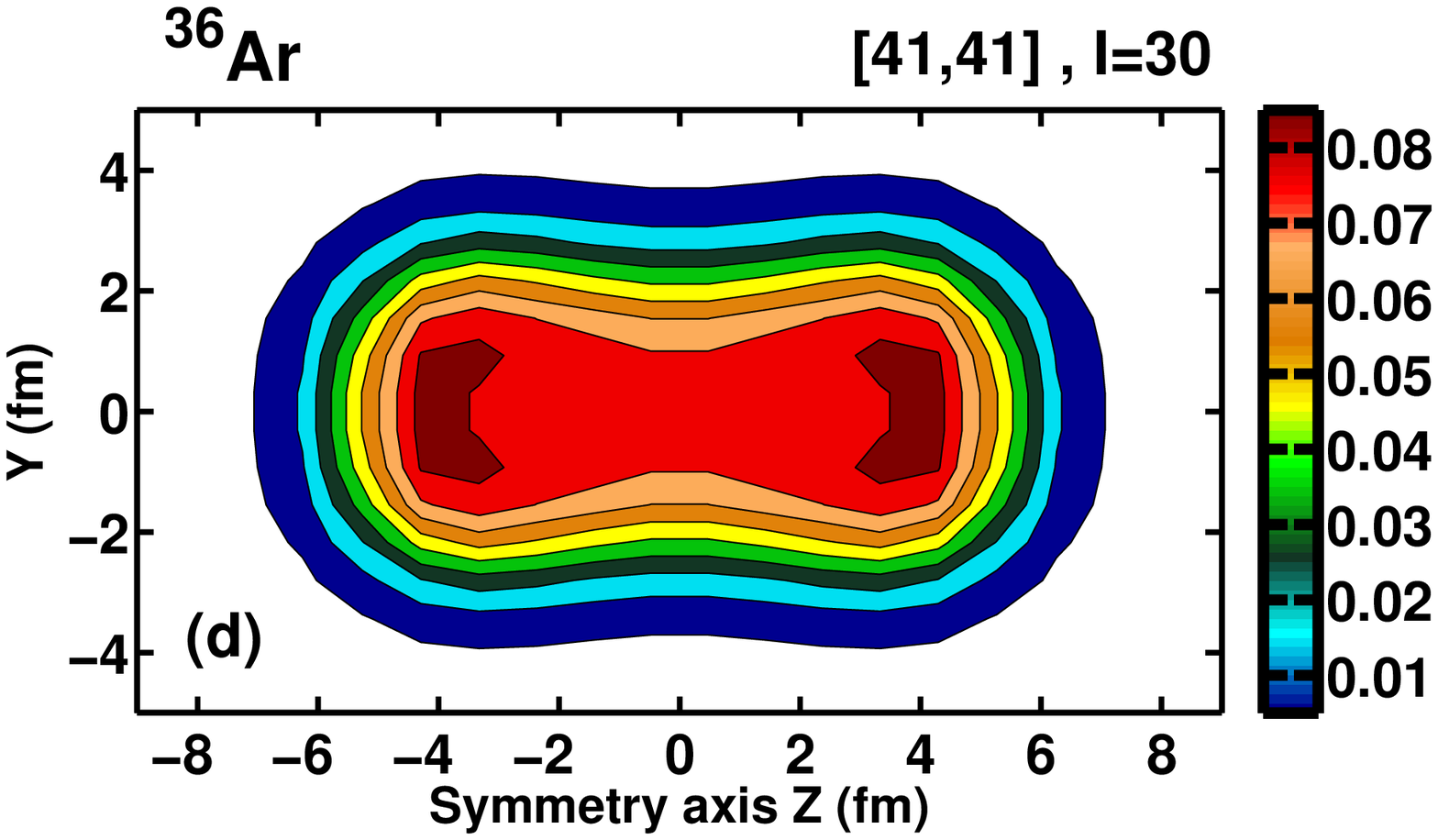}
\caption{(Color online) The same as Fig.\ \ref{density-Ca40} 
          but for $^{36}$Ar.}
\label{density-Ar36}
\end{figure*}

  The transition quadrupole moments $Q_t$ and $\gamma$-deformations of the
calculated configurations are summarized in Fig.\ \ref{Ar36-qt-gamma}. One
can see that the $Q_t$ values of the [4,4] and [31,31] MD configurations are 
by approximately 66\% and 100\% larger than the normalized transition quadrupole 
moment $Q_t^{norm}$ for the SD shapes. Proton density distributions of the
SD [2,2] (at low spin), HD [4,4] and MD [31,31] and [41,41] configurations
are shown in Fig.\ \ref{density-Ar36} (see also Table \ref{table-ratio}
for the density semi-axis ratios).  
Three-dimensional representation of the proton density distribution in the 
MD [31,31] configuration is shown in Fig.\ 1c of the supplemental material
(Ref.\ \cite{Sup-A40}).
The SD shapes are characterized by more compact 
(with higher average density in the interior of the nucleus) density 
distribution as compared with the HD and MD shapes (Fig.\ \ref{density-Ar36}).
The formation of the necking degree of freedom is clearly seen in the
MD [31,31] and [41,41] configurations.

  The results obtained in the cranked Nilsson-Strutinsky (CNS) approach are 
very similar to the CRMF ones (see Fig. 6 in Ref.\ \cite{Pingst-A30-60}). 
Indeed, the MD [31,31] configurations are yrast above spin $I=18\hbar$ also 
in the CNS calculations. Considering that both experimental data in this nucleus 
extends up to $I=16^+$ and yrast or near yrast higher spin configurations 
are formed from HD and/or MD ones, the calculations in the CRMF and CNS 
frameworks clearly indicate this nucleus as one of the best candidates for 
the observation of the hyper- and megadeformations. In simple words, if
it will be possible to bring higher (than $16\hbar$) angular momentum into 
this system, the population of the HD and MD states is the most likely 
outcome of this process.

\subsection{$^{38}$Ar nucleus}

\begin{figure}[ht]
\includegraphics[angle=0,width=8.8cm]{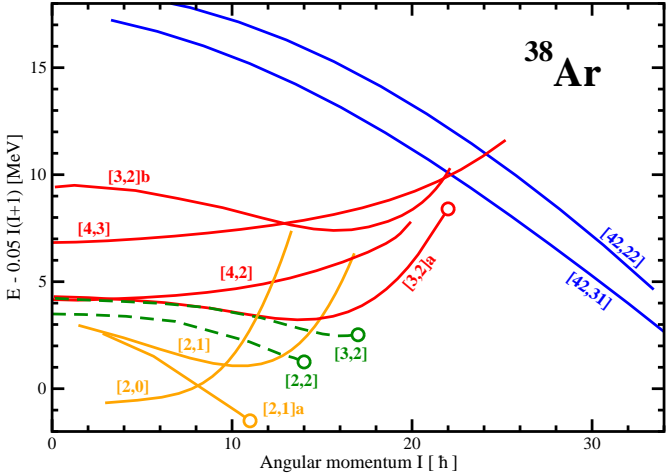}
\caption{(Color online) The same as Fig.\ \ref{Ca40-eld} but for 
          $^{38}$Ar.}
\label{Ar38-eld}
\end{figure}

 Low-spin yrast line in this nucleus is built from terminating configurations
(Fig.\ \ref{Ar38-eld}). The lowest SD configuration [3,2]a is close to the yrast 
line at medium spin and it becomes yrast above $I=17\hbar$ (Fig.\ \ref{Ar38-eld}).
Similar to the observed SD band in $^{36}$Ar, it terminates in non-collective state 
but at higher spin of $I=22\hbar$. Its proton density distribution at the medium 
spin of $I=12\hbar$ is shown in Fig.\ 3a of the supplemental material (Ref.\ 
\cite{Sup-A40}). The yrast line above $I=22\hbar$ is built from the [42,31] MD 
signature partner configurations with small energy splitting (Fig.\ \ref{Ar38-eld}). 
They differ in the occupation by third proton of different signatures of the 3/2[321] 
orbital; there is almost no signature splitting between the different signatures of 
this orbital (see Fig.\ \ref{routh-md-40ca}). 
An interesting feature of this configuration is substantial impact of rotation 
on the density distribution leading to a larger elongation and more pronounced  
necking with increasing spin from $I=24\hbar$ to $I=32\hbar$ (see Figs.\ \ref{density-rot}a
and b below). This however is not associated with the 
substantial change of transition quadrupole moment $Q_t$ (Fig.\ \ref{Ar38-qt-gamma}).

\begin{figure}[ht]
\includegraphics[angle=0,width=8.8cm]{fig-26.eps}
\caption{(Color online) The same as Fig.\ \ref{Ca40-Qt-gamma} but for $^{38}$Ar.}
\label{Ar38-qt-gamma}
\end{figure}

\subsection{$^{32}$S nucleus}

  The SD configurations were predicted in $^{32}$S long time ago in Refs.\ 
\cite{SRN.72,LL.75}. The SD bands built on such structures have been studied both 
in non-relativistic cranked DFTs based on the Gogny \cite{TNI.01,RER.00} and 
Skyrme \cite{YM.00,MDD.00} forces and in the CRMF calculations with the NL3 CEDF 
in Ref.\ \cite{Pingst-A30-60}. The detailed structure of the yrast spectra of
this nucleus has also been investigated in the cranked Nilsson-Strutinsky (CNS) 
approach in Ref.\ \cite{Pingst-A30-60}. Note that contrary to more microscopic 
studies which are limited to collective structures, this CNS study considers also
terminating/aligned states along the yrast line which is important for a proper
description of the yrast line at low and medium spins.

\begin{figure}[ht]
\includegraphics[angle=0,width=8.8cm]{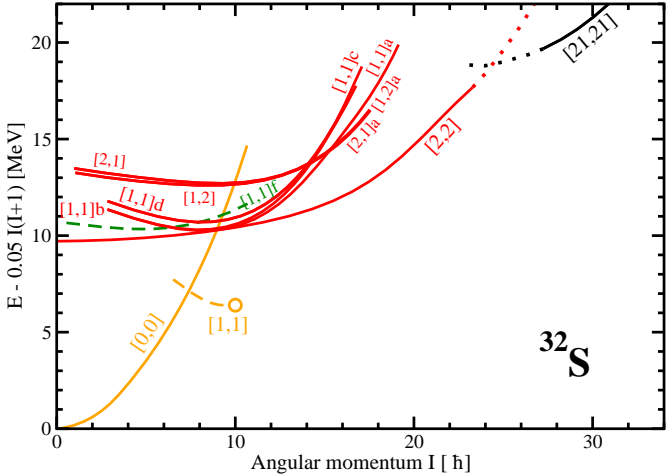}
\caption{ (Color online) The same as Fig.\ \ref{Ca40-eld} but for 
$^{32}$S. Dotted lines show expected diabatic continuations of the [2,2] 
and [21,21] configurations beyond the spin range where the convergence 
has been obtained.}
\label{S32-eld}
\end{figure}

  Fig.\ \ref{S32-eld} shows the high-spin structures in $^{32}$S. The lowest 
SD configuration with the structure [2,2] is yrast above spin $I=10\hbar$. 
The same result has also been obtained in other models quoted above. Above
spin $I=24\hbar$, the occupation of the lowest $N=4$ hyperintruder proton
and neutron orbitals leads to the HD [21,21] configuration. Note that this
induces an unpaired band crossing the consequence of which is the 
impossibility to trace in the calculations the SD band above $I\sim 23\hbar$ 
and HD band below $I\sim 27\hbar$. This problem could be avoided if diabatic 
orbitals would be built using an approach of Ref.\ \cite{PhysRep-SBT}; the expected 
diabatic continuations of the SD [2,2] and HD configurations [21,21] are shown 
by dotted lines in Fig.\ \ref{S32-eld}. Note that the CNS calculations of Ref.\ 
\cite{Pingst-A30-60} also suggest that the lowest HD configuration has the [21,21] 
structure and becomes yrast at similar spins. The same HD configuration has also 
been obtained in cranked Skyrme HF calculations of Ref.\ \cite{YM.00}; it also 
become yrast around $I\sim 25\hbar$ in the calculations with SIII and SkM* Skyrme 
forces.
 
 At spin $I=0$, the calculated $Q_t$ values for the [2,2] SD configuration
are 50\% larger than $Q_t^{norm}$ (Fig.\ \ref{S32-qt-gamma}). This would even 
allow to describe this band as HD. However, at this spin the [2,2] SD 
configuration is located around 10 MeV above the ground state which 
prevents its observation. The rotation and the limited angular momentum content 
in low deformation configurations brings this SD configuration to the yrast 
line. However, it also triggers the decrease of the collectivity (as measured 
by $Q_t$) so this configuration is more properly described as SD in the spin 
range where it is yrast. The occupation of the lowest $N=4$ proton and neutron orbitals 
leading to the [21,21] HD configuration triggers substantial increase of $Q_t$; 
at spin $I=31\hbar$ it is by 60\% larger than the $Q_t^{norm}$. Density distributions
of the [2,2] and [21,21] configurations at spins of interest are shown in Fig.\ 
\ref{density-S32}.
 Note that many [1,1] configurations are of transitional type; they are SD only 
at very low spins (Fig.\ \ref{S32-qt-gamma}) and are only highly-deformed at 
higher spins. Truly SD configurations are obtained with additional occupation of 
the $N=3$ orbital leading to either [2,1] or [1,2] configurations  (Fig.\ 
\ref{S32-qt-gamma}).

  Present calculations indicate large gap between the yrast SD [2,2] 
configuration and excited configurations in the spin range $I=16-22\hbar$ 
(Fig.\ \ref{S32-eld}). Although this would favor the population of
this configuration, all experimental attempts to observe this band
undertaken in the beginning of the last decade have failed.
  
\begin{figure}[ht]
\includegraphics[angle=0,width=8.8cm]{fig-28.eps}
\caption{(Color online) The same as Fig.\ \ref{Ca40-Qt-gamma} but for $^{32}$S.}
\label{S32-qt-gamma}
\end{figure}

\begin{figure*}[ht]
\centering
\includegraphics[angle=0,width=8.8cm]{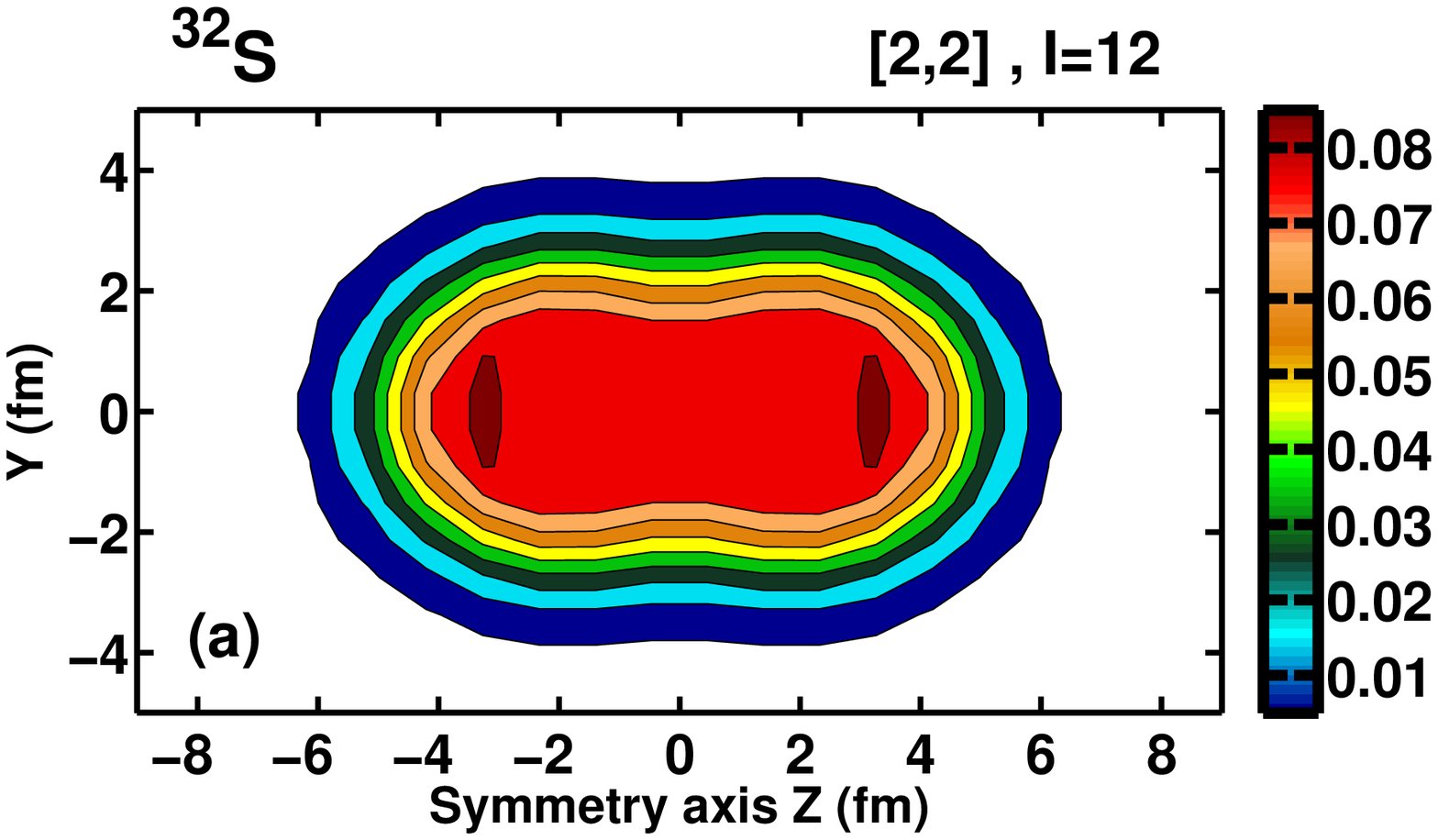}
\includegraphics[angle=0,width=8.8cm]{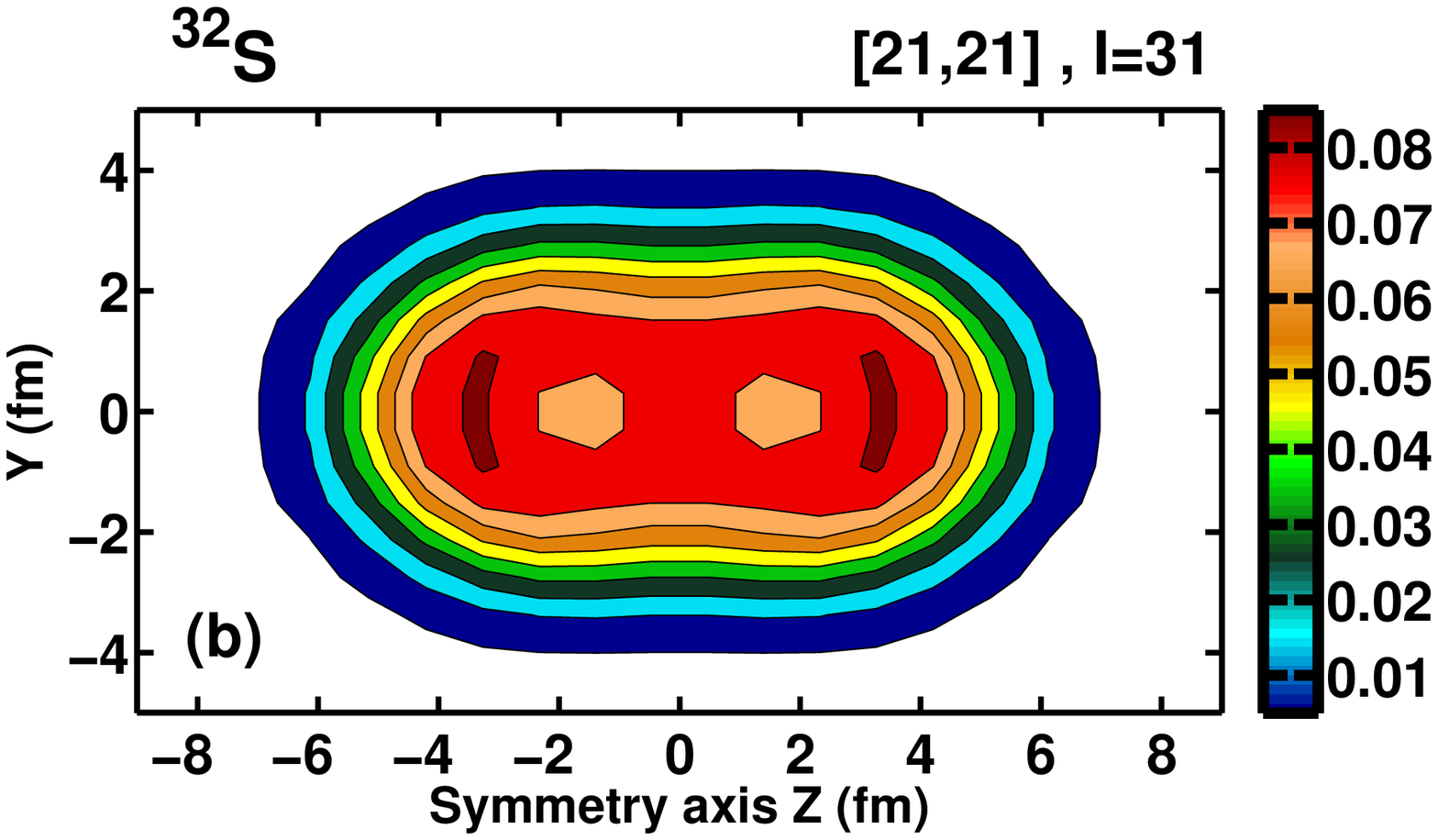}
\caption{(Color online) The same as Fig.\ \ref{density-Ca40} 
          but for $^{32}$S.
}
\label{density-S32}
\end{figure*}

\subsection{$^{34}$S nucleus}

\begin{figure}[ht]
\includegraphics[angle=0,width=8.8cm]{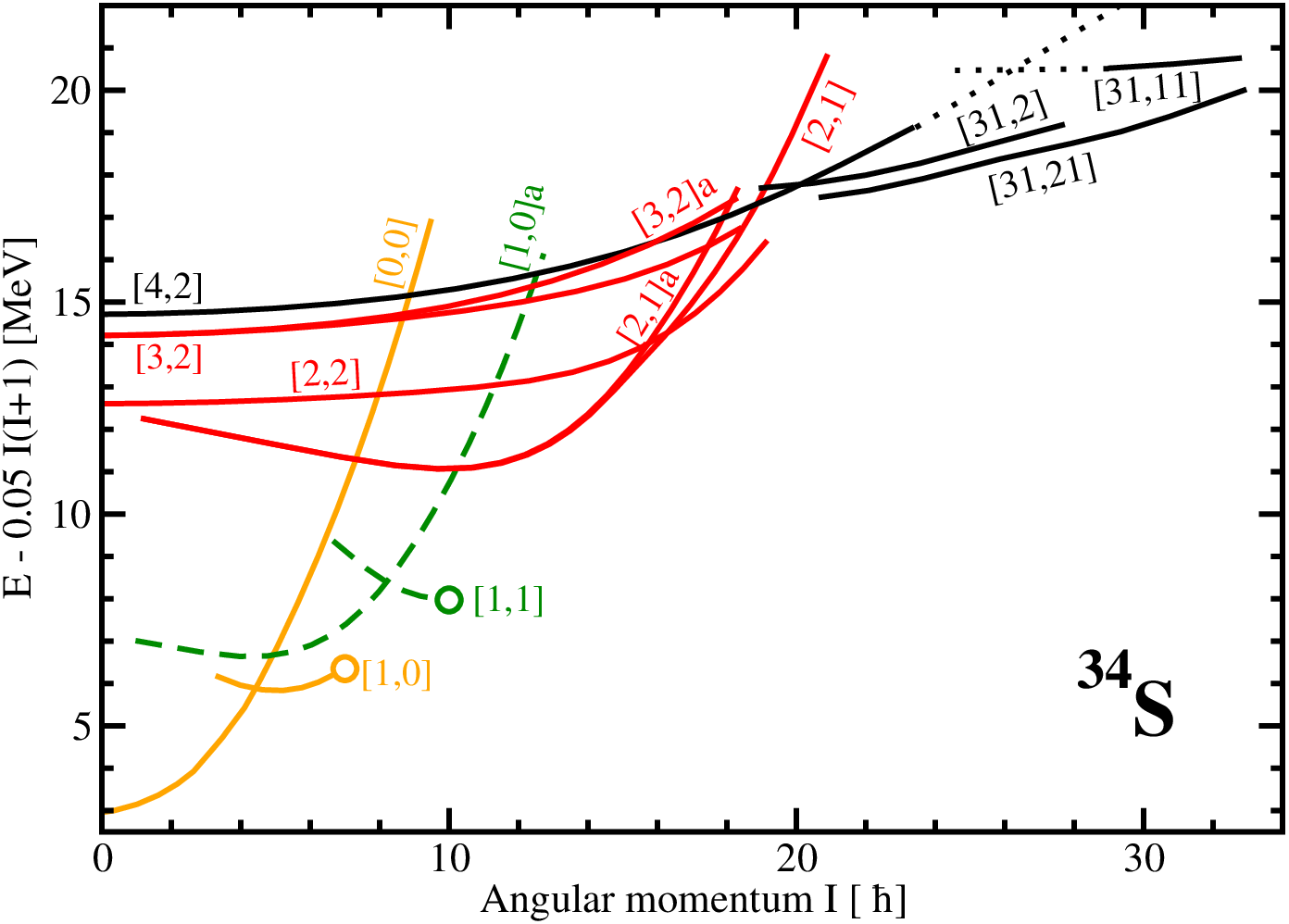}
\caption{(Color online) The same as Fig.\ \ref{Ca40-eld} but for $^{34}$S.
}
\label{S34-eld}
\end{figure}

  The configurations in the $^{34}$S nucleus are formed from the ones in $^{32}$S 
by adding two neutrons in respective orbitals. The [2,1] configurations in $^{34}$S 
are similar to the [1,1] ones in $^{32}$S; they are SD at low spin but lose the 
collectivity with increasing spin so that they are better described as highly-deformed 
at the highest calculated spins (Fig.\ \ref{S34-qt-gamma}).  Indeed, their density 
distributions at medium and high spins are characterized by rather modest semi-axis 
ratio; this is seen in Fig.\ 3b of the supplemental material (Ref.\ \cite{Sup-A40})
on the example of the 
[2,1] configuration which has semi-axis ratio of 1.32 (see Table \ref{table-ratio}) 
at $I=14\hbar$. Similar to $^{32}$S truly SD shapes are formed with the occupation 
of at least two $N=3$ protons and two $N=3$ neutrons. They are represented by the
[2,2] SD configuration (which is yrast above $I=16\hbar$ [see Fig.\ \ref{S34-eld}]) 
and by the excited [3,2] and [3,2]a SD configurations. The occupation of the lowest 
$N=4$ neutron and proton orbitals leads to the HD configurations [31,21] and 
[31,11]. The first configuration is yrast at spin $I=20\hbar$ and above (Fig.\ 
\ref{S34-eld}) and its proton density distribution is shown in Fig.\ 
\ref{density-S34}.
 
\begin{figure}[ht]
\includegraphics[angle=0,width=8.8cm]{fig-31.eps}
\caption{(Color online) The same as Fig.\ \ref{Ca40-Qt-gamma} but for $^{34}$S.}
\label{S34-qt-gamma}
\end{figure}

\begin{figure}[ht]
\includegraphics[angle=0,width=8.8cm]{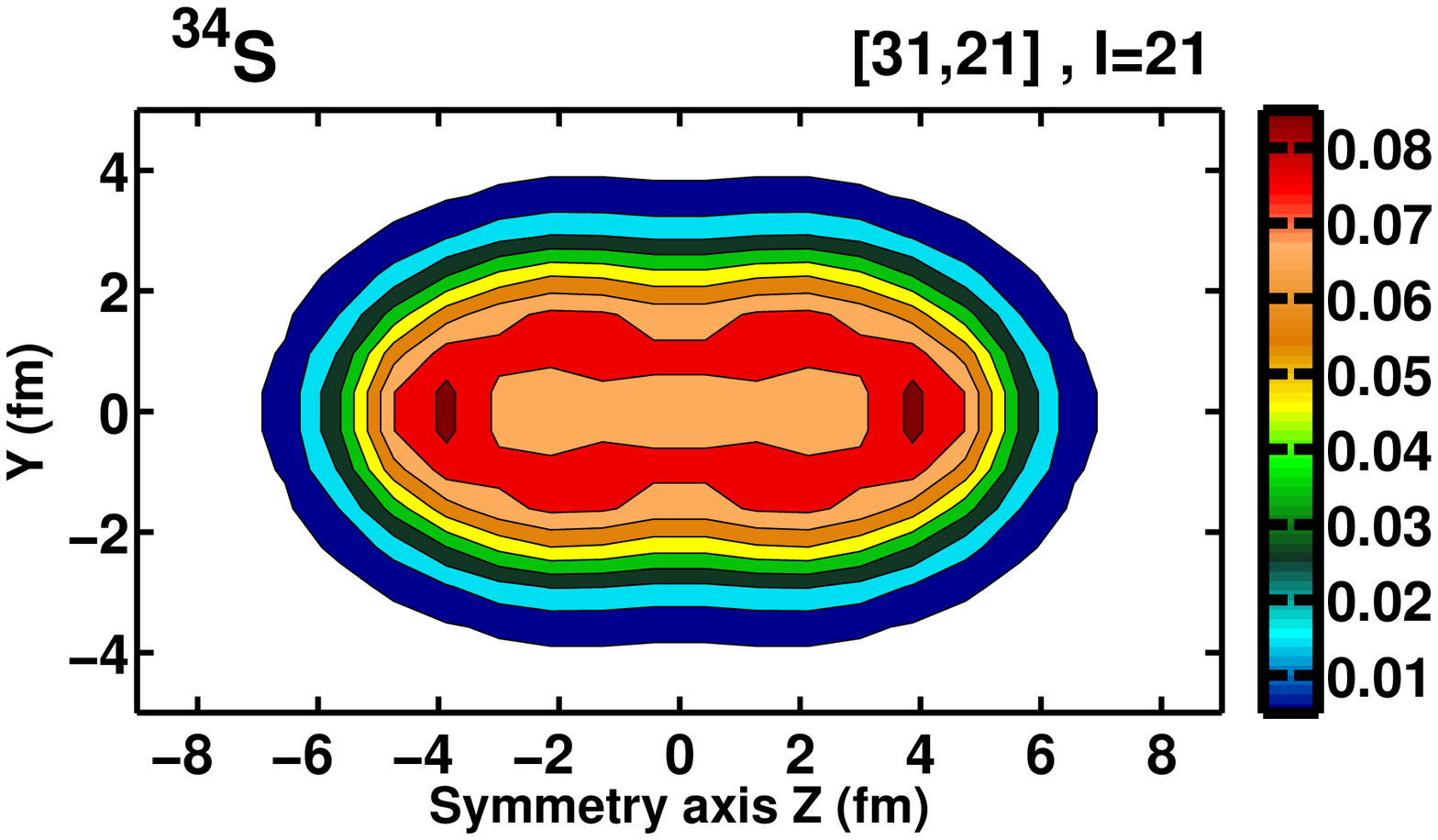}
\caption{(Color online) The same as Fig.\ \ref{density-Ca40} 
          but for $^{34}$S.
}
\label{density-S34}
\end{figure}

\section{Clusterization and molecular structures}
\label{sec-clus}

\begin{figure*}[ht]
\centering
\includegraphics[angle=0,width=8.8cm]{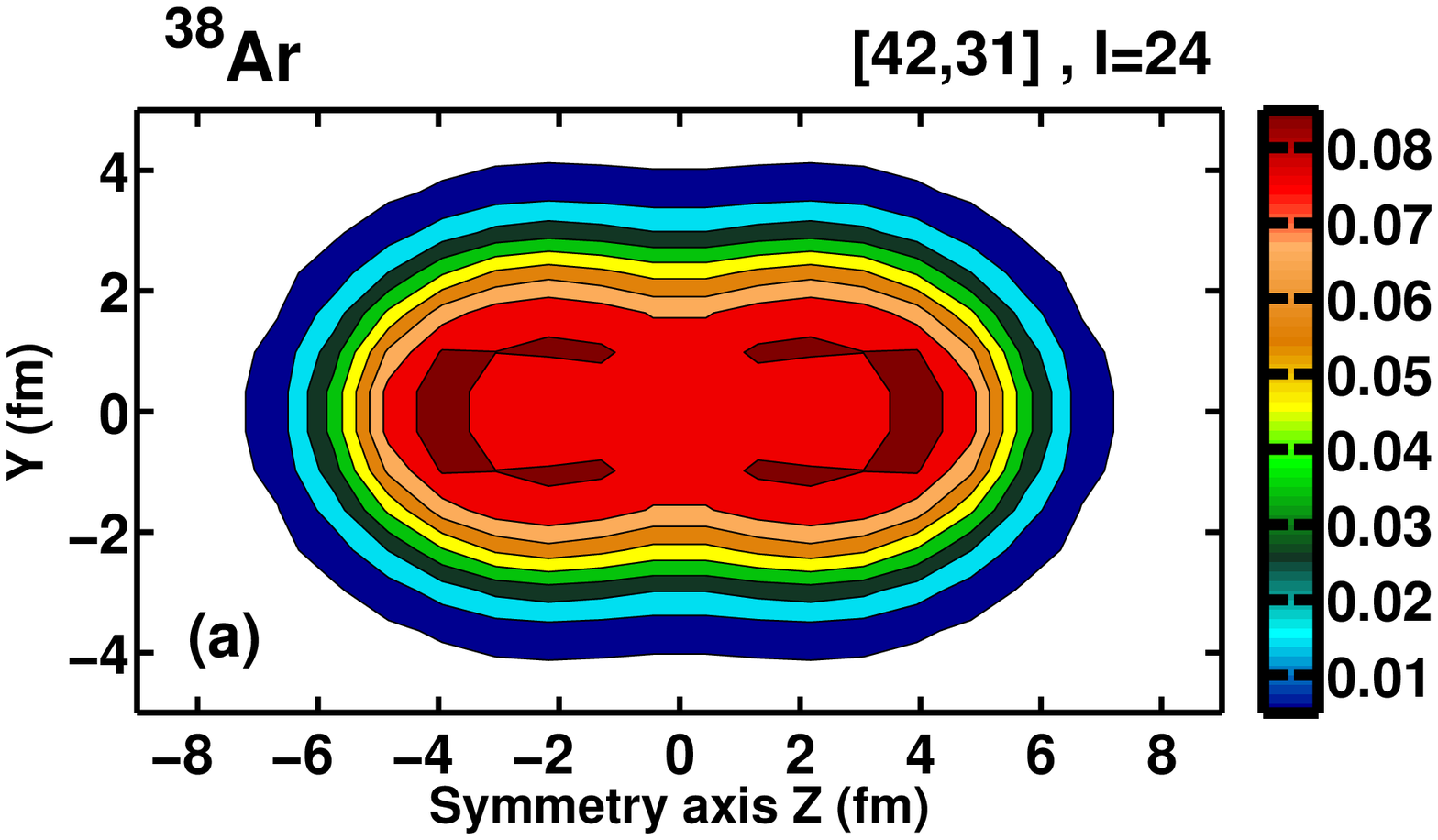}
\includegraphics[angle=0,width=8.8cm]{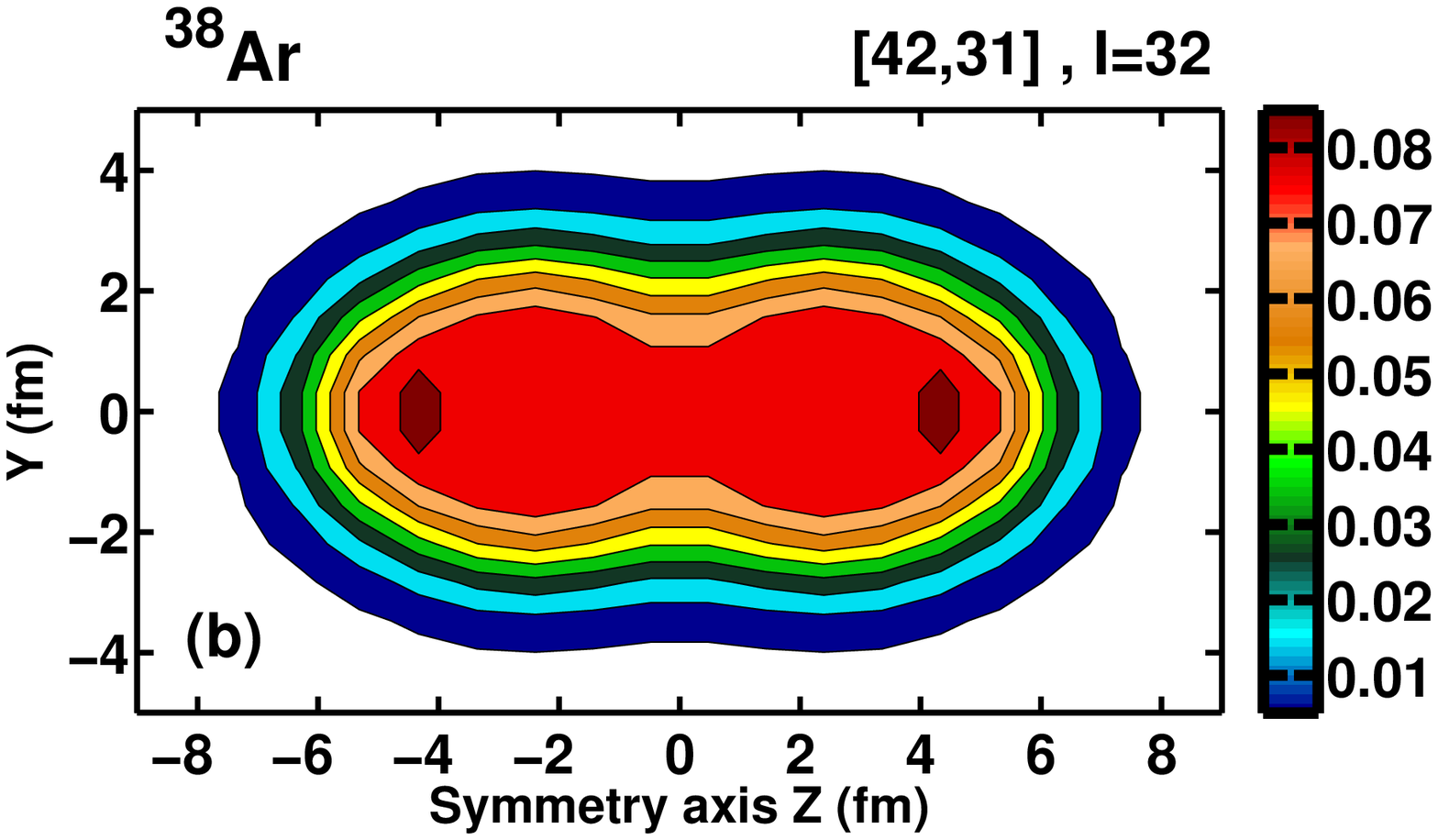}
\includegraphics[angle=0,width=8.8cm]{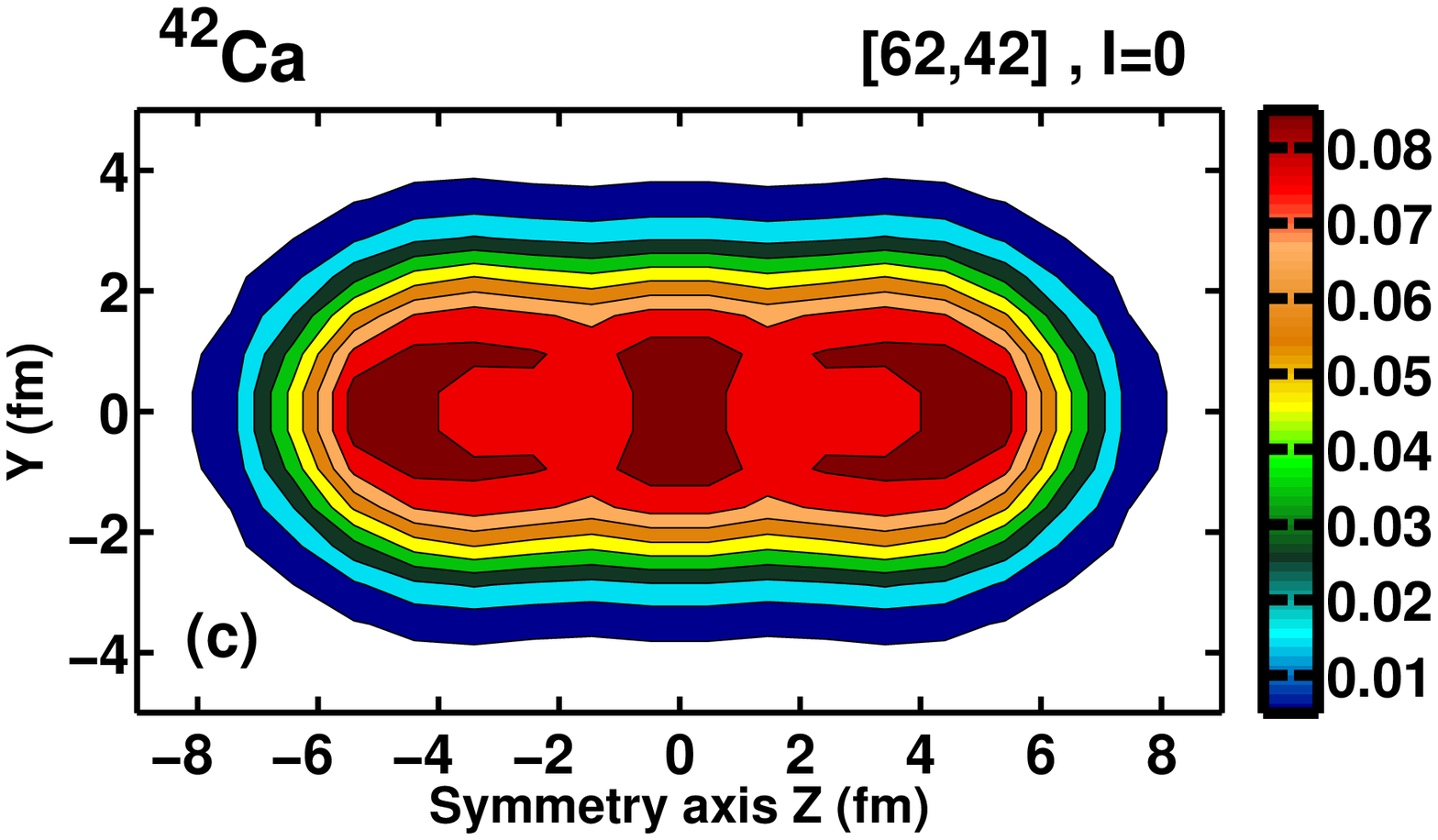}
\includegraphics[angle=0,width=8.8cm]{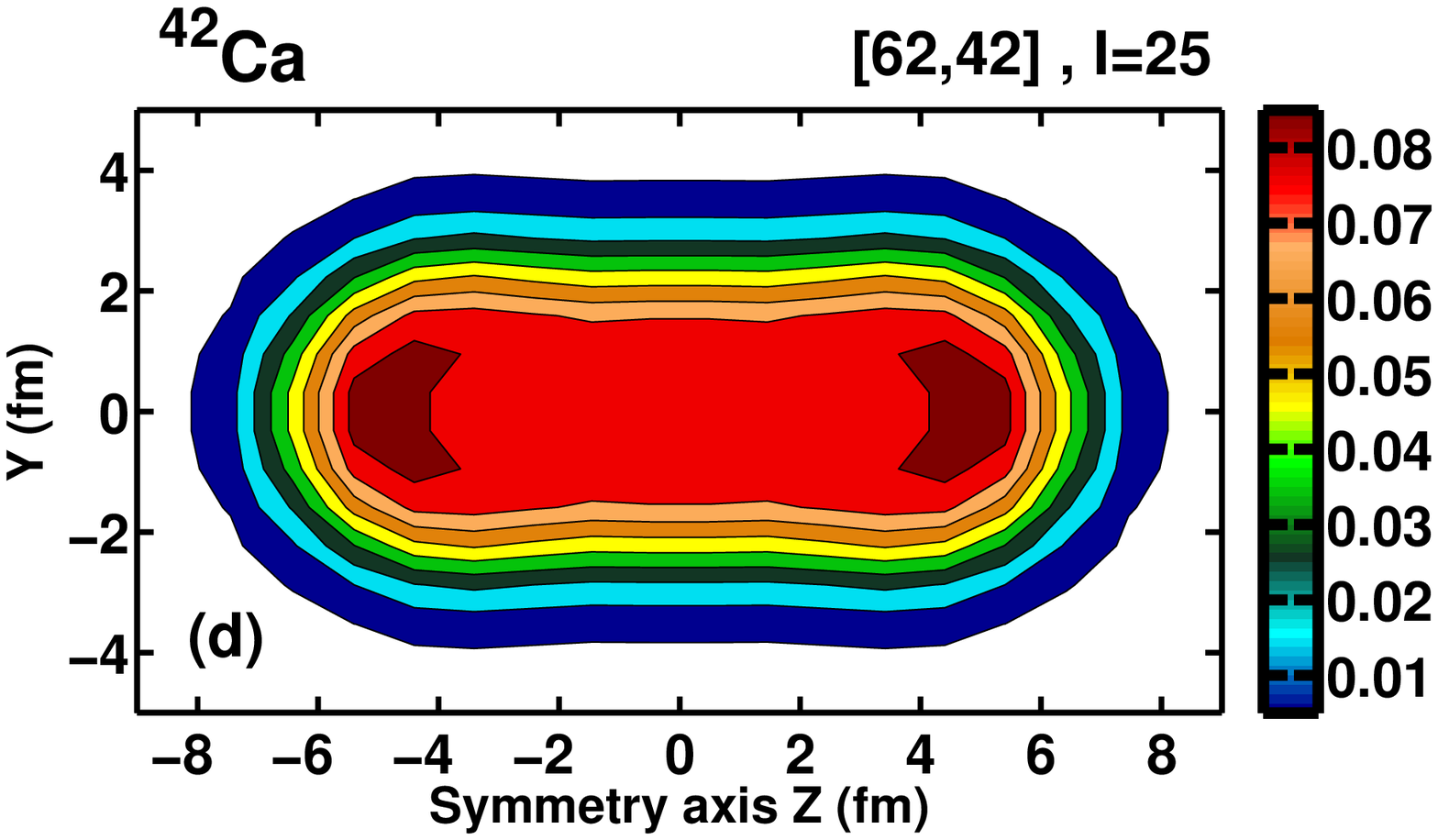}
\includegraphics[angle=0,width=8.8cm]{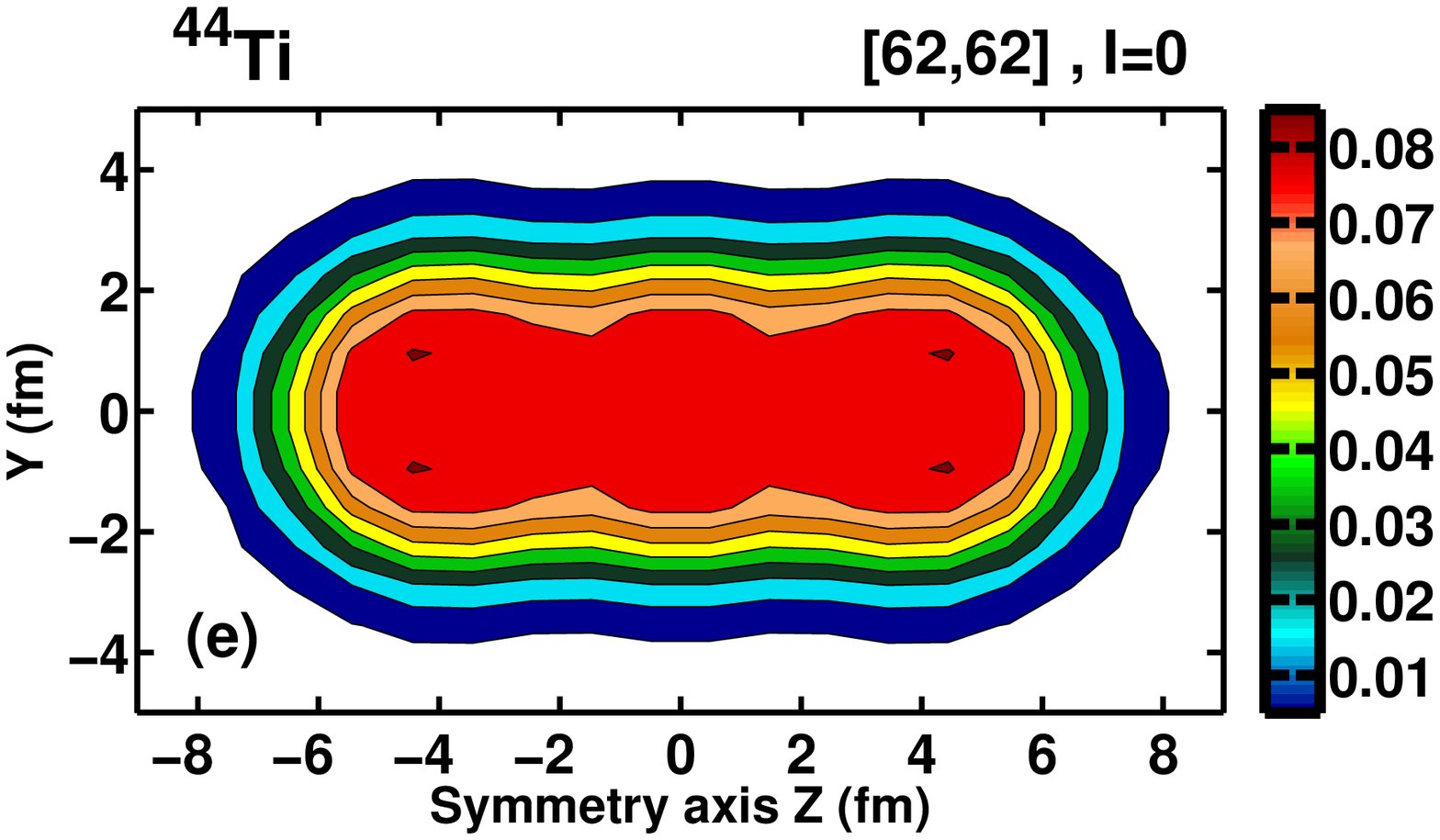}
\includegraphics[angle=0,width=8.8cm]{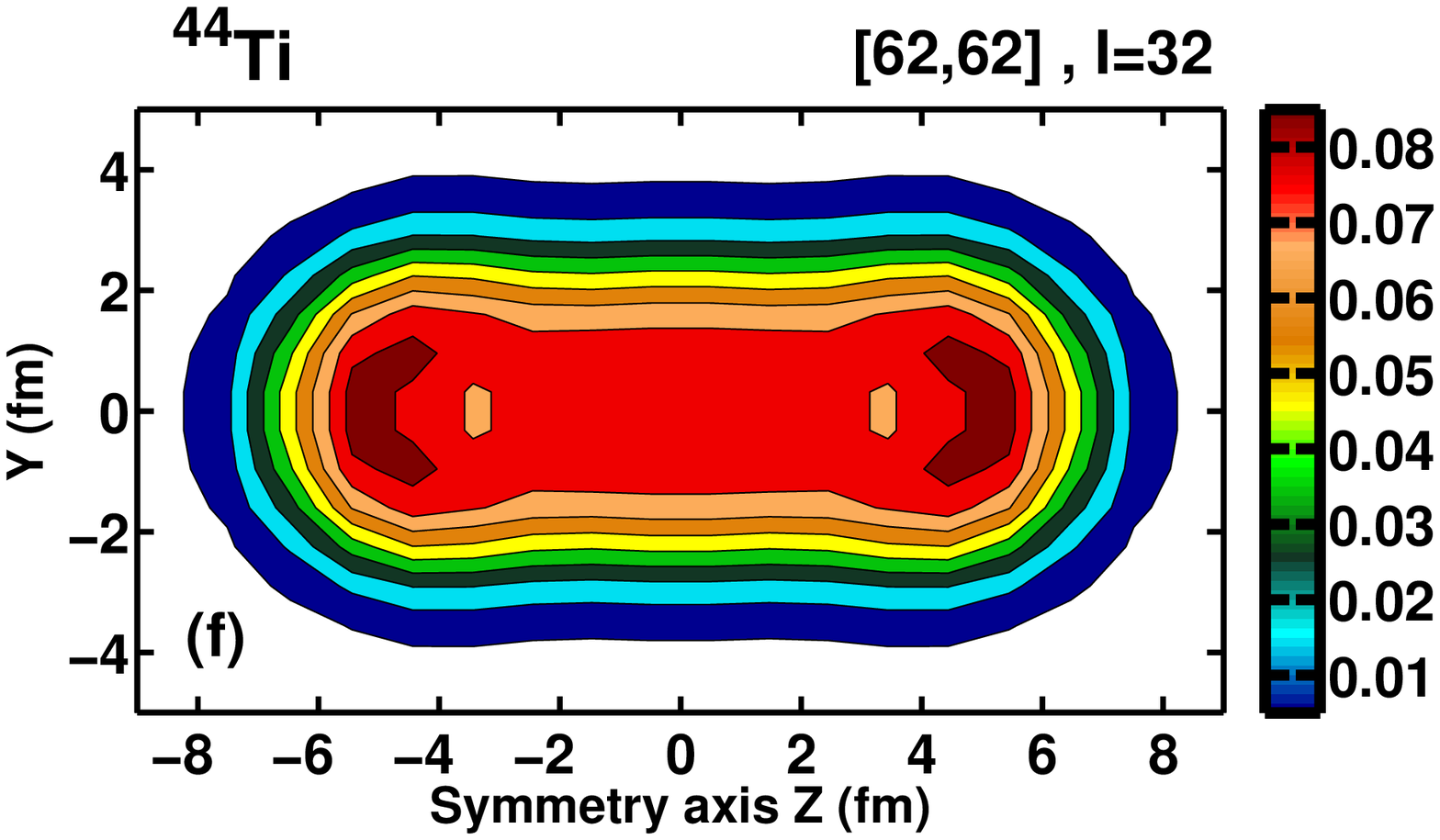}
\caption{(Color online) The impact of rotation on the proton 
         density distribution in selected configurations.}
\label{density-rot}
\end{figure*}

 One of the main goals of the present paper is the search 
for possible candidates showing clusterization and molecular 
structures in the near-yrast region of the nuclei under study. 
Different single-particle states have different spatial
density distributions which are dictated by their underlying 
nodal structure of their wavefunctions (see Fig.\ 9 in Ref.\ 
\cite{TO-rot}); the centers of the density distribution are 
found at the nodes and peaks of the oscillator eigenfunctions.
The total density distribution is built as a 
sum of the single-particle density distributions of occupied 
single-particle states. Thus, for some specific occupations of the 
single-particle states at some deformations one may expect the 
effects in the density distributions which could be interpreted in  
terms of clusterization and molecular structures. Note that the 
structure of  the wavefunctions of the single-particle orbitals 
is affected by rotation (see discussion in Sec.\ V of Ref.\ 
\cite{TO-rot}); this could lead to a modification of the
single-particle density distributions (see Fig.\ 9 in Ref.\ 
\cite{TO-rot}). For some of the orbitals the effect of rotation
on single-particle density distributions is quite substantial, 
while it has very little impact on the single-particle density 
distributions for others. This could lead either to a destruction 
or the emergence/enhancement of the clusterization and molecular 
structures with rotation.


 Well known case of molecular structure in this mass region is the superdeformed 
configuration [2,2] in $^{32}$S; according to Refs.\ \cite{MKKRHT.06,MH.04} the 
wavefunction of this band contains significant admixture of the molecular 
$^{16}$O+$^{16}$O structure.  Indeed, the development of neck is seen in its 
density distribution (Fig.\ \ref{density-S32}a). Similar neck exists also in 
the HD [21,21] configuration of $^{32}$S (Fig.\ \ref{density-S32}b) but the 
presence of  density depressions at $z \sim \pm 2$ fm may suggest more complicated 
structure than the pair of two $^{16}$O. In addition, the neck is also present in 
the density distribution of the [31,21] configuration in $^{34}$S but this 
configuration is characterized by an unusual density distribution with density 
depression in the highly elongated central region which is surrounded by the region 
of maximum density (Fig.\ \ref{density-S34}). Note that the SD configurations, which 
have the structure of $^{16}$O+$^{16}$O+two valence neutrons  in molecular orbitals, 
have recently been predicted in the AMD+GCM calculations of Ref.\ 
\cite{S-34-mol.14}.

  Present study reveals also a number of other interesting molecular 
structures which are discussed below. We were able to trace 
some of such configurations in an extended spin range starting from 
spin zero (or from very low spin), at which they are located at $20-30$ 
MeV excitation energy above the ground state, up to very high spin
where they are either yrast or close to the yrast line. These 
are the [42,31] and [42,22] MD configurations in $^{38}$Ar (Fig.\ 
\ref{Ar38-eld}), the [31,31] and [41,41] MD configurations in $^{36}$Ar
(Fig.\ \ref{Ar36-eld}), the [42,42] MD configuration in $^{40}$Ca
(Fig.\ \ref{Ca40-eld}), the [62,42] MD configuration in $^{42}$Ca
(Fig.\ \ref{Ca42-eld}),  the [62,62] MD configurations in $^{44}$Ti
(Fig.\ \ref{Ti44-eld}) and [52,52] and [421,421] MD configurations 
in $^{42}$Sc (Fig.\ \ref{Sc42-eld}). These examples allowed us to study 
the impact of rotation on clusterization.

  The molecular structures become well pronounced in the [31,31] and 
[41,41] MD configurations of $^{36}$Ar (Fig.\ \ref{density-Ar36}c and d) 
which are characterized by well established neck.  They are also seen in 
the [42,31] configuration of $^{38}$Ar (Figs.\ \ref{density-rot}a and b); 
note that in this case the rotation increases the separation of the 
fragments and makes the neck much more pronounced.

\begin{figure}[ht]
\includegraphics[angle=0,width=8.8cm]{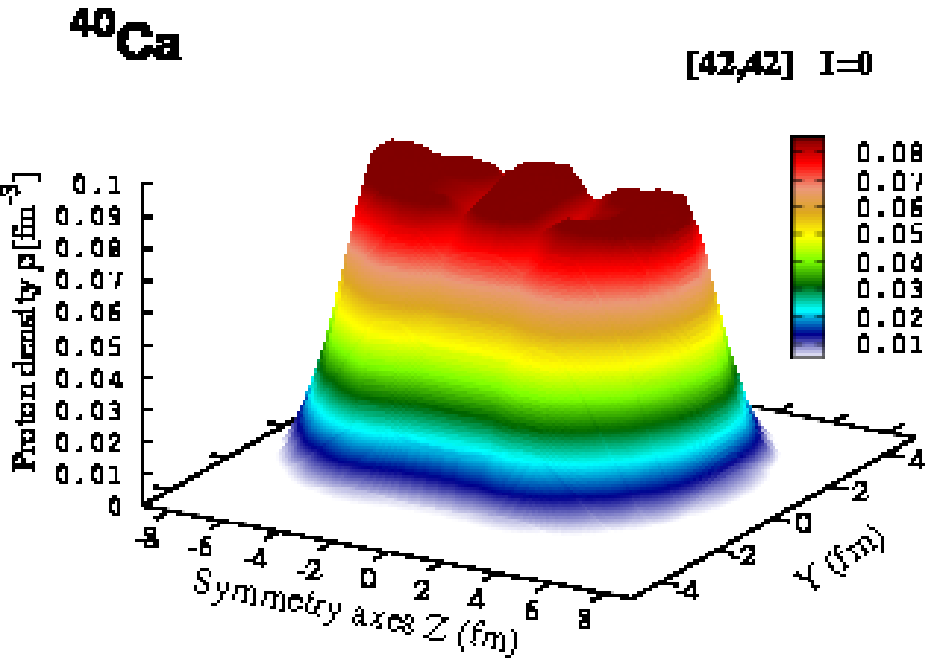}
\includegraphics[angle=0,width=8.8cm]{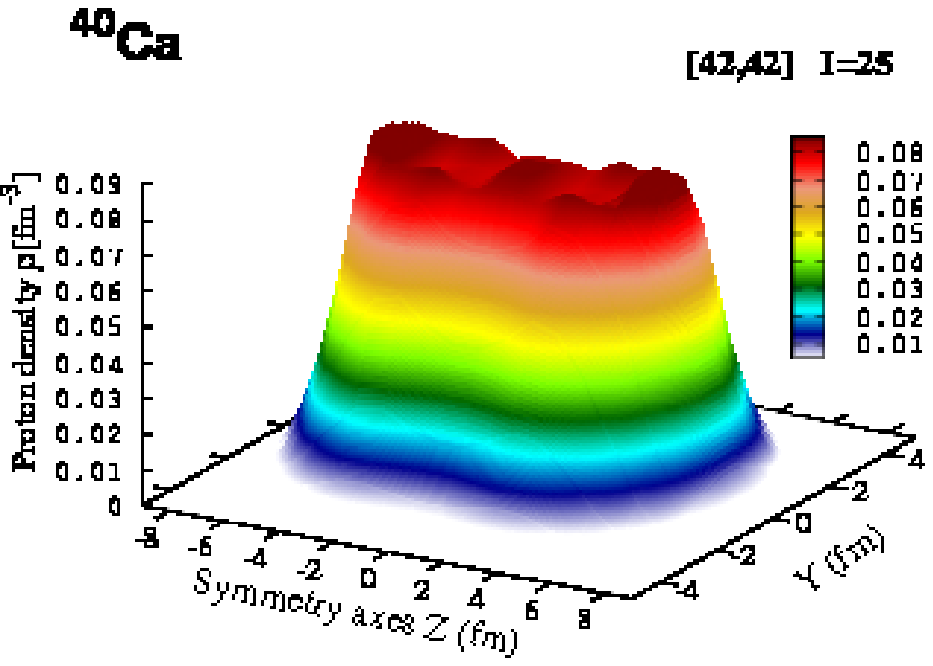}
\caption{(Color online) Three-dimensional representation
of the evolution or proton density distribution with spin
in the [42,42] configuration of $^{40}$Ca.}
\label{40ca-rot-den}
\end{figure}

  In $^{40}$Ca, the density distribution of the MD [42,42] configuration at 
spin zero shows a triple-humped structure (top panel of Fig.\ \ref{40ca-rot-den}). 
Similar configuration has been analyzed in Ref.\ \cite{MKKRHT.06} and it was 
concluded that $\alpha$-cluster interpretation becomes quite fuzzy. Alternatively, 
one may consider this configuration as a $^{12}$C+$^{16}$O+$^{12}$C chain built of 
distorted $^{16}$O and $^{12}$C nuclei. The validity of such interpretation should 
be verified in future by comparison with the results of the cluster and/or 
antisymmetrized molecular (AMD) calculations similar to the ones presented in 
Ref.\ \cite{MKKRHT.06}. The comparison of the density distributions for this 
configurations at  $I=0\hbar$ and $I=25\hbar$ (see Fig.\ \ref{40ca-rot-den})
shows that the rotation hinders the tendency for clusterization. Indeed, the 
central hump becomes less pronounced and the depressions in the density 
distributions develop in central parts of the left and right segments at 
$I=25\hbar$ (bottom panel of Fig.\ \ref{40ca-rot-den}).

 Similar effects are also seen in the MD [42,22] configuration of $^{38}$Ar 
(see Fig.\ 7 in the supplemental material (Ref.\ \cite{Sup-A40}) which 
could be considered as the 
MD [42,42] configuration of $^{40}$Ca with two proton holes in the $N=3$ 
orbitals. The addition of two neutrons to the MD [42,42] configuration 
of $^{40}$Ca creates the MD [62,42] configuration in $^{42}$Ca which has 
the features in the proton density distribution (see Figs.\ 
\ref{density-rot}c and d) similar to the ones seen in Fig.\ \ref{40ca-rot-den}.

  These results show that in few configurations discussed above
the rotation tries to suppress the features of the density distribution 
which could be attributed to the clusterization. However, the density 
modifications induced by rotation definitely depend on the nucleonic 
configuration.  For example, the density distribution of the [62,62] 
configuration in $^{48}$Cr is modified only modestly by rotation 
(see Figs.\ 2f and g in the supplemental material (Ref.\ \cite{Sup-A40}).
Note that this configuration does not show the 
features typical for  clusterization.  On the other hand, with 
increasing spin the separation of the fragments becomes larger 
and the neck becomes more pronounced in the [42,31] configuration 
of $^{38}$Ar  (Figs.\ \ref{density-rot}a and b) and the 
[421,421] configuration of $^{42}$Sc (Fig.\ \ref{density-42Sc}c 
and d).

  Another interesting case of possible clusterization is the [62,62] MD 
configuration in $^{44}$Ti (Fig.\ \ref{density-rot}e and f). Three fragments 
are clearly seen in the density distribution at $I=0\hbar$ indicating possible 
$^{16}$O+$^{12}$C+$^{16}$O chain of nuclei. Note that with rotation the central 
fragment dissolves but two outer segments became slightly more pronounced. It 
is interesting that similar three fragments structure survives in the [52,52] 
MD configuration up to very high spins in $^{42}$Sc (Fig.\ \ref{density-42Sc}b). 
This configuration could be considered as built from the [62,62] one in $^{44}$Ti 
by creating proton and neutron holes in the $N=3$ orbital.

  Very interesting case of molecular structures is seen on the example of 
the [421,421] MD configuration in $^{42}$Sc (Figs.\ \ref{density-42Sc}c and 
d and  and Fig. 1a in the supplemental material (Ref.\ \cite{Sup-A40}). 
This system could probably be described as a combination of two prolate 
deformed $^{20}$Ne cores located in tip-to-tip arrangement with extra proton 
and neutron.

  It is necessary to understand that suggested interpretations of molecular
structures are based on the consideration of only density distributions. 
Their validity should be verified in future by the analysis of the 
wavefunctions of the underlying configurations (and their overlaps with mean 
field solutions) obtained in the cluster and/or antisymmetrized molecular dynamics
calculations similar to the ones presented in Refs.\ \cite{MH.04,MKKRHT.06}.

\section{Rotational properties of extremely deformed configurations}
\label{sec-j2j1}

 Most important physical observables characterizing the SD, HD and
MD structures are kinematic ($J^{(1)}$) and dynamic ($J^{(2)}$) moments of 
inertia and transition quadrupole moments $Q_t$. The latter provides
a direct information on the deformation of the charge distributions and that
was a reason why the calculated $Q_t$ values were presented earlier. It is 
however necessary to recognize that previous history of the experimental 
investigation of the SD bands clearly shows that the $Q_t$ quantity is 
measured in dedicated experiment and thus it is available only for small 
fraction of the SD bands.

 Thus it is expected that in future experiments it will be easier to 
obtain the information on rotational properties of the bands which are 
described in terms of kinematic and dynamic moments of inertia using the 
following expressions
\begin{eqnarray}
J^{(1)}(\Omega_x) & = & J \left( \frac{dE}{dJ} \right)^{-1}=\frac{J}{\Omega_x}, \\
J^{(2)}(\Omega_x) & = & J \left( \frac{d^2E}{dJ^2} \right)^{-1} = \frac{dJ}{d\Omega_x},
\end{eqnarray}
where
\begin{equation}
\Omega_x=\frac{dE}{dJ},
\end{equation}
defines the rotational frequency  and $E$ and $J$ are total energy and the
expectation value of total angular momentum on the axis of rotation,
respectively. Their experimental counterparts are extracted from the observed 
energies of  the $\gamma$-transitions within a band according to  the prescription given 
in Sec. 4.1 of Ref.\ \cite{PhysRep-SBT}. Note that the kinematic moment 
of inertia depends on the absolute values of spins, while only the differences 
$\Delta I=2$ enter the definition of dynamic moment of inertia.

  The SD bands observed in the $A\sim 40$ mass region are exception from 
the general rule that the SD bands are very seldom linked to the low-spin 
level scheme. Thus, contrary to absolute majority of the SD bands in 
nuclear chart their spins are known; it is quite likely that some SD 
bands which will be observed in this mass region in future will follow 
this pattern. On the contrary, it is expected that the spins of the HD 
and MD bands will be difficult to define in future experiments. For such
bands, only dynamic moment of inertia will be available for comparison
with the results of calculations.

\begin{figure*}[ht]
\includegraphics[angle=0,width=17.6cm]{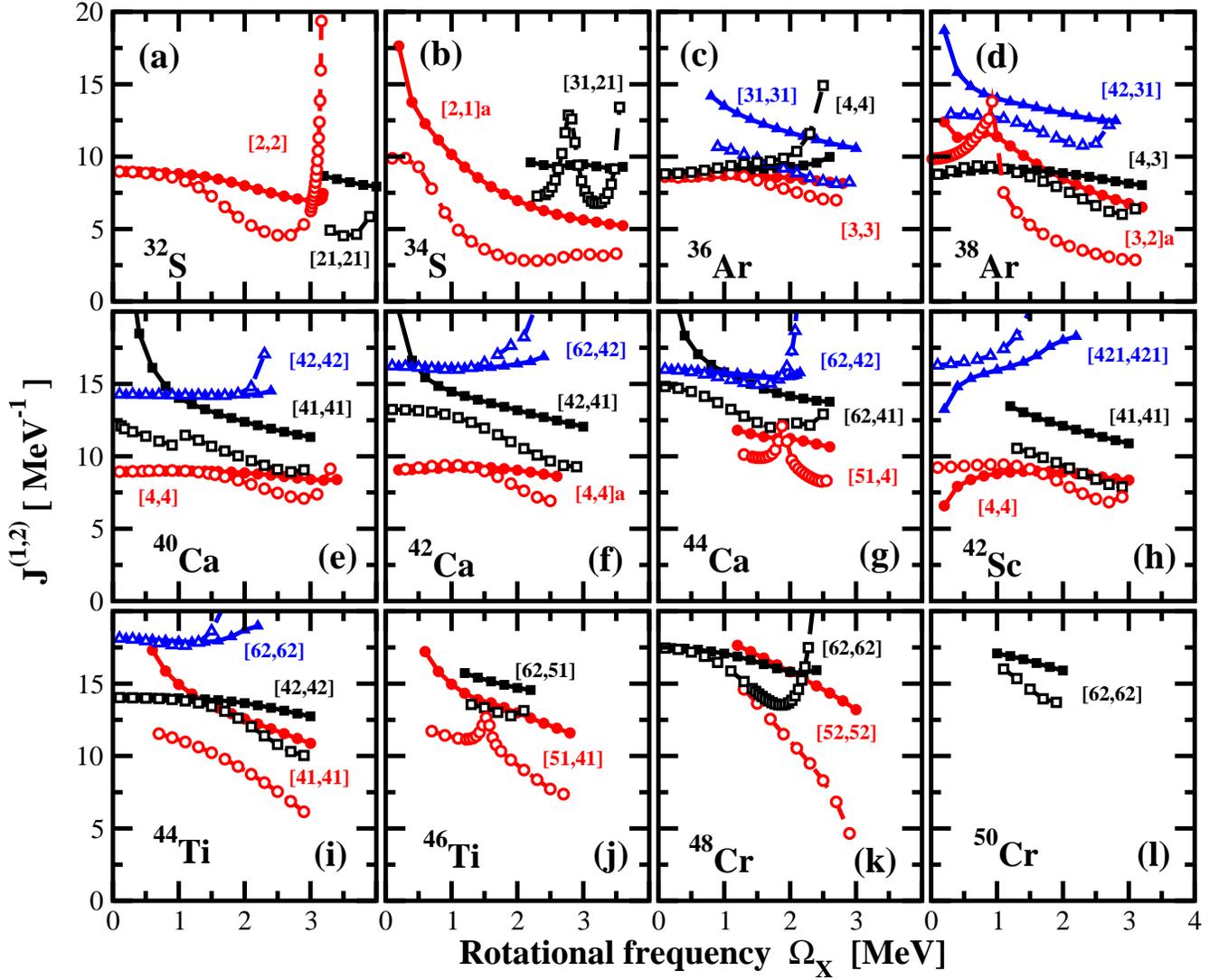}
\vspace{0.2cm}
\caption{ (Color online) Kinematic ($J^{(1)}$) and dynamic ($J^{(2)}$) moments 
of inertia of typical SD, HD and MD configurations in indicated nuclei.
The calculated $J^{(1)}$ and $J^{(2)}$ values are shown by solid and open
symbols, respectively. Red circles, black squares and blue triangles are 
used for the SD, HD and MD configurations, respectively.}
\label{j2j1-sys}
\end{figure*}

  The kinematic and dynamic moments of inertia of the (typically lowest in energy) 
SD, HD and MD bands are presented in Fig.\ \ref{j2j1-sys} for each nucleus under 
study. For majority of the SD and HD bands one observes that the following condition 
$J^{(1)} \geq J^{(2)}$ is satisfied at medium and high frequencies.
As discussed in Ref.\ \cite{PhysRep-SBT} this condition
is valid for the rotational bands in unpaired regime. This condition is not 
valid in the region of unpaired band crossing with weak interaction
where $J^{(2)}$ grows rapidly
with increasing rotational frequency. This takes place at the highest calculated
frequencies in the [2,2] SD configuration of $^{32}$S (Fig.\ \ref{j2j1-sys}a),
[4,4] SD configuration in $^{40}$Ca (Fig.\ \ref{j2j1-sys}e), and [62,62] HD
configuration in $^{48}$Cr (Fig.\ \ref{j2j1-sys}k). Note also that such situation
is seen at medium spin in the [31,21] HD configuration of $^{34}$S  (Fig.\ 
\ref{j2j1-sys}b) and the [51,4] SD configuration of $^{44}$Ca (Fig.\ 
\ref{j2j1-sys}g).

 The moments of inertia of the MD bands show three different patterns of behavior. 
Some of the MD bands 
undergo a centrifugal stretching that result in an increase of the transition 
quadrupole moments $Q_t$ with increasing rotational frequency. This process 
also reveals itself in the moments of inertia: the kinematic moments of inertia 
are either nearly constant or slightly increase with increasing rotational 
frequency, whereas the dynamic moments of inertia show two patterns of behavior.
In one of them the dynamic moment of inertia is almost the same as kinematic one
at low to medium rotational frequencies  but then $J^{(2)}$ becomes bigger than
$J^{(1)}$ and the difference between them gradually increases with frequency.
These are the MD configurations shown in Figs.\ \ref{j2j1-sys}e, f, g and i. The 
pattern of the behavior of the [421,421] MD configuration in $^{42}$Sc is very 
different (Fig.\ \ref{j2j1-sys}h); both moments increase with increasing rotational 
frequency but $J^{(2)} \geq J^{(1)}$ at all calculated frequencies. Note that this 
configuration has most elongated density distribution among studied in the present 
paper with clear indication of molecular structure (see Sec.\ \ref{sec-ca40}.)
The rotational properties of above discussed MD bands are very similar to the HD 
ones in the $Z=40-58$ mass region investigated in Ref.\ \cite{AA.08,AA-HD.09}. On the 
other hand, the [31,31] MD configuration in $^{36}$Ar (Fig.\ \ref{j2j1-sys}c) and 
[42,31] MD configuration in $^{38}$Ar (Fig.\ \ref{j2j1-sys}d) show the relative 
properties of the two moments similar to the ones seen in the  majority of the SD 
and HD bands shown in Fig.\ \ref{j2j1-sys}.

 The examples shown in Fig.\ \ref{j2j1-sys} clearly indicate strong dependence
of the calculated $J^{(1)}$ and $J^{(2)}$ values on the nucleonic configuration and 
frequency. In most of the cases, at medium and high rotational frequencies there 
is a correlation between the moments of inertia and deformation so that the moments 
of inertia increase with increasing deformation. However, there are exceptions from 
this observation. For example, the dynamic moments of inertia of the [4,4] SD and
[41,41] HD configurations in $^{42}$Sc are quite similar (see Fig.\ \ref{j2j1-sys}h)
despite substantial difference in the transition quadrupole moments (see Fig.\ 
\ref{Sc42-beta-gamma}a). Even more striking example is the similarity of dynamic 
moments of inertia of the [3,3] SD and [31,31] MD configurations in $^{36}$Ar (Fig.\ 
\ref{j2j1-sys}c). Such similarities are also seen for the
kinematic moments of inertia as illustrated by the case of the [52,52] SD and
[62,62] HD bands in $^{48}$Cr (Fig.\ \ref{j2j1-sys}k).  Thus, the decision on the nature of the band
(SD, HD or MD) observed in experiment cannot be based solely on the measured
values of dynamic or kinematic moments of inertia; only the measurement of
the transition quadrupole moment can reveal the true nature of the band.

\section{Conclusions}
\label{concl}

 A systematic search for extremely deformed structures in the 
$N\sim Z$ $A\sim 40$ nuclei has been performed for the first time in 
the framework of covariant density functional theory. The aim of 
this study was to define at which spins such structures become yrast, 
their properties and to find the configurations showing the fingerprints 
of clusterization and molecular structures. The main results 
can be summarized as follows.

\begin{itemize}

\item
  Present investigation shows that extremely deformed structures
inevitably become yrast with increasing spin in the nuclei under
study. This is because normal and highly-deformed configurations
forming the yrast line at low and medium spins have limited angular 
momentum content. The key question is at which spin the transition 
from  terminating to extremely deformed configurations takes place.
This is basically defined by the maximum spin which could be
built in terminating configurations with limited number of 
particle-hole excitations across the respective spherical shell 
gaps. This spin is quite limited for particle-hole excitations
across the proton $Z=20$ and neutron $N=20$ spherical shell
gaps. As a result, the nuclei most favored for the observation 
of extremely deformed structures are located in the vicinity of 
$^{36}$Ar and $^{40}$Ca. For example, present calculations suggest
that in $^{36}$Ar the increase of spin above measured  $I=16\hbar$
state  is only possible by the population of the hyperdeformed band. 
On the contrary, the configurations built on particle-hole excitations 
across the spherical $N=28$ and $N=28$ gaps, which bring substantial 
amount of angular momentum, dominate the yrast line at medium spin  
(up to $I\sim 30\hbar$) in the Cr nuclei.   As a result, only at 
higher spins extremely deformed configurations become yrast. 

\item
Similar to previous studies in the medium mass nuclei \cite{AA.08,AA-HD.09},  
present  calculations indicate that the $N=Z$ nuclei are better candidates 
for the observation of extremely deformed structures as compared with the 
nuclei which have an excess of neutrons over protons since the transition 
to extremely deformed structures takes place at lower spins.

\item
  The above discussed consideration of the most favored candidates
for experimental observation of extremely deformed structures is based 
on the spins at which they become yrast in model calculations. However, 
it is expected that experimental observations will also depend on
employed combination(s) of the target(s) and projectile(s) and 
respective cross-sections of the reactions. Taking this factor and 
related uncertainties into account the $N=Z$ and $N=Z+2$ S, Ar, Ca, and 
Ti isotopes should be considered as good candidates for experimental 
observation of extremely deformed structures. On the other hand, the 
experimental observation of such structures in the $^{48,50}$Cr isotopes 
is clearly disfavored by the present analysis as compared with above 
mentioned $N=Z$ and $N=Z+2$ isotopes.

\item
 The underlying single-particle structure of nucleonic configurations
with specific nodal structure of the single-particle density distribution 
leads to a clusterization in the form of molecular structures. The
calculations suggest that in some nuclei such structures are either
yrast or close to yrast at high spin. Thus, their observation with 
new generation of $\gamma$-tracking detectors such as GRETA and AGATA 
may be possible in near future. The calculations with cluster or/and 
antisymmetrized molecular dynamics models are definitely needed in order
to establish the weights of those clusters in the structure of total
wavefunction.

\item
  The impact of rotation on the density distribution and clusterization
(molecular nature) depends sensitively on nucleonic configuration. The 
density distributions of some configurations are only weakly affected by 
rotation. The features typical for clusterization, which are present at 
zero spin, are washed away by rotation in other configurations. On the 
other hand, the clusterization is enhanced by rotation in some specific 
configurations; with increasing spin the separation of the fragments 
becomes larger and the neck becomes more pronounced.

\item
  There is a strong dependence of the calculated kinematic and dynamic
moments of inertia on the configuration and frequency. In most of the 
cases the moments of inertia increase with increasing deformation at medium 
and high rotational frequencies. However, there are exceptions from this 
observation. As a result, the decision on the nature of the band (SD, HD or 
MD) observed in experiment cannot be based solely on the measured values of 
dynamic or kinematic moments of inertia; only the measurement of transition 
quadrupole moments will reveal the true nature of the band.

\end{itemize}

\section{Acknowledgements}

  This material is based upon work supported by the U.S. 
Department of Energy, Office of Science, Office of Nuclear 
Physics under Award Number DE-SC0013037.

\bibliography{references14}

\end{document}